\documentclass[final,5p,number]{elsarticle}

\usepackage{amssymb}
\usepackage{amsmath}

\usepackage{graphicx}
\PassOptionsToPackage{hyphens}{url}
\usepackage[hidelinks]{hyperref}
\usepackage{graphicx}

\usepackage{caption}
\usepackage{subcaption}
\usepackage[export]{adjustbox}

\usepackage{stix,bbding,pifont,utfsym,fontawesome}

\usepackage{tabularray}
\usepackage{tabularx}
\usepackage{lscape} 

\usepackage{pifont}
\newcommand{\xmark}{\ding{55}}%

\usepackage{acro}[=v2]
\DeclareAcronym{ML}{short=ML, long=machine learning}
\DeclareAcronym{FTM}{short=FTM, long=fine timing measurement}
\DeclareAcronym{ToA}{short=ToA, long=time of arrival}
\DeclareAcronym{ToT}{short=ToT, long=time of transmission}
\DeclareAcronym{ToD}{short=ToD, long=time of departure}
\DeclareAcronym{TDoA}{short=TDoA, long=time difference of arrival}
\DeclareAcronym{DoA}{short=DoA, long=direction of arrival}
\DeclareAcronym{ToF}{short=ToF, long=time of flight}
\DeclareAcronym{RTT}{short=RTT, long=round-trip time}
\DeclareAcronym{RTP}{short=RTP, long=round-trip phase}
\DeclareAcronym{GPS}{short=GPS, long=Global Positioning System}
\DeclareAcronym{GPR}{short=GPR, long=Gaussian process regression}
\DeclareAcronym{RSSI}{short=RSSI, long=received signal strength indicator}
\DeclareAcronym{RSS}{short=RSS, long=received signal strength}
\DeclareAcronym{LOS}{short=LOS, long=line of sight}
\DeclareAcronym{NLOS}{short=NLOS, long=non-line of sight}
\DeclareAcronym{RF}{short=RF, long=random forest}
\DeclareAcronym{LS-SVM}{short=LS-SVM, long=least-squares support vector machine}
\DeclareAcronym{LS}{short=LS, long=least-squares}
\DeclareAcronym{SVM}{short=SVM, long=support vector machine}
\DeclareAcronym{NN}{short=NN, long=neural network}
\DeclareAcronym{ANN}{short=ANN, long=artificial neural network}
\DeclareAcronym{CNN}{short=CNN, long=convolutional neural network}
\DeclareAcronym{RNN}{short=RNN, long=recurrent neural network}
\DeclareAcronym{DNN}{short=DNN, long=deep neural network}
\DeclareAcronym{UWB}{short=UWB, long=ultra-wideband}
\DeclareAcronym{LSTM}{short=LSTM, long=long short-term memory}
\DeclareAcronym{KNN}{short=KNN, long=k-nearest neighbor}
\DeclareAcronym{WKNN}{short=WKNN, long=weighted k-nearest neighbor}
\DeclareAcronym{LR}{short=LR, long=logistic regression}
\DeclareAcronym{AP}{short=AP, long=access point}
\DeclareAcronym{CSI}{short=CSI, long=channel state information}
\DeclareAcronym{GNSS}{short=GNSS, long=global navigation satellite system}
\DeclareAcronym{P2P}{short=P2P, long=point-to-point}
\DeclareAcronym{PMF}{short=PMF, long=protected management frames}
\DeclareAcronym{GB}{short=GB, long=gradient boosting}
\DeclareAcronym{MUSIC}{short=MUSIC, long=MUltiple SIgnal Classification}
\DeclareAcronym{GP}{short=GP, long=Gaussian process}
\DeclareAcronym{RT}{short=RT, long=regression trees}
\DeclareAcronym{BF}{short=BF, long=Bayesian filter}
\DeclareAcronym{KF}{short=KF, long=Kalman filter}
\DeclareAcronym{UKF}{short=UKF, long=unscented Kalman filter}
\DeclareAcronym{COTS}{short=COTS, long=commercial-off-the-shelf}
\DeclareAcronym{PSO}{short=PSO, long=particle swarm optimization}
\DeclareAcronym{MM}{short=MM, long=motion model}
\DeclareAcronym{BLE}{short=BLE, long=Bluetooth low energy}
\DeclareAcronym{PF}{short=PF, long=particle filter}
\DeclareAcronym{FF}{short=FF, long=federated filter}
\DeclareAcronym{WPM}{short=WPM, long=wearable particle monitor}
\DeclareAcronym{MCS}{short=MCS, long=modulation and coding scheme}
\DeclareAcronym{SNR}{short=SNR, long=signal to noise ratio}
\DeclareAcronym{PoC}{short=CoF, long=proof of concept}
\DeclareAcronym{TS}{short=TS, long=Thompson sampling}
\DeclareAcronym{ES}{short=ES, long=exponential smoothing}
\DeclareAcronym{TWR}{short=TWR, long=two-way ranging}
\DeclareAcronym{MAP}{short=MAP, long=map information}
\DeclareAcronym{CB}{short=CB, long=closest beacon}
\DeclareAcronym{EDM}{short=EDM, long=Euclidean distance matrix} 
\DeclareAcronym{LSE}{short=LSE, long=Least Square Error}
\DeclareAcronym{SPSO}{short=SPSO, long=standard particle swarm optimization}
\DeclareAcronym{DBSCAN}{short=DBSCAN, long=density-based spatial clustering of applications with noise}
\DeclareAcronym{PDR}{short=PDR, long=pedestrian dead reckoning}
\DeclareAcronym{NLS}{short=NLS, long=nonlinear least squares}
\DeclareAcronym{MLE}{short=MLE, long=maximum-likelihood estimation}
\DeclareAcronym{OBSS}{short=OBSS, long=overlapping basic service set}
\DeclareAcronym{FUSIC}{short=FUSIC, long=combination of FTM and MUSIC}
\DeclareAcronym{NB}{short=NB, long=naive Bayes}
\DeclareAcronym{OFDM}{short=OFDM, long=orthogonal frequency division multiplexing}
\DeclareAcronym{OFDMA}{short=OFDMA, long=orthogonal frequency division multiple access}
\DeclareAcronym{DQN}{short=DQN, long=deep Q-network}
\DeclareAcronym{LTE}{short=LTE, long=long-term evolution}
\DeclareAcronym{VPC}{short=VPC, long=virtual positioning client}
\DeclareAcronym{DT}{short=DT, long=decision tree}
\DeclareAcronym{AoA}{short=AoA, long=angle of arrival}
\DeclareAcronym{UAV}{short=UAV, long=unmanned aerial vehicle}
\DeclareAcronym{MEMS}{short=MEMS, long=micro-electromechanical systems}
\DeclareAcronym{GMP}{short=GMP, long=geomagnetic positioning}
\DeclareAcronym{SoA}{short=SoA, long=state of the art}
\DeclareAcronym{ASAP}{short=ASAP, long=as soon as possible}
\DeclareAcronym{QR}{short=QR, long=quick response}
\DeclareAcronym{INS}{short=INS, long=inertial navigation system}
\DeclareAcronym{IoT}{short=IoT, long=Internet of things}
\DeclareAcronym{SL}{short=SL, long=supervised learning}
\DeclareAcronym{DL}{short=DL, long=deep learning}
\DeclareAcronym{CDF}{short=CDF,  long=cumulative probability function}
\DeclareAcronym{PDF}{short=PDF,  long=probability density function}
\DeclareAcronym{ACK}{short=ACK,  long=acknowledgment}
\DeclareAcronym{LTF}{short=LTF,  long=long training field}
\DeclareAcronym{MIMO}{short=MIMO, long=multiple input multiple output}
\DeclareAcronym{WLAN}{short=WLAN, long=wireless local area network}
\DeclareAcronym{RSRP}{short=RSRP, long=reference signal received power}
\DeclareAcronym{RSRQ}{short=RSRQ, 
long=reference signal received quality}
\DeclareAcronym{WPL}{short=WPL, long=weighted path loss}
\DeclareAcronym{KKT}{short=KKT, long=Karush-Kuhn-Tucker}
\DeclareAcronym{RFID}{short=RFID, long=radio-frequency identification}
\DeclareAcronym{NFC}{short=NFC, long=near-field communication }
\DeclareAcronym{CALM}{short=CALM, long=camera-based AP-integrated localization mechanism}
\DeclareAcronym{DTDoA}{short=DTDoA, long=differential time difference of arrival}
\DeclareAcronym{WSN}{short=WSN, long=wireless sensor network}
\DeclareAcronym{PoA}{short=PoA, long=phase of arrival}
\DeclareAcronym{MLP}{short=MLP, long=multilayer perceptron}
\DeclareAcronym{RMSE}{short=RMSE, long=root mean square error}
\DeclareAcronym{EKFRW}{short=EKF-RW, long=extended Kalman filter with a random walk motion model}
\DeclareAcronym{EKFSH}{short=EKF-SH, long=extended Kalman filter with a step-and-heading-based filter}
\DeclareAcronym{CRF}{short=CRF, long=conditional random field}
\DeclareAcronym{DCCA}{short=DCCA, long=deep canonical correlation analysis}
\DeclareAcronym{SM}{short=SM, long=statistical modeling}
\DeclareAcronym{SIR}{short=SIR, long=sequential importance resampling}
\DeclareAcronym{MDS}{short=MDS, long=multidimensional scaling}
\DeclareAcronym{GMC}{short=GMC, long=gaussian mean clustering}
\DeclareAcronym{ICT}{short=ICT, long=information and communication technology}
\DeclareAcronym{BTCS}{short=BTCS, long=Bluetooth channel sounding}
\DeclareAcronym{HADM}{short=HADM, long=high accuracy distance measurement}
\DeclareAcronym{NDP}{short=NDP, long=null data packets}
\acsetup{
}
\ExplSyntaxOn
\prop_new:N \l__acro_foreign_format_prop
\ExplSyntaxOff

\usepackage{booktabs}
\usepackage{siunitx}
\usepackage{amsmath}

\begin{document}

\begin{frontmatter}



\title{Indoor Positioning with Wi-Fi Location: A Survey of IEEE~802.11mc/az/bk Fine Timing Measurement Research}


\author[agh]{Katarzyna Kosek-Szott}
\ead{katarzyna.kosek-szott@agh.edu.pl}
\author[agh]{Szymon Szott}
\ead{szymon.szott@agh.edu.pl}
\author[agh]{Wojciech Ciezobka}
\ead{wojciech.ciezobka@agh.edu.pl}
\author[agh]{Maksymilian Wojnar}
\ead{maksymilian.wojnar@agh.edu.pl}
\author[agh]{Krzysztof Rusek}
\ead{krzysztof.rusek@agh.edu.pl}
\author[intel]{Jonathan Segev}
\ead{jonathan.segev@intel.com}

\address[agh]{AGH University of Krakow, Poland}
\address[intel]{Intel Corporation, USA}

\begin{abstract}
Indoor positioning is an enabling technology for home, office, and industrial network users because it provides numerous \ac{ICT} and \ac{IoT}  functionalities such as indoor navigation, smart meter localization, asset tracking, support for emergency services, and detection of hazardous situations. The IEEE 802.11mc fine timing measurement (FTM) protocol (commercially known as Wi-Fi Location) has great potential to enable indoor positioning in future generation devices, primarily because of the high availability of Wi-Fi networks, FTM's high accuracy and device support. Furthermore, new FTM enhancements are available in the released (802.11az) and recently completed (802.11bk) amendments. Despite the multitude of literature reviews on indoor positioning, a survey dedicated to FTM and its recent enhancements has so far been lacking. We fill this gap by classifying and reviewing over 180 research papers related to the practical accuracy achieved with FTM, methods for improving its accuracy (also with machine learning), combining FTM with other indoor positioning systems, FTM-based applications, and security issues. Based on the conducted survey, we summarize the most important research achievements and formulate open areas for further research.

\end{abstract}




\end{frontmatter}



\section{Introduction}
Outdoor positioning systems based on one of the available \acp{GNSS} have achieved wide\-spread success.
The \ac{GPS}, the most widely used one, provides localization\footnote{We use the terms localization and positioning interchangeably.} with a
\qtyrange{2}{3}{\meter} accuracy. 
However, attenuation does not allow satellite signals to traverse building structures. Therefore, there is an urgent need to provide precise indoor localization to support a variety of \ac{ICT} and \ac{IoT} applications: indoor navigation, tracking of people and objects, detection of hazardous situations, geofencing, and support of emergency services \cite{mendoza2019meta}.

There are many competing technologies in this area, including \ac{UWB}~\cite{dabove2018} and Bluetooth~\cite{tekler2020}.
However, we focus on IEEE 802.11, commercially known as Wi-Fi, because of its ubiquitous availability, high performance, low costs, and reusability.
Within the IEEE 802.11 standard, \ac{FTM} (defined in the IEEE 802.11mc amendment), commercially known as Wi-Fi Location, provides the necessary functionality to perform precise indoor positioning.
Besides positioning, \ac{FTM} enables time synchronization of devices (including smart meters) \cite{kang2021time} and can support time-sensitive applications \cite{8021as}. 
Although other indoor positioning technologies are also considered, researchers are already determining if \ac{FTM} can replace them \cite{nkrow2023wifi}.
These and other considerations presented in this article support the likely outcome of \ac{FTM} becoming the dominant indoor localization method in the future.

Historically, initial research on positioning in \acp{WLAN} focused on \ac{RSS} measurements. 
Such measurements are passive, but inaccurate, and usually require a prior fingerprinting phase. Additionally, the antenna pattern and device orientation significantly influence RSS (due to, e.g., human body absorption or blockers such as metal parts or glass screens found on hand-held devices). Furthermore, since the transmit power is not always known or fixed, inference based on previously received frames is challenging. 
In 2016, an active, time-based, \ac{P2P} protocol for positioning (\ac{FTM}) was proposed in IEEE 802.11 REVmc (also known as IEEE 802.11-2016) \cite{ieee80211-2016}.
\ac{FTM} determines the range between an initiating station and a receiving station by measuring the \ac{RTT} in a sequence of frame exchanges.
Therefore, this method is often referred to in the literature as \ac{RTT}-based positioning\footnote{Other alternative names are \ac{TWR} and \ac{ToF}, the latter being half of \ac{RTT}.}.
The idea of measuring distance using \ac{RTT} is because ``for most indoor environments the signal propagation speed can be assumed to be constant, as the total travel distance in non-air media is usually negligibly short compared to the travel distance in air'' \cite{bullmann2020comparison}.
To achieve meter-level positioning accuracy, \ac{FTM} requires a timing resolution of the order of nanoseconds.
Fortunately, \ac{FTM} does not require clock synchronization between devices, which not only reduces complexity but also enhances reliability compared to other timing-based approaches \cite{guo2019indoor}.

\subsection{Motivation, Contributions, and Impact}

\begin{figure}
    \centering
    \includegraphics[width=0.9\columnwidth]{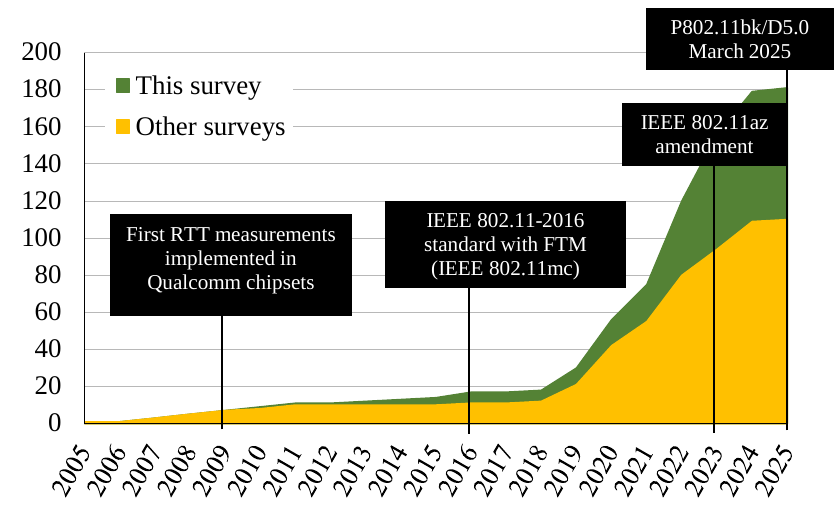}
    \caption{Cumulative number of FTM-related research papers as covered herein as well those surveyed so far. Papers covered by other surveys are spread over ~30 different literature reviews.}
    \label{fig:cumulative_papers}
\end{figure}

\ac{FTM} is extended with the IEEE 802.11az amendment, which brings improved indoor navigation by leveraging \ac{MIMO} for \ac{NLOS} conditions. It also adds security features, scalability, location-based link adaptation, power efficiency, and support for \SI{160}{\mega\hertz} channels \cite{80211-promo}. 
Ultimately, 802.11az improves \ac{FTM}'s 
\qtyrange{1}{2}{\meter} 
accuracy to below \qty{1}{\meter}.
The IEEE 802.11 Working Group recently completed the IEEE 802.11bk amendment, which allows measurements in \SI{320}{\mega\hertz} channels, which will further improve accuracy, below \SI{0.1}{\meter} (``product on shelf'' accuracy) \cite{80211-promo1}.
This high degree of standardization activity confirms the growing interest in Wi-Fi-based positioning among vendors, manufacturers, and operators. 
Likewise, we observe an explosion of research papers related to \ac{FTM} (Fig.~\ref{fig:cumulative_papers}), which deal with such indoor positioning challenges as multi-path propagation (fading), \ac{NLOS} environments (shadowing), clock drift, variable processing delays, and signaling overhead.
%
Therefore, ongoing standardization efforts, increasing availability of devices, and abundance of FTM-related research papers
motivate our work.

Our main contribution is the focused survey of research related to the \ac{FTM} protocol of IEEE 802.11 and its successors. After a tutorial introduction to \ac{FTM}, we review approximately 180 research papers directly related to FTM and organize them into five categories based on the following research questions:
\begin{itemize}
    \item What is the practical accuracy of \ac{FTM} positioning? 
    \item How can \ac{FTM}'s accuracy be improved?
    \item How can \ac{FTM} be used with other positioning systems?
    \item How can \ac{FTM} be used to build user applications?
    \item What are the security issues related to \ac{FTM} operation?
    \item How is \ac{FTM} evolving?
\end{itemize}
For each category, we provide tabular comparisons and summarize the main research trends. 
Additionally, we provide references to datasets available online, as well as to simulation tools that implement FTM.
This literature review allows us to draw conclusions regarding \ac{FTM}-related research and identify potential future research areas.

We foresee that our survey can have the following impact. 
First, it serves as a tutorial introduction to \ac{FTM} operation, indoor positioning methodologies, methods of improving localization accuracy, multi-source data fusion, and relevant security issues.
Second, it improves the visibility of using \ac{FTM} for indoor localization, which is inline with both standardization efforts and manufacturing trends.
Third, it can inspire research on designing next-generation indoor positioning systems: either based solely on \ac{FTM} or using \ac{FTM} as part of a multi-source data-fusion system. 
Fourth, by showing that IEEE 802.11mc FTM implementations are prone to cyberattacks, we highlight the need for more security analysis in this area and present current methods and the upcoming solutions defined in the latest IEEE 802.11 amendments to address this challenge. This aspect is typically omitted in competing surveys on indoor positioning (Table~\ref{tab:surveys}).
Finally, by describing how \ac{ML} is applied to \ac{FTM} research, this survey can help researchers in the proper use of \ac{ML} in future generations of FTM-based localization methods.

\begin{table*}
\footnotesize
\caption{Comparison of previously published surveys which overlap with our survey.}
\label{tab:surveys}
\begin{tabular}{ccp{10cm}p{6em}p{6em}}
\toprule
\textbf{Ref.} & \textbf{Year} & \textbf{Scope} & \textbf{Level of FTM analysis} & \textbf{Security analysis} \\ 
\midrule
\cite{yang2013from} & 2013 & Indoor localization based on channel response, human detection. & None & Yes \\
\cite{makki2015survey} & 2015 & Time-based indoor positioning. &  None & No \\
\cite{maghdid2016seamless} & 2016 & Outdoor-indoor localization methods implemented on smartphones. & None  & No\\
\cite{oguntala2018indoor} & 2018 & Real-time indoor IoT-based applications.  & None & No \\
\cite{laoudias2018survey}  & 2018  & Enabling technologies  for network 3D localization, tracking, and navigation. & Low & Yes  \\
\cite{kimgeok2020review}  & 2020 & Indoor positioning.  & Low  & No \\
\cite{asaad2022comprehensive} & 2022 & Indoor and outdoor localization solutions for IoT. & Low & No \\
\cite{hayward2022survey} & 2022 & Indoor positioning systems and services for cyber-physical systems. & Low & No \\
\cite{nessa2020survey} & 2020 & ML techniques for indoor positioning. & None & No \\
\cite{subedi2020survey} & 2020 & Smartphone-based indoor positioning. & Low & No \\
\cite{yang2021survey} & 2021 & ML techniques for indoor positioning. & Low & No\\
\cite{najarro2022fundamental} & 2022 & Target node localization in wireless sensor networks. & None & No \\
\cite{panja2022survey} & 2022 & Smartphone-based indoor localization. & None & No\\
\cite{retscher2022indoor} & 2022 & Review of user requirements and   smartphone localization. & Low & No \\
\cite{sesyuk2022survey} & 2022  & 3D indoor localization systems and technologies. & Moderate & No \\
\cite{dai2023survey}  & 2023 & Wi-Fi-assisted indoor positioning. &  Moderate & Only privacy \\
\cite{isaia2023review} & 2023 & Wide range of wireless positioning techniques and technologies. & Moderate & No \\
\cite{leitch2023indoor}  & 2023  & Indoor localization. & Moderate & No \\
\cite{naser2023smartphone} & 2023 & Smartphone-based indoor localization. &  Low & No \\
\cite{qiao2023trip}  & 2023 & Wi-Fi indoor localization. &  None & No\\
\cite{sartayeva2023survey} & 2023  & Indoor positioning, security- and   privacy-oriented. & Low & Yes \\ 
\cite{li2024wireless} & 2023 & Wireless positioning.&  Low & No\\
\cite{wang2024survey}&2024&Deep learning-oriented. Wi-Fi-based fingerprinting.& Low & No\\
\cite{yang2024positioning}&2024&Enabling technologies  (e.g., ML, reconfigurable surfaces) for positioning. &  Low & Yes\\
\cite{retscher2024experiences} & 2024 & Review of  techniques and sensors embedded in smart mobile devices used for localization & Low & No 
\\
\cite{chataut20246g} &	2024 & Wireless communication.	Synergy between 6G and AI technologies. & None & Yes \\
\cite{Wu2025WiFi} & 2025 & FTM-oriented (only 802.11mc) & High & Only privacy\\
\textbf{This survey }& \textbf{2025} & \textbf{FTM-oriented (802.11mc/az/bk)} & \textbf{High} & \textbf{Yes} \\
\bottomrule
\end{tabular}
\end{table*}

\subsection{Competing Surveys}
\label{sec:competing}
In the literature, there has been an abundance of surveys on different radio technologies that enable indoor positioning. 
In fact, there have been so many, that there even exists a meta-survey (survey of surveys) \cite{mendoza2019meta}.
Most of them focus on indoor positioning in general and not on a single technology. Only several of them consider security and privacy issues (Table~\ref{tab:surveys}) and only three signal the security and privacy issues of time-based distance measurements \cite{yang2024positioning, sartayeva2023survey}. 
Furthermore, most surveys focusing solely on Wi-Fi either do not study FTM at all or only signalize its availability as one of many available techniques (Table \ref{tab:surveys}).
However, we have observed an increase in interest in FTM-based positioning since 2019 (Fig.~\ref{fig:cumulative_papers}). This is mainly due to the increased number of FTM-supporting products (e.g., smartphones with Android 9 or higher), which enable evaluating the operation of FTM in practice and observing not only its advantages over other ranging options, but also its shortcomings and security gaps. As we show in our survey, most of the observed obstacles can be addressed by adjusting FTM's operation (as proposed in 802.11az and 802.11bk) or supporting FTM ranging and localization techniques with machine learning or in fusion with other radio technologies.

Another justification of our survey is that of all the papers we cover, 39\% have not been described in any other survey. 
This percentage increases to 48\% if we consider articles published in the last four years, when interest in \ac{FTM} has noticeably increased (Fig.~\ref{fig:cumulative_papers}).
Furthermore, the papers that have already been surveyed are described in about 30 different surveys (Table \ref{tab:surveys}), which complicates their comparative analysis. 
Furthermore, as shown in Table \ref{tab:surveys}, only one other survey concentrates solely on FTM \cite{Wu2025WiFi}. However, the competing survey is complementary to ours. It explains in detail the operation of different localization methods (trilateration, fingerprinting, etc.) and is more general. Among others, the survey in \cite{Wu2025WiFi}
\begin{itemize}
    \item does not provide a detailed analysis of FTM parameters and their impact on FTM ranging; 
    \item does not describe the successors of 802.11mc (802.11az and 802.11bk), which provide important updates in the areas of security (e.g., protected frames, protected \ac{LTF}), privacy (i.e., passive ranging with reception only mode), scalability (by integrating MU-MIMO and up to 320 MHz wireless channels) and increased accuracy; 
    \item does not provide a detailed review of existing FTM datasets. 
\end{itemize}
Therefore, we conclude that a detailed literature review of \ac{FTM}-related research is still needed as an answer to the fast-increasing interest in FTM both by standardization bodies (as evidenced by the development of 802.11az and 802.11bk) and device manufacturers 
(support for 802.11az in Android 15 and the release of 802.11az chipsets scheduled for 2025).

\subsection{Brief Comparison of Localization Techniques}
To help readers understand the surveyed area, we briefly introduce other prominent indoor localization and tracking techniques and explain their main disadvantages compared to the studied FTM. 

\emph{\ac{UWB}-based localization}
 uses \ac{ToF} measurements but needs more exchanges than \ac{FTM} due to channel access limitations.  \ac{UWB} operates on a wide frequency range of at least \SI{500}{\mega\hertz} but at the noise floor level (because its maximum power is limited by regulations). It is highly accurate (centimeter-level accuracy), but it also requires specialized equipment (UWB anchors and tags) and is therefore more expensive than FTM, which uses existing Wi-Fi infrastructure. Furthermore, it is sensitive to \ac{NLOS} conditions.
 
\emph{Bluetooth-based localization}
 typically uses \ac{RSSI} measurements and obtains meter-level accuracy. However, environmental factors such as multipath fading and noise degrade its accuracy \cite{maduranga2024improved}. The recently released \ac{BTCS} feature 
 significantly improves accuracy by using an elaborate design: phase-based ranging is consecutively done on 40 adjacent \SI{2}{\mega\hertz} channels. Furthermore, all Bluetooth-based solutions also require deploying dedicated devices (called beacons).
 
\emph{\Ac{RFID}-based localization}
 uses \ac{RFID} tags and readers, which may be active or passive, and obtains meter-level accuracy. However, it is susceptible to signal interference and multipath propagation, which make it less reliable than FTM.

\begin{table*}[t!]
\caption{Comparison of UWB, \ac{BTCS}, and Wi-Fi 802.11 standards \cite{80211-promo2}.}
\footnotesize
\centering
    \begin{tabular}{lp{4cm}p{4cm}p{4cm}}
    \toprule
    \textbf{Standard:} & \textbf{UWB} & \textbf{\ac{BTCS}} & \textbf{802.11} \\ 
    \midrule
    \textbf{Bandwidth} & 500 MHz, 1 GHz & 40 x 2 MHz & 20--320 MHz \\ 
    \textbf{Data rates} & 110 kb/s -- 27.24 Mb/s & 2 M/s & +40 Gb/s \\ 
    \textbf{PSD [dBm/MHz]} & -41.3, -31.3 & 17 & 17 \\ 
    \textbf{Noise floor [dBm]} & -87  & -111  & -101--89 \\ 
    \textbf{Standalone radio} & No & Yes & Yes \\ 
    \textbf{LOS accuracy [m]} & \textless0.1 & 1–2 & 0.4 802.11az, \textless0.1 802.11bk \\ 
    \textbf{LOS range [m]} & \textless15 & 10 & 50+ \\ 
    \textbf{Services} & Ranging, sensing & Low-rate data, ranging (no sensing) & High-rate data, ranging, sensing \\ 
    \textbf{Data/ranging co-existence} & Preemption/none & Preemption/none & Native QoS/preamble detection and energy detection \\ 
    \bottomrule
    \end{tabular}
    \label{tab:comparison_standards}
\end{table*}

\begin{figure*}
\captionsetup[subfigure]{justification=centering}
     \centering
     \begin{subfigure}[b]{0.475\textwidth}
\includegraphics[width=\textwidth]{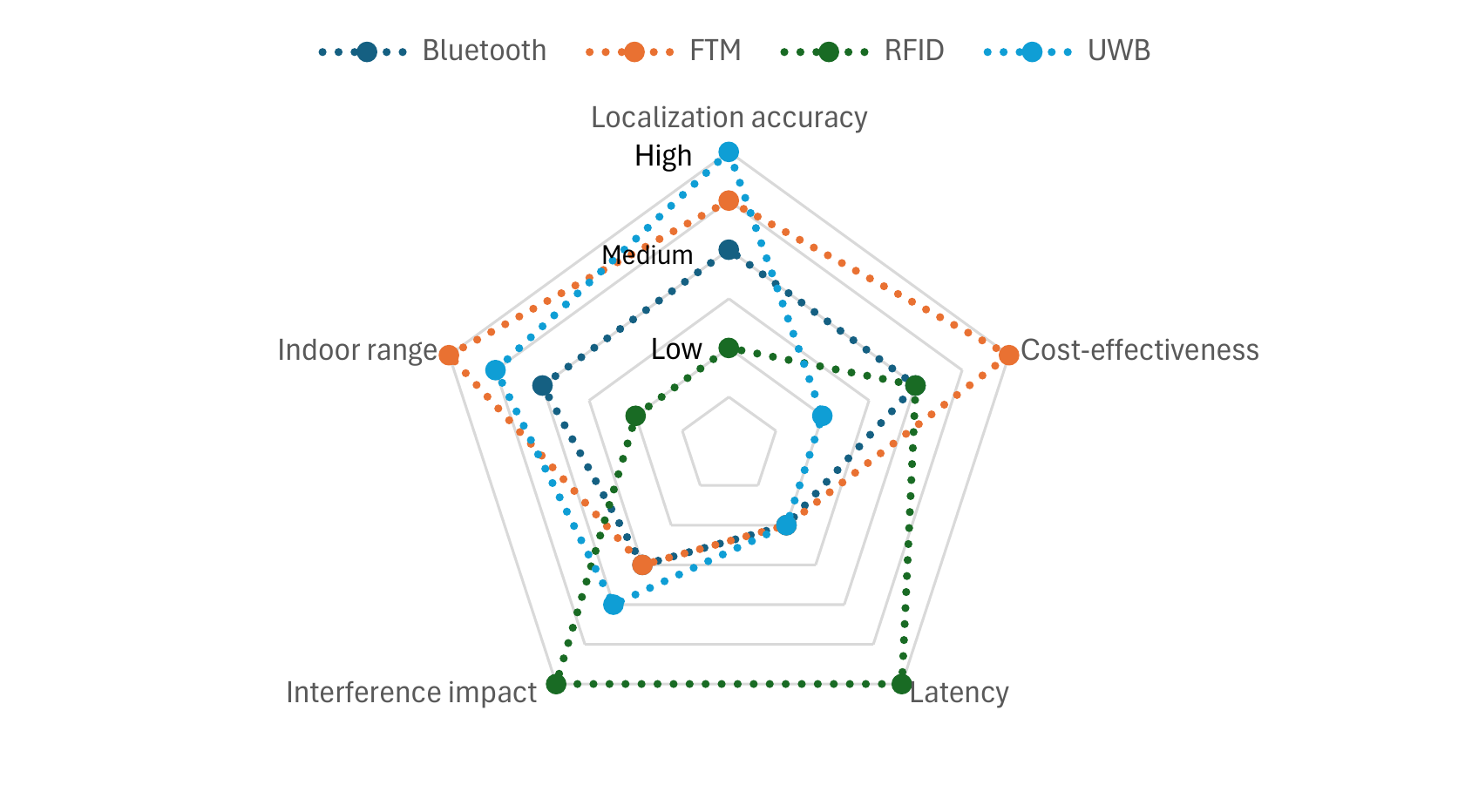}
    \caption{Features}
    \label{fig:comparison}
     \end{subfigure}
     \begin{subfigure}[b]{0.475\textwidth}
\includegraphics[width=\textwidth]{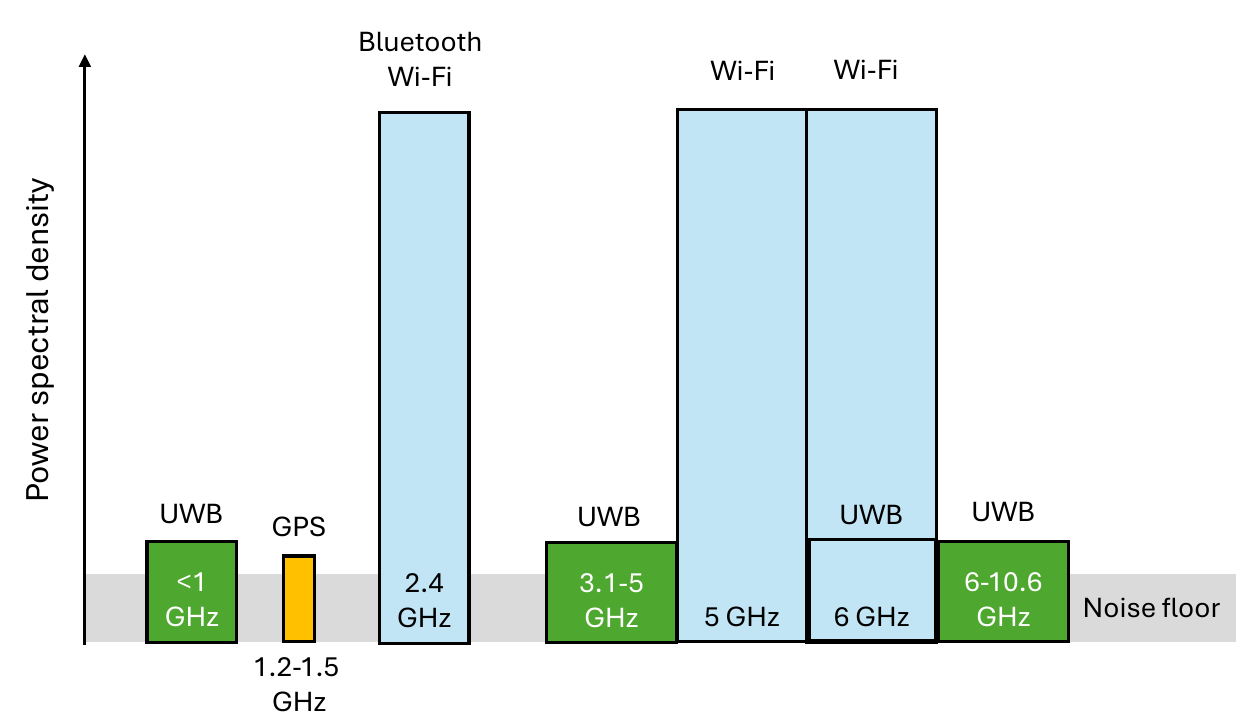}
        \caption{Power spectral density \cite{80211-promo2}}
        \label{fig:power-spectral-density}
     \end{subfigure}
\caption{Comparison of localization technologies (when used for indoor navigation). Note that GPS can operate under the noise floor but requires longer integration.}
\label{fig:overall_comparison}
\end{figure*}



In Fig.~\ref{fig:overall_comparison}, we compare the main characteristics of localization technologies in terms of features and power spectral density. As can be seen, FTM (based on 802.11) is a technology with many advantages. Its only disadvantage in comparison to UWB is its worse accuracy, which, however, will become comparable in the upcoming 802.11bk amendment. Furthermore, UWB has a much lower power spectral density, which can result in a lower range and a dependence on device orientation \cite{80211-promo2}, whereas Wi-Fi is widely available, has a high power spectrum density and operates in a wide frequency range. 
For more details on these various localization techniques, we refer readers to Table~\ref{tab:comparison_standards} and the surveys in Table~\ref{tab:surveys}.  
On the other hand, competing technologies can be used together with FTM to provide better localization and tracking accuracy, as we describe in Section~\ref{sec:data-fusion}.

\subsection{Organization}
\begin{figure}
    \centering  \includegraphics[width=0.9\columnwidth]{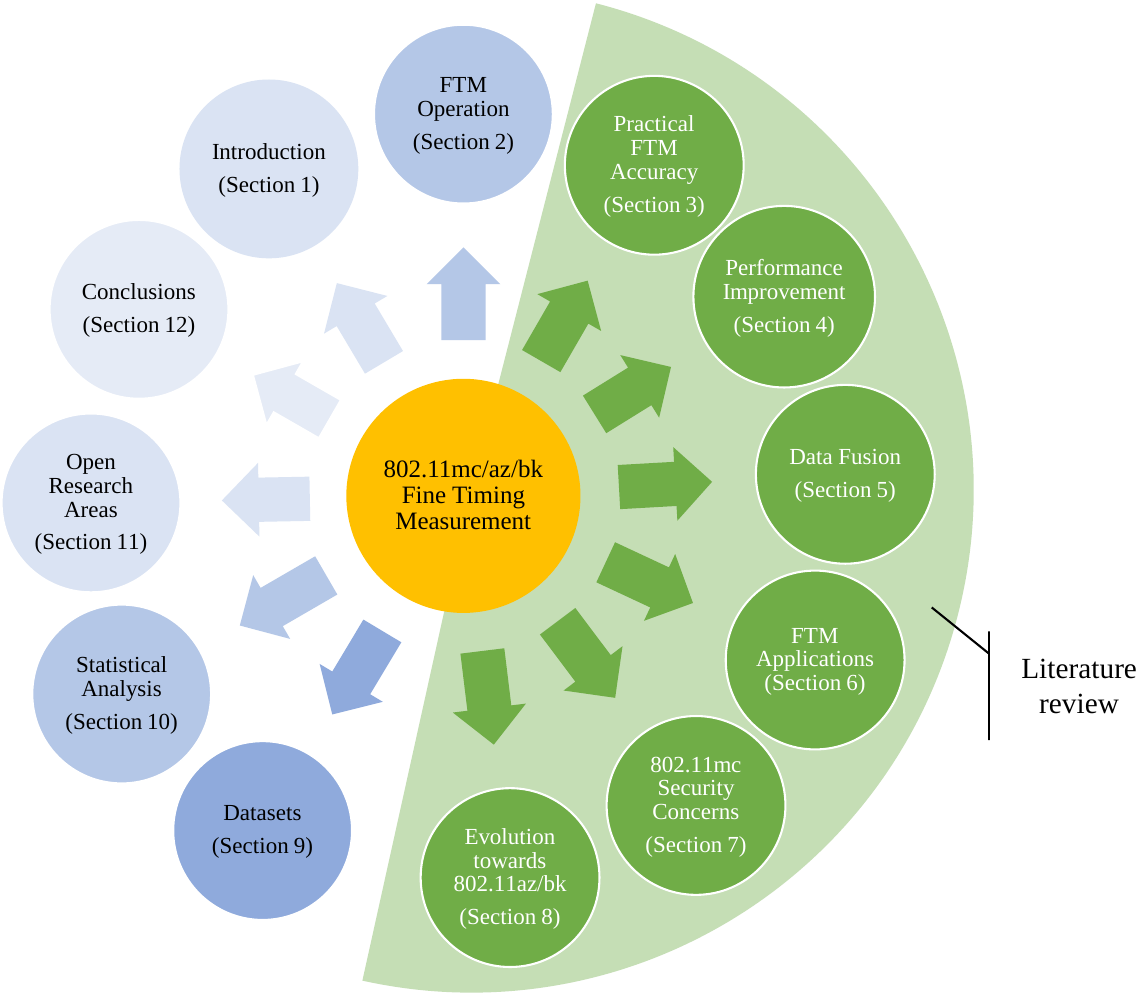}
    \caption{Survey organization.}
    \label{fig:paper-organization}
\end{figure}

This survey is organized as follows (Fig.~\ref{fig:paper-organization}).
First, in Section~\ref{sec:ftm-operation}, we describe the operation of \ac{FTM} and explain how it can be extended to support positioning. 
The next several sections contain the literature review. 
We begin by describing papers which evaluate the  performance of \ac{FTM} (Section~\ref{sec:performance-evaluation}). 
Then we describe FTM enhancements proposed in the literature. In particular, we cover: improved FTM 
ranging and localization (Section~\ref{sec:performance-improvement}), increased positioning accuracy with data fusion (Section~\ref{sec:data-fusion}), possible FTM applications (Section~\ref{sec:enabling-technology}), FTM-related security aspects (Section~\ref{sec:security}), and the successors of IEEE 802.11mc FTM (Section~
\ref{sec:ftm_successors}). Then we list the available datasets (Section~\ref{sec:datasets}) and provide a statistical analysis of the \ac{SoA} papers (Section~\ref{sec:statistical-analysis}). 
In Section~\ref{sec:open-research} we present open research areas. Finally, we conclude our work in Section~\ref{sec:conclusions}. 

\section{Operation of Fine Timing Measurement}
\label{sec:ftm-operation}
\begin{figure}
\captionsetup[subfigure]{justification=centering}
     \centering
     \begin{subfigure}[b]{\columnwidth}
         \centering
    \includegraphics[width=0.9\columnwidth]{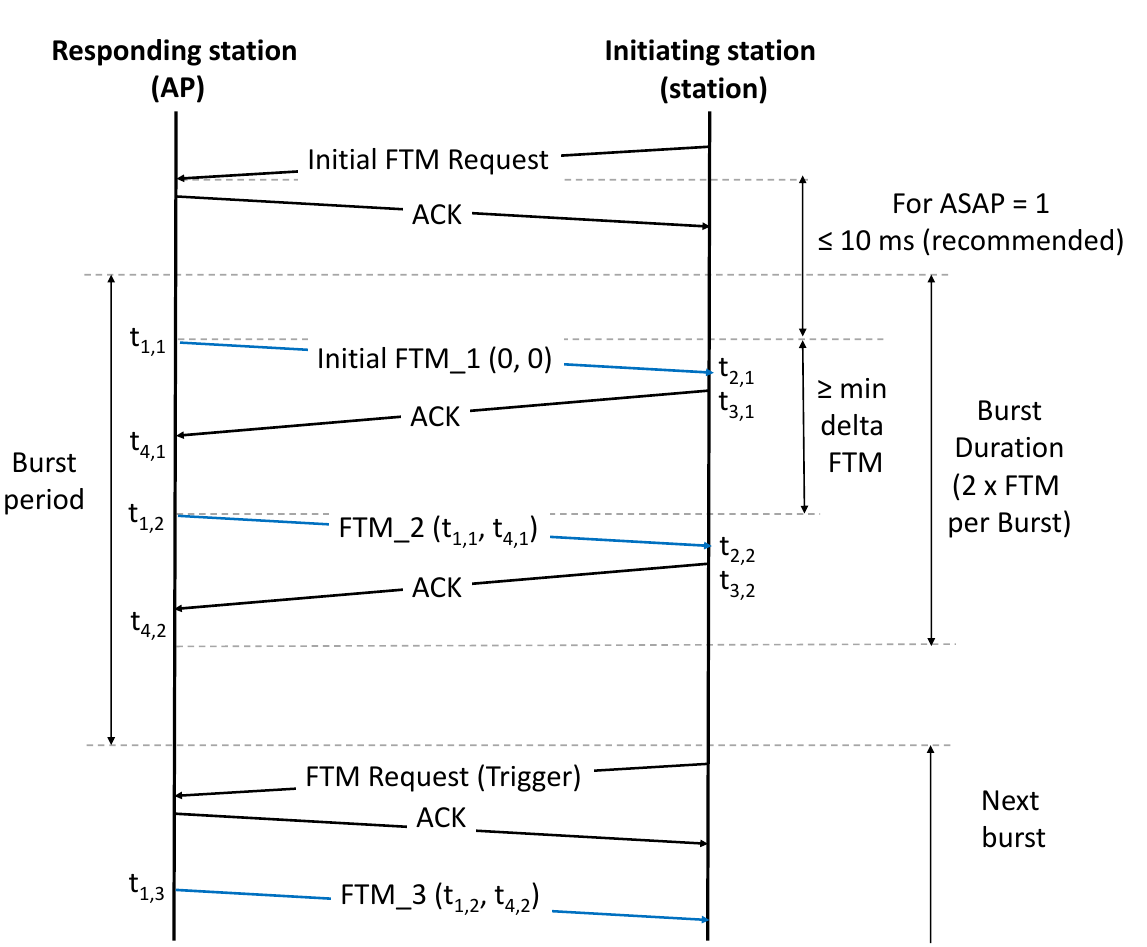}
         \label{fig:ftm-asap}
      \end{subfigure}
\caption{FTM operation for two FTM exchanges per burst 
\cite{ieee802112020}.}
\label{fig:ftm}
\end{figure}

We begin by providing a tutorial overview of \ac{FTM} operation.
\ac{FTM} is a standalone procedure to estimate the distance between communicating Wi-Fi devices using \ac{RTT} measurements. It was introduced in IEEE 802.11 REVmc and remains part of the current standard \cite{ieee802112020}.
The operation of FTM is based on \ac{ToF}, which is calculated as the difference between \ac{ToA} (estimated using sophisticated first path detection) and \ac{ToD} (accurately provided by the hardware).
The FTM procedure is organized as a burst of frame exchanges (Fig.~\ref{fig:ftm}). 
Upon initialization, the \textit{initiating station} sends a trigger frame (FTM Request) to the responding station. 
Then, a set of FTM frames and \acp{ACK} are exchanged between the responding station and the initiating station within a burst. 
As a result, the initiating station obtains four timestamps (all measured locally) for the $i$-th frame exchange ($t_{1,i}$ to $t_{4,i}$). 
These timestamps are used to compute the \ac{RTT}:
\begin{equation}
\label{eq:rtt}
    RTT_i = (t_{4,i} - t_{1,i}) - (t_{3,i} - t_{2,i}),
\end{equation}
where $t_{1,i}$ is the time at which the $i$-th FTM frame is transmitted by the responding station, $t_{2, i}$ is the time at which the $i$-th FTM frame is received by the initiating station, $t_{3, i}$ is the time at which the initiating station transmits the acknowledgment of the successful $i$-th FTM frame reception, and $t_{4,i}$ is the time at which the acknowledgment of the $i$-th FTM frame is received by the responding station \cite{ieee802112020}. 

FTM requires $n$ frame exchanges to calculate $n-1$ RTTs.
Therefore, the minimum number of FTM frames exchanged within a burst is $n=2$, as presented in Fig.~\ref{fig:ftm}, where a single RTT is measured in the burst and the $t_{1,1}$ and $t_{4,1}$ timestamps are transferred by the responding station to the initiating station in the second FTM frame (FTM\_2). 

The $i$-th distance estimation ($\rho^{RTT}_i$) can be calculated from the $i$-th RTT as
\begin{equation}
    \rho^{RTT}_i = \frac{RTT_i}{2} c, 
\end{equation}
where $c$ is the speed of light.




The signaling overhead required to perform an FTM operation depends on different parameters: duration of a burst instance, minimum time between consecutive FTM frames (called ``min delta FTM''), burst period length, number of FTM frames per burst, and number of burst instances.
According to the standard \cite{ieee802112020}, the FTM burst duration can be set from \SI{250}{\micro\second} to \SI{128}{\milli\second}. 
Min delta FTM is expressed in \SI{100}{\micro\second} units and the burst period length in units of \SI{100}{\milli\second}.
To further reduce overhead, the initiating station can set the \ac{ASAP} parameter to indicate that it requests that the initial FTM frame be used as a measurement frame in addition to its regular role of session parameter assignment. Furthermore, to reduce channel access delay, FTM frames are transmitted in the voice access category \cite{ieee80211-2016}. 
Table~\ref{tab:devices} presents FTM parameters supported by different vendors. Additionally, exemplary values of min delta FTM and retransmission limits are available in \cite{schepers2022privacypreserving}, the analysis of the impact of the FTM burst length on ranging accuracy for different smartphones (Pixel 3a, Pixel 4a, Pixel 6 Pro, Xiaomi Mi 10T) and \acp{AP} (Google Nest, Linksys Velop) is given in \cite{zola2025assessing}, while more information on devices that support FTM is given in \cite{ftmdevices}.
In general, the availability of \ac{FTM} ranging matches that of Wi-Fi coverage, while its current accuracy is \SI{1}{\meter}-\SI{2}{\meter}, to be improved by 802.11az to below \SI{1}{\meter} and by 802.11bk to below \SI{0.1}{\meter}.

\begin{table*}[t!]
  \centering
  \caption{FTM parameters supported by major chipset vendors.}
  \footnotesize
		\begin{tabularx}{\textwidth}{p{6em}p{5em}p{5em}p{5em}p{5em}p{5em}p{5em}p{5em}p{5em}p{5em}}
    \toprule
          \textbf{Type} & \textbf{Burst length} & \textbf{FTMs per burst} & \textbf{No. of bursts} & {\textbf{ASAP}} & {\textbf{Retry limit}} & {\textbf{Min delta FTM}} & {\textbf{Frequency channel}} & {\textbf{Channel bandwidth}} & {\textbf{One sided RTT}} \\
    \midrule
    ESP32\textsuperscript{*} & Yes   & Yes   & N/A   & Yes   & N/A   & N/A   & N/A   & Yes   & N/A \\
    Intel\textsuperscript{$ \dag  $} & Yes   & Yes   & Yes   & Yes   & Yes   & N/A   & Yes   & Yes   & Yes \\
    Qualcomm\textsuperscript{$ \ddag  $} & Yes   & Yes   & Yes   & Yes   & Yes   & Yes   & Yes   & Yes   & N/A \\
    Broadcom\textsuperscript{\S} & Yes   & Yes   & Yes   & Yes   & Yes   & Yes   & N/A   & Yes   & N/A \\
    \bottomrule
    \end{tabularx}
  \label{tab:devices}
  \flushleft
  \scriptsize{\textsuperscript{*} \url{https://github.com/espressif/esp-idf}, supported targets ESP32-C2, ESP32-C3, ESP32-C6, ESP32-S2, ESP32-S3} \break
  \scriptsize{\textsuperscript{$ \dag  $} \url{https://github.com/HappyZ/iw_intel8260_localization}}
  \break
  \scriptsize{\textsuperscript{$ \ddag  $} \url{https://github.com/qca/sigma-dut}}
  \break
  \scriptsize{\textsuperscript{\S} \url{https://github.com/StreamUnlimited/broadcom-bcmdhd-4359}}
\end{table*}%

\begin{figure*}
\captionsetup[subfigure]{justification=centering}
     \centering
     \begin{subfigure}[b]{0.3\textwidth}
		\includegraphics[width=\linewidth]{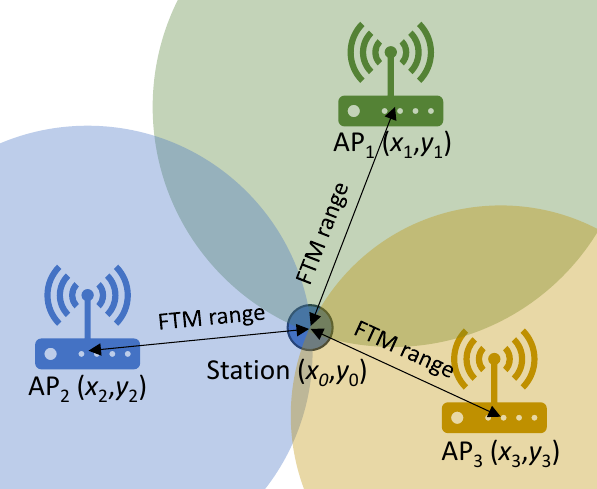}
		\caption{}
		\label{fig:ftm_localization}
     \end{subfigure}
     \begin{subfigure}[b]{0.65\textwidth}
		\includegraphics[width=\linewidth]{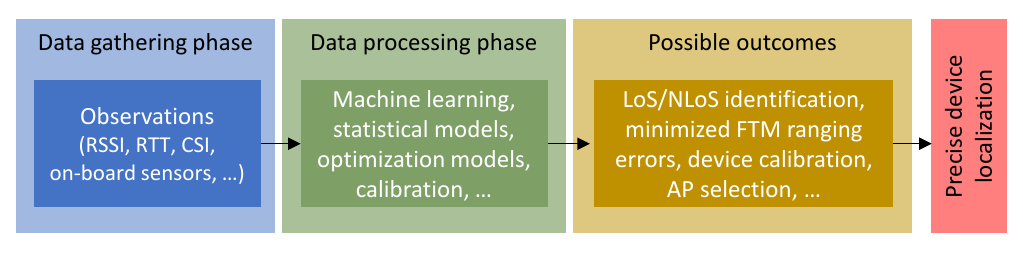}
		\caption{}
		\label{fig:ftm_localization_process}
     \end{subfigure}
\caption{FTM-based localization: (a) trilateration, (b) classification steps to improve accuracy.}
\label{fig:localization_and_process}
\end{figure*}



Next, we explain how \ac{FTM} ranging can serve as a basis for 2D localization. 
Using trilateration, 
or multilateration, a station can estimate its own position ($x_0$,$y_0$), relative to the fixed \acp{AP} with known positions ($x_i$,$y_i$), by solving (e.g., using numerical methods) the following set of equations:
\begin{equation}
    (x_0-x_i)^2-(y_0-y_i)^2=(ToF_i \cdot c)^2,
\end{equation}
where $i\in\{1,2,3,\ldots\}$ (at least three \acp{AP} are needed as shown in Fig. \ref{fig:ftm_localization}) and \ac{ToF} can be calculated 
using \ac{RTT} measurements:
\begin{equation}
\ac{ToF}_i = \frac{\ac{RTT}_i}{2}.    
\end{equation}
Increasing the number of \acp{AP} used for \ac{FTM} measurements naturally improves the localization accuracy~\cite{sun2022geomagnetic}.

\begin{table*}[htbp]
  \centering
  \caption{Definitions of time measures relevant to ranging and positioning.}
  \footnotesize
    \begin{tabular}{p{5em}p{12cm}p{9em}}
    \toprule
    \textbf{Type} & \textbf{Definition} & \textbf{Usage} \\
    \midrule
    RTT & Round trip time of a frame exchange between a sender and a destination station (i.e., how fast the FTM frame is transmitted and acknowledged). & FTM \\
    ToD & Time of departure represents the time at which the start of the preamble of a given FTM frame appeared at the transmit antenna connector. & FTM \\
    ToA & Time of arrival represents the time at which the first path signal arrives at the receive antenna connector. & FTM \\
    ToF & Time of flight of frames exchanged between two stations or between a given station and different APs. 
    & FTM, multilateration 
    \\
    TDoA & Difference in times of arrival of FTM frames from multiple sources. & Multilateration \\
    \bottomrule
    \end{tabular}%
  \label{tab:time}%
\end{table*}%



FTM ranging-based localization can be further improved by integrating additional information and using \acl{SM} or \acl{ML} techniques, as summarized in Fig. \ref{fig:ftm_localization_process} and later disused in Sections \ref{sec:performance-improvement} and \ref{sec:data-fusion}. In general, the process of more precise FTM ranging-based localization can be divided into the following phases: 
\begin{itemize}
    \item \textit{Data gathering phase} --- multiple sources of information can be considered, including channel information (e.g., \acl{CSI}, \ac{RSS}) and other technologies (e.g., on-board sensors, Bluetooth/UWB data), as well as map information. Data can be gathered either locally on the device that performs localization or taken from remote sources (e.g., APs).
    \item \textit{Data processing phase} --- processing can be done using either \ac{SL}, \ac{ML} or other optimization models. 
    \item \textit{Using possible outcomes} --- this phase comprises correcting FTM errors and device calibration, distinguishing between the \ac{LOS} and \ac{NLOS} conditions for improved ranging as well as the selection of the appropriate \acp{AP}.
\end{itemize}

Localization methods that constitute an alternative to FTM's ToF-based operation include \ac{TDoA} (in which there is one sender and multiple receivers) or differential TDoA (in which there is one sender, one reference node, and multiple receivers). For more information on the particulars of these alternatives, we refer  readers to \cite{makki2015survey}. Meanwhile, a brief summary of these different time measures is given in Table~\ref{tab:time}.



\section{Practical Accuracy of FTM Positioning}
\label{sec:performance-evaluation}
We begin our literature review by looking at performance evaluations of \ac{FTM} and note that initial studies of \ac{RTT} performance in IEEE 802.11 networks are actually pre-\ac{FTM}.
Wi-Fi \ac{RTT} measurements are first considered in \cite{gunther2005measuring}, while a similar approach demonstrates the advantages of \ac{RTT} measurements over a signal strength method \cite{golden2007sensor}.
Further pre-\ac{FTM} performance analyses of \ac{ToA} or \ac{RTT} are numerous \cite{ciurana2007ranginga,llombart2008scalability, hoene2008four, ciurana2009softoa, martinez2009indoor, bahillo2010accurate, ciurana2011comparative, koenig2011multipath,  banin2013next, marcaletti2014filtering} and a survey of research in this area up to 2015 is provided in \cite{makki2015survey}.
For completeness, we note that some of these works, such as \cite{ciurana2011comparative, koenig2011multipath}, refer to a network management protocol known as Timing Measurement (defined in 802.11v) which pre-dates \ac{FTM} and is inaccurate due to several design limitations.
With the release of IEEE 802.11 REVmc \cite{ieee80211-2016} in 2016 and the availability of supporting devices \cite{diggelen2018how, kovacshazy2024performance}, researchers began to investigate the performance of \ac{FTM} itself.

\begin{table*}
\footnotesize
\caption{Summary of research in the area of evaluating FTM performance. The ML/SM column indicates application of machine learning or statistical methods. Evaluation methods: E -- experiments, S -- simulations.}
\label{tab_perf_eval}
    \begin{tabularx}{\linewidth}{ccp{4em}p{4em}p{34em}cc}
\toprule
\textbf{Key}  & \textbf{Year} & \textbf{Technology} & \textbf{Scenario} & \textbf{Contribution} & \textbf{ML/SM} & \textbf{Method}       \\ 
\midrule
    \cite{ibrahim2018verification}    & 2018 & FTM        & Indoor, 
    outdoor & Open FTM software platform, FTM performance evaluation: meter-level ranging is possible in outdoor environments after calibration.  & --- & E 
    \\
    \cite{bai2020comparative}         & 2020 & FTM        & Outdoor                          & Comparison of Wi-Fi RTT and GPS-based positioning. FTM is shown to be more accurate and stable than GPS. &  --- & E                 \\
    \cite{bullmann2020comparison}     & 2020 & FTM        & Indoor 
    & Comparison of FTM- and RSS-based indoor positioning. FTM is shown to be more accurate than RSS-based method.                                                & --- & E                 \\
    \cite{gentner2020wifirtt}         & 2020 & FTM        & Indoor 
    & Derivation of \acs*{FTM} error model: Gaussian mixture more accurate than standard Gaussian for error estimation. & SM & E                 \\
    \cite{martin-escalona2020ranging} & 2020 & FTM        & Indoor 
    & Evaluation of \acs*{FTM} performance with \acs*{COTS} devices: ranging accuracy depends on device type and frequency band.           & --- & E                 \\
    \cite{ogawa2020measurement}       & 2020 & FTM        & Indoor, outdoor                  & Evaluation of ranging accuracy of \acs*{FTM}: that high accuracy can be obtained with \acs*{COTS} devices.              & --- & E                 \\
    \cite{zola2021ieee}               & 2021 & FTM        & Indoor 
    & Evaluation of \acs*{FTM} performance in \acs*{NLOS} settings: ranging accuracy depends on device orientation.                    & --- & E                 \\
    \cite{dong2022error}              & 2022 & FTM        & Outdoor                          & Comparison of \acs*{FTM}-supporting devices: each device type has own bias. &  --- & E                 \\
    \cite{feng2022analysis}           & 2022 & FTM        & Indoor 
    & \acs*{FTM} performance evaluation in \acs*{NLOS} with \acs*{ML}-based   fingerprinting: better than \acs*{RSS}-based fingerprinting and RTT trilateration.                                         & Both & E                 \\
    \cite{guo2022impact}              & 2022 & FTM        & Outdoor                          & Clock drift reduces \acs*{FTM} accuracy.                                        & --- & E                 \\
    \cite{zubow2022ftmns3}            & 2022 & FTM        & Outdoor                          & \acs*{FTM} module for ns-3; multipath reduces \acs*{FTM} accuracy.
    & --- & S, E              \\
    \cite{nkrow2023wifi}           & 2023 & FTM        & Indoor 
    & Comparison of \acs*{UWB} and \acs*{FTM}: \acs*{UWB} is better in \acs*{LOS} while \acs*{FTM} -- in \acs*{NLOS} conditions.     
    & --- & E                 \\
    \cite{orfanos2023testing}         & 2023 & FTM        & Indoor 
    & Evaluation of \acs*{FTM} performance considering user orientation: user dynamics impact \acs*{FTM} accuracy.
    & --- & E                 \\
    \cite{sugiyama2023study}          & 2023 & FTM        & Outdoor                          & Evaluation of \acs*{FTM} performance for drone positioning: signaling interference reduces \acs*{FTM} accuracy.
    & --- & E                 \\
    \cite{zubow2023simulation}        & 2023 & FTM        & Indoor 
    & Improved \acs*{FTM} module for ns-3; wider channel bandwidth improves \acs*{FTM} accuracy   & --- & S, E      \\
    \cite{picazo2023ieee}        & 2023 & mmWave \acs*{FTM}        & Indoor                & Enhanced and secured positioning in mmWave band: 802.11az is competitive with other positioning technologies.     & --- & E     \\
    \cite{vur2024portable} & 2024 & FTM & Indoor &  Estimation of workers position in a factory using multilateration: FTM provides satisfactory level of localization accuracy.  & --- & E \\
\bottomrule
		\end{tabularx}
\end{table*}

Typically, performance analyses of Wi-Fi features are dominated by simulation-based studies.
However, this is not the case with \ac{FTM}.
An open \ac{FTM} software platform is provided in \cite{ibrahim2018verification}, 
while later papers investigate FTM operation using COTS Wi-Fi equipment.
The lack of available simulation models is remedied by a dedicated ns-3 module \cite{zubow2022ftmns3,zubow2023simulation}. 
An additional simulation tool is the Matlab implementation of the 802.11az amendment \cite{matlab}, which enables the analysis of improved FTM indoor positioning as well as human presence and router impersonation detection. A detailed evaluation of the  MATLAB 2024a indoor positioning toolboxes is available in \cite{Famili2025Unlocking}.

A review of the performance-focused studies of \ac{FTM} (Table~\ref{tab_perf_eval}) leads to the following main conclusions.
Most importantly, \ac{FTM} is in principle a good approach to positioning.
Indoors it outperforms alternative \ac{RSS}-based methods (especially in \ac{LOS} conditions), while outdoors it can surpass \ac{GPS}-based methods.

However, \ac{FTM} requires calibration to exhibit its superior performance. This calibration is usually in the form of a fixed, per-\ac{AP} offset.
Another important observation is that \ac{FTM} positioning can be impacted by multiple errors:
hardware-dependent bias (arising from device manufacturers), blocker-dependent bias (in \ac{NLOS} environments), fluctuations (due to multipath fading), and outliers (combination of \ac{NLOS} and mobility effects) \cite{dong2022error}.
Additionally, \ac{FTM} performance strongly depends on the quality of the clock deployed in each device \cite{guo2022impact}. 
Therefore, in the next section we show how researchers address this and other \ac{FTM}-related challenges.

\section{Improving the Accuracy of FTM}
\label{sec:performance-improvement}

Improving the performance of \ac{FTM} has been a key focus of numerous research efforts. This section explores three main areas of improvement. One considered aspect is to enable the support of multiple devices, replacing the FTM's \ac{P2P} operation with passive reception of FTM frame exchanges (Table~\ref{tab:PerfImp_p2p_broadcasting}). Then, there is the area of improving the performance of ranging when done actively (Table~\ref{tab:PerfImp_ranging}).  Finally, research is also being done in the area of improving FTM-based localization (Tables~\ref{tab:PerfImpr_Localization} and \ref{tab:PerfImpr_Localization_cont}).

\begin{table*}[t!]
\footnotesize
\caption{Summary of research in the area of improving passive FTM ranging. The ML/SM column indicates application of machine learning or statistical methods. Evaluation methods: E -- experiments, S -- simulations.}
\label{tab:PerfImp_p2p_broadcasting}
\begin{tabularx}{\linewidth}{ccp{3em}p{33em}ccc}
    \toprule
    \textbf{Key} & \textbf{Year} & \textbf{Scenario} & \textbf{Contribution} & 
    \textbf{ML/SM} & \textbf{Conditions} & \textbf{Method} \\
    \midrule
    \cite{banin2019scalable} & 2019  & Indoor & Method for privacy-preserving indoor localization and passive positioning; distributed \acs*{FTM} supports more clients.    & --- & \acs*{NLOS}   & E \\
    \cite{martin-escalona2020passive} & 2020 & Indoor & Passive \acs*{TDoA} algorithm combined with \acs*{FTM} measurements: scalable localization solution for future \acs*{IoT} and \acs*{ICT} deployments. & --- & NLOS & S \\   
    \cite{fujii2022study} & 2022  & \acs*{FTM} pair
    & Passive \acs*{FTM} ranging: many rovers can position themselves by listening to \acs*{FTM} broadcast frames. & --- & \acs*{LOS}    & E \\
    \cite{schepers2022privacypreserving} & 2022  & Indoor
    & Privacy-preserving passive indoor positioning for dense networks.    & --- & \acs*{LOS}   & E \\
    \cite{busnel2023ftmbroadcast} & 2023  & Indoor   & Extending \acs*{FTM} with broadcast communication to provide ranging. 
    & --- & LOS   & S \\
    \cite{lin2023new} & 2023  & Indoor 
    & Evaluation of collaborative positioning; centralized solution with distributed MAC and passive ranging reduces overhead.  & --- &   \acs*{LOS}    & S \\
    \cite{mohsen2023privacypreserving} & 2023  & Indoor 
    & Privacy-preserving indoor positioning with passive measurements:  \acs*{DNN} maps \acs*{TDoA} fingerprints with user locations for improved accuracy. 
    & ML & \acs*{NLOS}   & E \\
    \bottomrule
    \end{tabularx}
\end{table*}

\begin{table*}[t!]
\footnotesize
\caption{Summary of research in the area of improving active FTM ranging. The ML/SM column indicates application of machine learning or statistical methods. Evaluation methods: E -- experiments, S -- simulations.}
\label{tab:PerfImp_ranging}
\begin{tabularx}{\linewidth}{ccp{4em}p{33em}ccc}
    \toprule
    \textbf{Key} & \textbf{Year} & \textbf{Scenario} & \textbf{Overview} & 
    \textbf{ML/SM} & \textbf{Conditions} & \textbf{Method} \\
    \midrule
    \cite{dvorecki2019machine} & 2019  & Indoor 
    & \acs*{DNN} for FTM ranging based on channel estimates as inputs,  better than SoA. & ML & Both  & E \\
    \cite{rea2019smartphone}  & 2019& Indoor
    & Single AP AoA combined with FTM for improved localization. MUSIC  to identify strongest AoA. Low 2D positioning error with single AP. & --- &--- &E 
    \\
    \cite{jiokeng2020when} & 2020  &  Indoor
    &  FUSIC-based ranging (FTM readings and CSI as inputs) as good in NLOS as FTM in LOS conditions. & --- & Both & E \\
    \cite{chigullapally2020wifi} & 2020  & Indoor & ML-based (NN, SVM, RF, and GB) FTM ranging error reduction. 
     RF is shown to be most efficient.   & ML & LOS   & E \\
    \cite{yu2020wi} & 2020 & Indoor, outdoor & Reduction of ranging error caused by clock deviation, NLOS, and multi-path propagation by using calibration, KF, and modeling.  
    & SM & Both & E 
    \\
    \cite{domuta2021two} & 2021 & Indoor 
    & Additional timestamp captures and new ToF computation method for clock offset compensation. 
    & ---
    & --- & S\\
    \cite{jung2021learning} & 2021  & FTM pair & DNN for finding generalized probabilistic distribution of ranging errors. 
     Improved accuracy in comparison to legacy FTM.  & ML & LOS & E \\
    \cite{barralvales2022fine} & 2022  & Indoor, outdoor & ML-based (\acs*{RT}, \acs*{SVM}, \acs*{GPR}, \acp{NN}) real-time distance estimation. Large range estimation errors in complex indoor scenarios.   & ML & Both  & E \\
    \cite{lu2024improving} & 2024  & Indoor 
    & DNN for FTM error calibration by distinguishing environmental characteristics and estimating propagation path lengths. 
    & ML & Both & E \\
    \cite{van2024multi} & 2024 & Indoor 
    & 
    Method for phase-coherent multichannel stitching (up to 320 MHz) for devices with low single-channel bandwidth (20 MHz) capabilities. & --- & LOS & E \\
    \cite{ratnam2024widra} & 2024 & Indoor &  Improved ranging accuracy with carrier phase measurements (supported by MUSIC) from the exchanged FTM frames. 
    & --- & LOS & S, E \\
    \cite{eleftherakis2024spring+} & 2024 & Indoor 
    & Positioning based on CSI (AoA) and FTM measurements supported by 2D-MUSIC. Smartphone localization with just a single AP. 
    & Both & Both &  S, E \\
    \bottomrule
    \end{tabularx}
\end{table*}

\subsection{Ranging}
\subsubsection{Passive Ranging}

\begin{figure}[t]
    \centering
    \includegraphics[width=0.9\columnwidth]{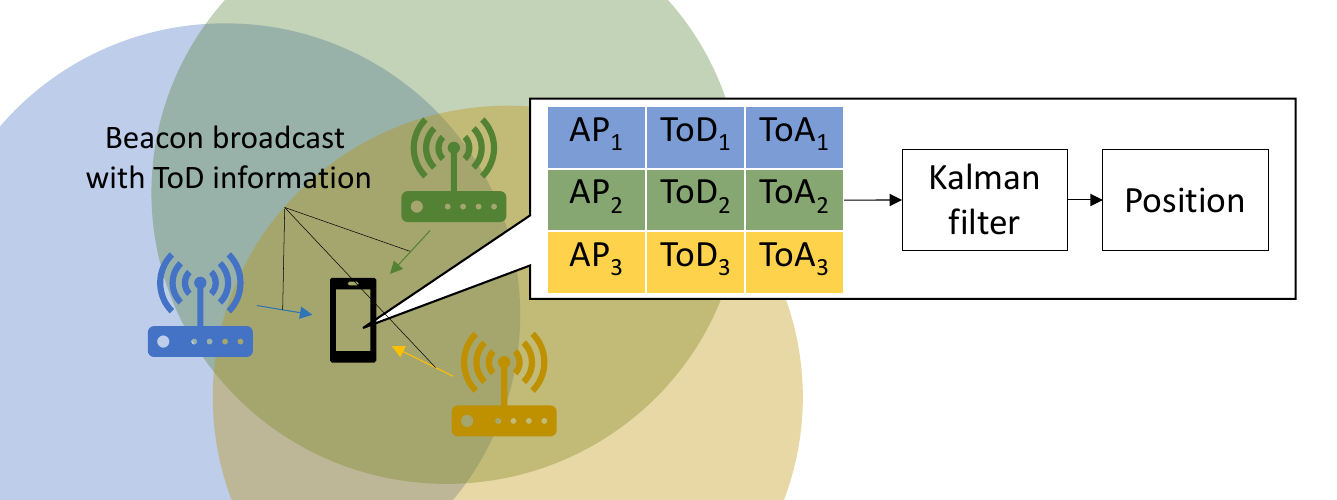}
    \caption{Exemplary broadcasting network for passive \ac{ToA} and \ac{ToD} measurement \cite{banin2019scalable}.}
    \label{fig:broadcast}
\end{figure}

\ac{FTM} operates as a \ac{P2P} protocol, limiting the number of stations that an FTM responder can handle. If a single measurement requires \SI{30}{\milli\second}, an AP can only manage 30 stations per second (excluding non-FTM traffic) \cite{banin2019scalable}. 
This limitation can be addressed by a network of broadcasting stations (Fig.~\ref{fig:broadcast}), forming a local geo-location network that mimics \ac{GNSS} operation.
Similar approaches are proposed in \cite{fujii2022study,schepers2022privacypreserving,busnel2023ftmbroadcast,lin2023new,martin-escalona2020passive}.  
These passive measurements can also be used to construct a fingerprint map and train a \ac{DNN} for further improvements in positioning accuracy \cite{mohsen2023privacypreserving}.
Table~\ref{tab:PerfImp_p2p_broadcasting} summarizes the research on shifting \ac{FTM} operation from \ac{P2P} to passive ranging. These approaches can significantly reduce \ac{FTM} overhead, making them particularly suitable for dense deployments.

\subsubsection{Active Ranging}
Legacy \ac{FTM} is based on active ranging, i.e., exchanging frames between two devices and estimating the distance between them.
Therefore, improving ranging performance is achieved by reducing errors in these estimates.
In this area, one subset of works focuses on errors caused by multi-path propagation \cite{rea2019smartphone, eleftherakis2024spring+,jiokeng2020when, yu2020wi, domuta2021two, van2024multi, ratnam2024widra, abdulrazzaq2024wireless}. Typically, a fusion of FTM and \ac{MUSIC} is implemented, in which apart from  \ac{ToF} estimates, \ac{CSI} is also considered to distinguish between \ac{LOS} and \ac{NLOS} paths. 
This fusion achieves ranging accuracy in \ac{NLOS} conditions comparable to \ac{FTM} performance in \ac{LOS} conditions.
A second subset of works applies \ac{ML} to improve ranging \cite{lu2024improving, barralvales2022fine, dvorecki2019machine, chigullapally2020wifi, jung2021learning}.
Example approaches include reducing FTM ranging errors with classification methods such as \ac{RT}, \ac{GPR} or \ac{GB} as well as calibrating \ac{FTM} with pre-trained \acp{NN}.
Research on improving the performance of FTM ranging is summarized in Table~\ref{tab:PerfImp_ranging}.
These methods typically do not change the \ac{FTM} protocol, but in some cases require gathering additional information (such as \ac{CSI}).

\begin{table*}[t!]
\footnotesize
\caption{Summary of research in the area of improving FTM-based localization (before 2022). 
The Cond. columns indicates LOS/NLOS conditions. Evaluation methods: E -- experiments, S -- simulations.}
\label{tab:PerfImpr_Localization}
\begin{tabularx}{\linewidth}{p{2em}p{2em}p{3em}p{33em}ccccc}
    \toprule
    \textbf{Key} & \textbf{Year} & \textbf{Scenario} & \textbf{Contribution} & 
    \textbf{Dense} & \textbf{ML/SM} & \textbf{Cond.} & \textbf{Method} \\
    \midrule
    \cite{banin2016wifi} & 2016  & Indoor 
    & Indoor position estimation using \acs*{KF} and BF. Latter can integrate non-linear map information with FTM data.   & No    & SM & NLOS  & E \\
    \cite{batstone2016robust}, \cite{batstone2016robusta} & 2016  & Indoor & 
    Node position estimation from measured distance with missing data and outliers.    & No    & --- & LOS   & E \\
    \cite{choi2019unsupervised} & 2019  & Indoor & NN model for range estimation and indoor localization, which outperforms RSS- and RTT-based benchmarks.  & No   & Both & LOS & E \\
    \cite{jathe2019indoor} & 2019  & Indoor
    & Linear regression corrects ranging offset. Statistical analysis improves positioning.  
    & No    & --- & Both  & E \\
    \cite{han2019smartphonebased} & 2019  & Indoor 
    & \acs*{SVM} used for LOS/NLOS identification for RTT and RSS-based ranging.  SVM-FTM  better than SVM-RSS.  
    & No    & ML   & Both  & E \\
    \cite{neri2019indoor} & 2019 & Indoor 
    & FTM-based vehicle localization in LOS conditions (tunnel). & No & SM & LOS& S
    \\
    \cite{henry2020sensor} & 2020  & Indoor
    & Geometric approach for improved indoor localization better performance than techniques based on least-squares errors.   & No    & --- & NLOS  & E \\
    \cite{horn2020doubling} & 2020  & Indoor & Weighted averages of per-channel measurements.  Improved distance estimation accuracy with frequency diversity. & No    & SM & NLOS  & E \\
    \cite{horn2020observation, horn2024round} & 2020, 2024  & Indoor, outdoor  & 
    Bayesian cell update as a promising method for estimating indoor position based on RTT.  & No    & SM & NLOS  & E \\
    \cite{huilla2020indoor} & 2020  & Indoor 
    & FTM fingerprinting method which reduces errors compared to RSS fingerprinting and \acs*{KF}-processed FTM ranging.    & No    & SM & NLOS  & E \\
    \cite{cao2020indoor} & 2020  & Indoor 
    & RTT-based positioning with LOS detection and range calibration;  better than other least-squares algorithms.    & No    & ML & LOS   & E \\
    \cite{cao2020wifi} & 2020  & Indoor 
    & Known AP positions used for offline range difference fingerprinting with \acs*{PSO}-based estimation; 
    better than weighted centroid and least-squares algorithms.   & No    & ML & NLOS  & E \\
    \cite{shao2020accurate} & 2020  & Crowded indoor
    & Positioning estimation transformed into combined position optimization problem. 
    & No    & --- & NLOS  & E \\
    \cite{si2020wifi} & 2020  & Indoor & Gaussian model to identify NLOS. Least-squares algorithm for positioning based on LOS signals.  
    & No    & SM & Both  & E, S \\
    \cite{zeng2020driver} & 2020  & Indoor 
    & FTM range estimation joint knee point detection for LOS/NLOS. RTT ranging to predict the user proximity. & No    & ---    & LOS   & E \\
    \cite{seong2021highprecision} & 2021  & Indoor 
    & CNN-based RTT compensating network and RNN-based positioning network. 
    & No    & ML & NLOS  & E \\
    \cite{cao2021smartphones} & 2021  & Indoor 
    & 2D localization estimate provided by weighted centroid algorithm. PSO 
    for improved altitude estimation.  
    & No    & ML & LOS   & E \\
    \cite{henry2021geometric} & 2021  &  Indoor 
    & Geometric resolution of angle inaccuracies. GPS signals help with indoor localization. ML-based `wall remover'.  
    & No    & ML & Both  & E \\
    \cite{choi2021enhanced} & 2021  & Indoor 
    & FTM outputs (distance, standard deviation, RSS) as NN inputs for reduced positioning error.   & Semi  & Both & Both  & E
    \\
     \cite{garcia-fernandez2021accuracy} & 2021  & Outdoor 
    & Hardware AP bias estimation. 
    Near-window stations use GNSS to help locate indoor APs. 
    & No    & --- & LOS   & E \\
    \cite{chan2021transfer} & 2021  & Indoor  & \acs*{NN}-based prediction of AP location in \acs*{NLOS} conditions.  Solution better than least-squares and circular positioning.  
    & No    & ML   & NLOS  & E \\ 
    \bottomrule
    \end{tabularx}
\end{table*}

\subsection{Localization}

Device localization based on FTM ranging can be improved in numerous ways.
First, statistical models can reduce localization error by filtering out measurement noise. Typically used models include \acp{BF}, \acp{PF}, and \acp{KF} \cite{horn2022indoor, si2022adaptive, horn2020observation, banin2016wifi, huilla2020indoor, ma2022wifi}. 

\ac{ML} methods also emerge as a promising approach to improve the accuracy of FTM-based localization. Distance estimation errors can be corrected with unsupervised \acp{NN} \cite{choi2019unsupervised, choi2021enhanced}, \ac{PSO} \cite{cao2021smartphones}, and \acp{DNN} trained on FTM fingerprint radio maps \cite{numan2022dnnbased,perdana2023evaluation,seong2021highprecision}. \ac{ML} methods can also ``remove walls'' by analyzing angles within overlapping triangles formed by sets of APs \cite{henry2021geometric}.

One of the key challenges for FTM-based indoor positioning is differentiating between signals with a \ac{LOS} path and those experiencing \ac{NLOS} propagation. \ac{NLOS} signals, often caused by obstacles such as walls or parked cars \cite{jathe2019indoor,mathias2008efficient} can lead to significant ranging errors and compromise positioning accuracy, especially in mmWave bands \cite{blanco2022augmenting}.
One approach to mitigating \ac{NLOS} signals is to optimize the placement of APs  \cite{henry2020sensor,liu2023dqwiapdom}, although 
the main works in this field focus on LOS/NLOS identification using Gaussian models \cite{cao2020indoor,cao2020wifi,si2020wifi,cao2024los} as well as \ac{RF} and \ac{DNN} methods \cite{feng2022wifi,dong2022realtime}.
Finally, it is worth noting that there exists a distance after which \ac{FTM} ranging errors do not change regardless of the environment (LOS or NLOS) \cite{zeng2020driver}.

\begin{table*}[t!]
\footnotesize
\caption{Summary of research in the area of improving FTM-based localization (since 2022). 
The Cond. columns indicates LOS/NLOS conditions. Evaluation methods: E -- experiments, S -- simulations.}
\label{tab:PerfImpr_Localization_cont}
\begin{tabularx}{\linewidth}{p{2em}cp{4em}p{33em}ccccc}
    \toprule
    \textbf{Key} & \textbf{Year} & \textbf{Scenario} & \textbf{Contribution} & 
    \textbf{Dense} & \textbf{ML/SM} & \textbf{Cond.} & \textbf{Method} \\
    \midrule 
    \cite{feng2022wifi} & 2022  & Indoor & \acs*{RF}-based \acs*{LOS}/\acs*{NLOS} classification. 
    Less features needed for LOS/NLOS identification, compared to SoA. &  No    & ML   & LOS   & E \\
    \cite{dong2022realtime} & 2022  & Indoor 
    & LOS/NLOS identification for RTT- and RSS-based ranging using \acs*{ML}: \acs*{LS-SVM}, \acs*{RF}, \acs*{DNN}.  
    & No    & ML   & Both  & E \\
    \cite{aggarwal2022wifi} & 2022  & Indoor,  outdoor 
    & 
    Auto-calibration solution allows devices to self-calibrate FTM offsets. 
    & No    & ML & Both  & E \\
    \cite{horn2022indoor} & 2022  & Indoor 
    & Indoor localization using one-sided RTT with legacy APs.  One-sided RTT less accurate than two-sided RTT.    & No    & SM & NLOS  & E \\
    \cite{numan2022dnnbased} & 2022  & Indoor 
    & FTM-based fingerprints and FTM ranging average, variance as inputs for DNN-based indoor localization.
    & No    & ML   & NLOS  & E \\
    \cite{martin2022improving} & 2022 & Indoor 
& Assessing RTT for positioning with classification algorithms; RTT in fingerprinting not always better than RSS. & No & Both & LOS & E
    \\
 \cite{ma2022wifi} & 2022  & Indoor 
& Three-step-positioning to overcome ranging errors.  
& No    & SM   & LOS   & E \\
\cite{wang2022toward}& 2022& Indoor 
& AoA from single AP combined with FTM for improved localization. Requirement: at least two antennas at transmitter. 
    & No & --- & Both & E
    \\
\cite{si2022adaptive} & 2022  & Indoor 
& Error compensation model, \acs*{NB}-based ranging evaluation model, weighted least-squares positioning. 
& No    & SM   & Both  & E \\
\cite{blanco2022augmenting} & 2022 & Indoor,  
   outdoor 
& Joint mmWave and sub-6 GHz localization: CSI-based angle estimation and FTM-based ranging, respectively. 
& No & --- & Both & E
    \\
\cite{perdana2022evaluation} & 2022  & Indoor 
& ML classification (k-NN, RF, DT, NB, SVM) tested with an RTT fingerprinting map, better than RSS-based.
& No    & Both   & LOS   & S \\
\cite{jin2022offtheshelf} & 2022  & Indoor 
& 
Method for combining \acs*{AoA} and \acs*{FTM} to improve accuracy without the need for \acs*{FTM}'s offset/bias reduction.  & No    & ---   & LOS  & E \\
\cite{perdana2023evaluation} & 2023  & Indoor 
& DNN for indoor positioning. 
Improved positioning accuracy in comparison to \acs*{WKNN}.  & No    & ML   & NLOS  & E \\
\cite{shi2023decimeterlevel} & 2023  & Indoor & Training-free indoor localization based on round-trip phase measurements.  Better precision than RTT-based localization.    & No    & ---    & LOS   & E \\
\cite{guo2023framework} & 2023 & Indoor & 
Improved positioning accuracy obtained using a communication state monitoring algorithm and phase distortion error constraint model. & No & --- & LOS & E \\
\cite{cao2024los}& 2024 & Indoor 
&  Real-time LOS/NLOS recognition with SVM. 
Improved reliability of NLOS ranges. Calibration of LOS measurements. 
&  No & ML & Both & E\\
\cite{shida2024correction} & 2024 & Indoor 
& RTT measurement correction.  Selecting APs with large angular variance improves positioning accuracy. & No & --- & Both & E \\
\cite{shao2024moc} & 2024 & Indoor & 
    Factor graph model composed of FTM ranging and a priori motion knowledge. Dynamic estimation of FTM confidence.   & No & No & NLOS & E    \\
    \bottomrule
    \end{tabularx}
\end{table*}


Additionally, studies show that ranging error offsets vary per-devices. These errors can be mitigated through dedicated calibration techniques.
A notable example is using position fixes based on data provided by outside user terminals which have access to a \ac{GNSS} \cite{garcia-fernandez2021accuracy}.
Indoor \acp{AP}, outside of \ac{GNSS} coverage, can improve their position accuracy and enable indoor terminal localization (Fig.~\ref{fig:gnssftm}).
Other notable research in this area includes WiLoC \cite{aggarwal2022wifi}, a method for automatic offset calibration, and using \ac{CSI}-based \ac{AoA} information to eliminate the requirement of pre-calibrating \ac{FTM} offsets \cite{jin2022offtheshelf}.

Finally, the literature explores alternative approaches to improve FTM performance. 
These include performing measurements on different channels to leverage frequency diversity \cite{horn2020doubling}, using round-trip phase measurements from a \ac{CSI} exchange \cite{shi2023decimeterlevel}, 
reducing the impact of missing observations and outliers \cite{batstone2016robust, batstone2016robusta}, and selecting APs with a large angular variance \cite{shida2024correction}.

A summary of research that improves FTM-based localization performance is provided in Tables~\ref{tab:PerfImpr_Localization} and~\ref{tab:PerfImpr_Localization_cont}. Almost all of the presented solutions are validated through experimentation and many leverage \ac{SM} or \ac{ML} techniques. Although some methods exhibit high computational demands, others utilize simple statistical analysis, making them suitable for deployment on low-power devices. Furthermore, most approaches can adapt not only to \ac{LOS} conditions but also \ac{NLOS} paths, which are characteristic of indoor environments. Most works do not consider dense \acp{FTM} but rather focus on improving FTM-based positioning for a single station.

\begin{table*}
\footnotesize
\caption{Summary of research in the area of \acs*{FTM}-based data fusion. The ML/SM column indicates application of machine learning or statistical methods.}\label{tab_data_fusion}
		\begin{tabularx}{\linewidth}{lllp{4em}p{25em}p{4em}}
\toprule
\textbf{Key  } & \textbf{Year} & \textbf{FTM fusion with} & \textbf{Scenario} & \textbf{Outcome} & \textbf{ML/SM} \\ \midrule
\cite{chen2015fusion} & 2015 & PDR & Indoor 
&  Improved accuracy over either method used separately.    & SM  \\
\cite{guo2019indoor} & 2019 & RSS & Indoor 
& Improved scalability and accuracy in static and dynamic tests.  & SM \\
\cite{dumbgen2019multi}                          & 2019 & INS, BLE, camera      & Indoor 
& Algorithm to join passive and active measurements.                          & ML \\
\cite{xu2019locating}                            & 2019 & PDR, map                             & Indoor 
& Improved accuracy over \acs*{FTM}-only approach.                                          & SM \\
\cite{biehl2019achieving}&2019&Map&Indoor 
& Using map geometry improves room estimation. & SM\\
\cite{yu2019robust}                              & 2019 & PDR                                  & Indoor 
& Improved accuracy over either method used separately.                                   & SM \\
\cite{hashem2020deepnar,hashem2020winar}         & 2020 & RTT fingerpr.                   & Indoor 
& Improved accuracy over either method used separately.                  & ML \\
\cite{huang2020hpips}                            & 2020 & RTT fingerpr., PDR, map         & Indoor 
& Reduced positioning error.                         & Both \\
\cite{ibrahim2020wigo}                           & 2020 & GPS, odometer                        & Outdoor 
& Lane-level positioning in urban canyon.                                                 & SM \\
\cite{yu2020precise}                             & 2020 & RSS, INS                            & Indoor 
& Improved accuracy over \cite{yu2019robust,xu2019locating}.                              & SM \\
\cite{sun2020indoor}                             & 2020 & PDR                                  & Indoor 
& Tight coupling of FTM and PDR improves performance.                                     & SM \\
\cite{girgensohn2020radiofrequencybased} & 2020 & Motion model, map & Indoor & 
RTT has greater accuracy than BLE, even with fewer FTM sensors. & SM  \\
\cite{han2021exploiting} & 2021  & \acs*{PDR} & Indoor
  & 
  Better accuracy than for conventional multilateration techniques.  & ---  \\
\cite{choi2021calibrationfree}                   & 2021 & PDR                                  & Indoor 
& PDR with FTM outperforms PDR with RSS.                                                  & SM \\
\cite{liu2021kalman}                             & 2021 & PDR                                  & Indoor 
& Noise reduction methods required for good accuracy.                                     & SM \\
\cite{alvarez-merino2021wifi}                    & 2021 & UWB, LTE                             & Indoor 
& LTE can supplement FTM, UWB in sparse deployments.                                      & --- \\
\cite{hashem2021accurate}                        & 2021 & RTT and RSS fingerpr.           & Indoor 
& Improved accuracy over \cite{hashem2020winar}.                                         & --- \\
\cite{lopez-pastor2021wifi} & 2021  & RSS & Outdoor 
& 
\acs*{RSS} and \acs*{RTT} sufficient for 2D localization. 
& --- \\
\cite{kia2021rssbased,kia2022accurate}           & 2021 & UWB, RSS                             & Indoor 
& RSS-weighted fusion improves accuracy.                                                  & Both \\
\cite{bullmann2022data} & 2022 & Map, \ac {PDR} & Indoor
&  UWB with better location estimate than FTM.     & SM    \\

\cite{wang2021experimentations}&2022&PDR&Indoor 
&High positioning accuracy in LOS/NLOS conditions.&SM\\
\cite{chan2022fusionbased}                       & 2022 & PDR, map                             & Indoor 
& Crowdsourced information determines FTM infrastructure location.                    & Both \\
\cite{liu2022indoor}                             & 2022 & INS                                  & Indoor 
& Robust localization for \acp{UAV}.                                                      & SM \\

\cite{alvarez-merino2022wifi}                    & 2022 & UWB                                  & Outdoor 
& Overcoming propagation challenges arising from concrete. & --- \\
\cite{garcia-fernandez2022mitigation}            & 2022 & GNSS                                 & Outdoor 
& Data fusion benefits are device-dependent.                                              & SM \\
\cite{guo2022robust}                             & 2022 & RSS, PDR                             & Indoor 
& Positioning error reduced to 0.5 m.                                                     & SM \\

\cite{rizk2022robust}                            & 2022 & RSS and RTT fingerpr.           & Indoor 
& Better accuracy than three \acs*{SoA} methods.                                        & ML \\
\cite{sun2022geomagnetic,sun2022simultaneous}    & 2022 & Magnetometer data            & Indoor 
& Gaussian process improves localization performance.                                                   & --- \\
\cite{sun2022smartphonebased}                    & 2022 & RSS/RTT fingerpr., map      & Indoor 
& Good accuracy in \acs*{NLOS} settings.                                                    & SM \\
\cite{wang2022adaptive}                          & 2022 & RTT, PDR, magnetic fingerprinting & Indoor 
& Multi-source data fusion improves accuracy.                                             & SM\\
\cite{wu2022indoor}                              & 2022 & RSS, PDR                            & Indoor 
& Positioning achieved with single \acs*{AP}.                                               & SM\\
\cite{yu2022hwps}                                & 2022 & RSS fingerpr., PDR              & Indoor 
& Meter-level positioning achieved with data fusion.                                      & SM \\
\cite{yu2022precise}                             & 2022 & RSS, INS                            & Indoor 
& Improved indoor 3D positioning and trajectory optimization.                             & SM \\
\cite{alvarez2022victim}&2022&UWB&Indoor& Integration of UWB and Wi-Fi improves localization.&---\\
\cite{wan2022self}&2022&RSSI and RTT fingerpr. & Indoor 
& Meter-level positioning in multi-floor indoor areas. & Both\\
\cite{fetzer2023interacting}                     & 2023 & PDR                                  & Indoor 
& Improved \acs*{PF} reduces positioning error.                                             & SM \\
\cite{gonzalez2023assessing}                     & 2023 & \acs*{RTT} and \acs*{RSS} fingerpr. & Indoor 
& Performance-scalability tradeoff (no. of measurements). & Both \\
\cite{guo2023factor}                             & 2023 & RSS, PDR                             & Indoor 
& Factor graph optimization better than \acs*{KF}.                                          & ML \\
\cite{numan2023dropout,numan2023smartphonebased} & 2023 & RSS and RTT fingerpr.           & Indoor 
& \acs*{DNN} improves fingerprinting. & ML \\
\cite{park2023automated}                         & 2023 & RSS, PDR                            & Indoor 
& Fingerprint map based on measurements and estimations.                                  & ML \\
\cite{park2023gpsaided}                          & 2023 & PDR                                  & Indoor
& Automatic estimation of \acs*{AP} positions.                                              & SM \\
\cite{yu2023intelligent}                         & 2023 & RTT fingerpr., Bluetooth        & Indoor 
& Improved accuracy over other data fusion methods.                                       & Both \\
\cite{zhou2023wifi}                              & 2023 & INS                                  & Indoor 
& Advanced data fusion methods required to improve accuracy.                              & SM \\
\cite{irshad2023picture}                         & 2023 & Cameras                              & Indoor 
& Improved accuracy over FTM-only approach.                                               & --- \\ 
\cite{khatib2023designing} & 2023 & UWB, LTE & Indoor 
& Blueprint for future 6G location testbeds.
& --- \\
\cite{raja2023wifi}&2023&RTT&Indoor 
&Better positioning than basic LS. 
&SM\\
\cite{feng2024wifi}                             & 2024 & RSS                                  & Indoor 
& Performance improvements on shown on real-world datasets. & ML \\ 
\cite{jurdi2024where} & 2024 & Inertial sensors & Indoor 
& High accuracy, efficient for high and low-end devices. & SM  \\ 
\cite{sun2024smartphone} & 2024 & RSS, PDR, map & Indoor & Better accuracy and stability than SoA fusion approaches. & SM \\
\cite{rana2024enhanced} & 2024 & RSS & Indoor & Gaussian process regression for improved positioning in complex  environments. 
& ML
\\
\bottomrule
		\end{tabularx}
\end{table*}

\section{FTM with Other Positioning Technologies}
\label{sec:data-fusion}

\begin{figure*}
\captionsetup[subfigure]{justification=centering}
     \centering
     \begin{subfigure}[b]{0.3\textwidth}
    \includegraphics[width=\linewidth]{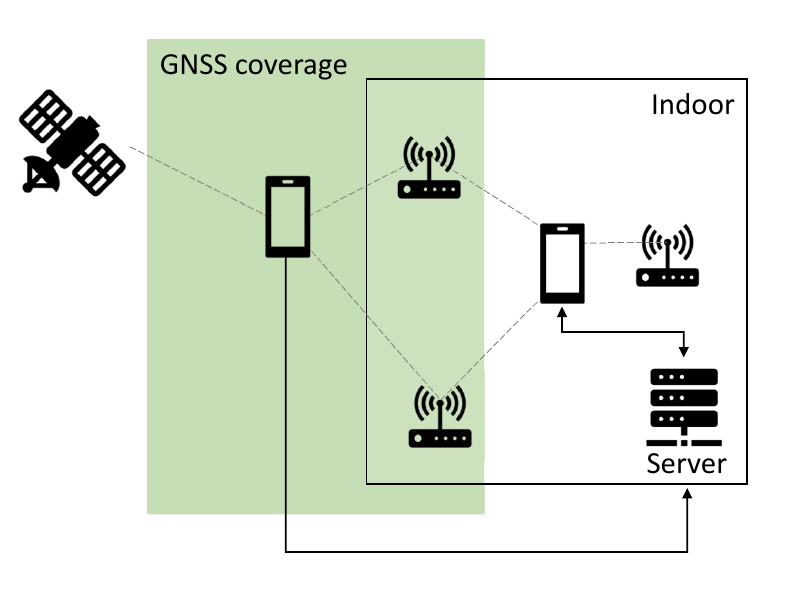}
    \caption{}
    \label{fig:gnssftm}
     \end{subfigure}
     \begin{subfigure}[b]{0.5\textwidth}
    \includegraphics[width=\linewidth]{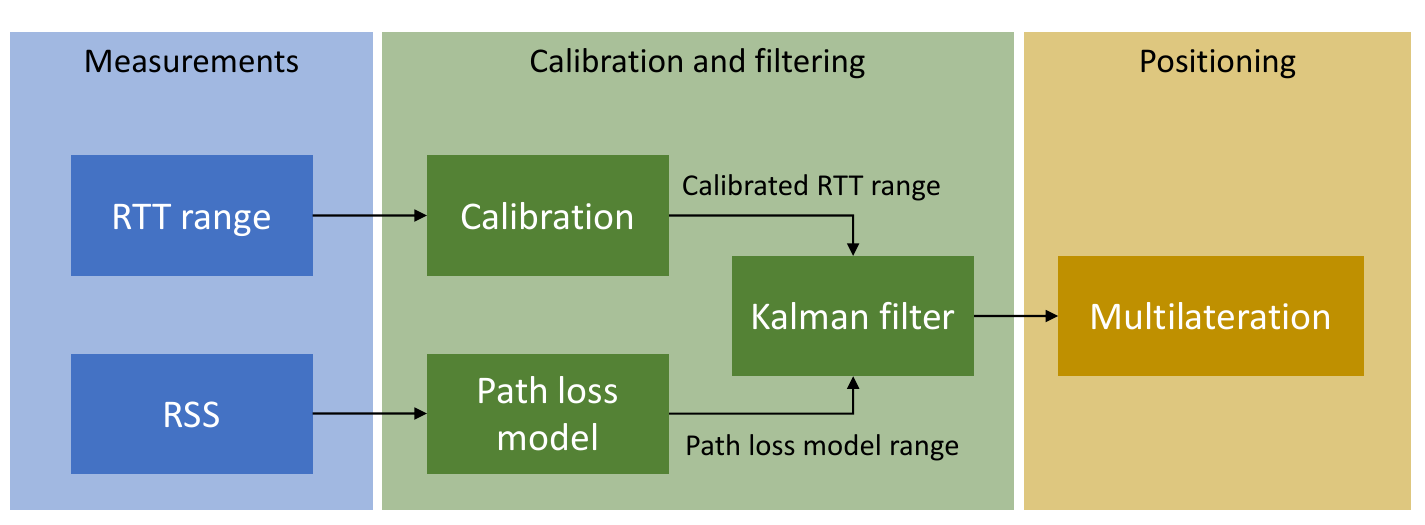}
    \caption{}
    \label{fig:datafusion}
     \end{subfigure}
\caption{Localization examples: (a) FTM with \ac{GNSS} \cite{garcia-fernandez2021accuracy}, (b) FTM data fusion with \ac{RSS} measurements \cite{guo2019indoor}.}
\label{fig:data_fusion_examples}
\end{figure*}

Outdoor smartphone positioning can achieve improved accuracy when multiple data sources are combined, e.g., \ac{GPS} information with cell tower triangulation.
Similarly, with the proliferation of devices supporting \ac{FTM}, researchers are using \ac{FTM} as one of many available data sources that can be combined into a hybrid positioning technology.
The extension of \ac{FTM} with other methods can alleviate some of \ac{FTM}'s inherent problems, such as the impact of multipath propagation and \ac{NLOS} conditions.
Data fusion typically proceeds as follows (Fig.~\ref{fig:datafusion}): measurements from two or more data sources are combined, usually with a \ac{KF} or \ac{PF}, and the 
end result is used for ranging or positioning.
Our classification of the methods being integrated with \ac{FTM} and presented in Table~\ref{tab_data_fusion} is as follows:
\begin{itemize}
    \item \ac{RSS} measurements --  methods in which only a single IEEE 802.11 interface is required,
    \item inertial motion sensors -- methods in which passive measurements by on-board sensors (e.g., gyroscopes) are required,
    \item fingerprinting -- methods in which an offline measurement campaign is required,
    \item multi-source -- methods which require data integration from multiple sources.
\end{itemize}
We also note that in commercial systems, vendors integrate \ac{FTM} and \ac{GPS} measurements (the latter being available to \acp{AP} installed at building perimeters) to provide self-location capabilities to the whole AP network inside the building. Examples of such solutions include Aruba's Open Locate\footnote{\url{https://www.arubanetworks.com/products/location-services/}} and Cisco's \ac{AP} Auto Location\footnote{\url{https://spaces.cisco.com/ap-auto-location/}}.


\subsection{FTM with RSS}

Before \ac{FTM}, signal strength measurements were used for pure Wi-Fi-based positioning.
\ac{RSS}-based localization can be performed passively, but has lower accuracy. In addition, if not carried out properly, distance measurements using RSS can be strongly affected by obstacles (such as walls), variations in transmit power, as well as antenna pattern and device orientation. Therefore, for better positioning, it is necessary to use additional methods, such as LOS/NLOS identification (as proposed, e.g., by \cite{si2020wifi, cao2024los}), and adjust the model accordingly.
However, \ac{RSS} is the first independent data source available on an 802.11 interface.

The cornerstone work of \citet{guo2019indoor}, one of the first to combine \ac{FTM} and \ac{RSS} positioning, emphasizes the need for filtering both \ac{RTT} and \ac{RSS} measurements (Fig.~\ref{fig:datafusion}).
Subsequent works in this field typically follow this approach \cite{raja2023wifi,feng2024wifi, lopez-pastor2021wifi, poveda-garcia2023wifi, lopez2023two, dong2023investigation}. 
No other papers integrate only \ac{RTT} and \ac{RSS} measurements. Further papers include more data sources and are covered in Section~\ref{sec:multisource}, e.g., the first approach mentioned above is extended by the authors in \cite{guo2022robust,guo2023factor} with the integration of \ac{PDR} measurements.

\subsection{FTM with On-board Sensors}

Smartphones have built-in inertial motion sensors (e.g., accelerometers, gyroscopes), referred to as \ac{MEMS}.
These sensors, combined as an \ac{INS}, can determine the position of the device. 
\ac{PDR} \cite{chen2011smart} is the most widely used algorithm; it combines such stages as gait recognition, step-length update, heading calculation, and location update.
This approach is passive and autonomous, but accurate only in the short term. 
Therefore, it is a good candidate to improve the performance of \ac{FTM} (rather than being used on its own).
Researchers typically combine \ac{FTM} with \ac{PDR} using a \ac{KF} or \ac{PF} variant \cite{chen2015fusion, yu2019robust, xu2019locating, sun2020indoor, girgensohn2020radiofrequencybased, choi2021calibrationfree, wang2021experimentations, liu2021kalman, han2021exploiting, chan2022fusionbased, liu2022indoor, bullmann2022data,   fetzer2023interacting, jurdi2024where}.
Other sensor-based approaches refrain from using \ac{PDR} and use raw \ac{INS} data \cite{zhou2023wifi}, use only magnetometer sensors for \ac{GMP} \cite{sun2022geomagnetic,sun2022simultaneous}, or use cameras  \cite{irshad2023picture,liu2022vi}.

\subsection{FTM with RSS Fingerprinting}

Fingerprinting is the process of first creating a radio map of a given indoor setting (typically with a dedicated and thorough measurement campaign, called an offline or recording phase) and then using this map during localization, comparing it with online measurements.
This approach can reduce positioning errors but has a high up-front cost. 
However, it is quite versatile as the radio map can be based on \ac{RSS} or \ac{RTT} measurements and is suitable for fusion with online \ac{FTM} measurements.
Research in this category confirms the requirement of an offline training phase (which can involve either \ac{RSS}, \ac{RTT}, or both types measurements), while data fusion is typically done with \ac{ML} methods such as \ac{DNN}
\cite{hashem2020winar, hashem2020deepnar, hashem2021accurate, rizk2022robust,  numan2023dropout, numan2023smartphonebased, sun2022smartphonebased, gonzalez2023assessing}.

\subsection{Multisource Approaches}
\label{sec:multisource}

The final category are methods that rely on multiple data sources during data fusion. 
These sources can be a combination of the previously mentioned methods (\ac{RSS}, sensors, fingerprinting) or introduce data from other positioning technologies as input, e.g., \ac{BLE} or \ac{UWB} sensors.
A typical data fusion combination, suitable for smartphones, involves \ac{FTM}, online \ac{RSS} measurements, and \ac{PDR} data \cite{yu2020precise, yu2022precise, wu2022indoor}.
Alternatively, \ac{FTM} and \ac{PDR} are combined with maps or \ac{RTT} fingerprints \cite{huang2020hpips, biehl2019achieving, yu2022hwps, wang2022adaptive, park2023automated, park2023gpsaided, wan2022self, yu2022ap}.
Other multi-source approaches use \ac{FTM} along with another positioning technology: \ac{BLE} \cite{dumbgen2019multi, yu2023intelligent}, \ac{UWB} \cite{alvarez-merino2021wifi, alvarez-merino2022wifi, kia2021rssbased, kia2022accurate, alvarez2022victim, khatib2023designing}, or \ac{GPS} \cite{ibrahim2020wigo, garcia-fernandez2022mitigation}.
As can be expected, the use of additional input data improves localization accuracy at the cost of system complexity and (in the case of \ac{BLE} and \ac{UWB}) additional infrastructure.

A review of research in this area (Table~\ref{tab_data_fusion}) leads to the following conclusions.
First, \ac{FTM} is a good candidate to include in data fusion for positioning, as studies have shown that it  improves localization accuracy.
Second, combining \ac{FTM} with data from sensor-based methods is a viable approach to localizing handheld devices such as smartphones due to the availability of measurements and their passive operation.
Finally, including alternative methods (such as fingerprinting or measurements from other network technologies) can further improve accuracy but usually has an up-front cost, making these approaches less promising for widespread adoption.

\begin{table*}[t!]
\footnotesize
  \centering
  \caption{Summary of research in the area of possible FTM applications. The columns names designate the following: L/P -- support for localization/positioning, ML/SM -- application of machine learning or statistical methods, EM -- method of evaluation. Evaluation methods: E -- experiments, S -- simulations.}
		\begin{tabularx}{\linewidth}{p{2em}cp{4em}p{25em}ccccc}
    \toprule
    \textbf{Key} & \textbf{Year} & \textbf{Scenario} & \textbf{Contribution} & 
    \textbf{L/P} & \textbf{Dense} & \textbf{ML/SM} & \textbf{Conditions} & \textbf{EM} \\
    \midrule
    \cite{ando2021combining} & 2021  & Indoor
    & FTM combined with particle monitor for dust concentration observation in a factory. & Yes   & No    & ---    & LOS   & E \\
    \cite{cao2022vitag} & 2022  & Indoor,  outdoor & Association of camera-detected subjects with their phone identifiers using an LSTM-based X-Translator. Online subject identification for indoor, outdoor, and crowded datasets. 
    & Yes   & Semi  & ML   & Both  & E \\
\cite{ashraf2022p2p} &2022&Outdoor&Remote navigation of GPS denied drones. 
FTM-based localization outperforms camera-based localization.&Yes&No&---&LOS&E
\\
\cite{shaikhanov2022falcon,shaikhanov2020autonomous} & 2022  & Indoor, outdoor
& FTM-based sensing of drone networks with target localization and tracking.  
& No    & No    & ---    & LOS   & E \\
    \cite{manabe2023performance} & 2023  & Indoor 
    & Local coordinate system constructed based on FTM measurements.   User positioning without initial database or exact information on AP location. 
    & Yes  & No  & ---  & Both  & E \\
    \cite{wang2023enabling} & 2023  & Indoor
    & FTM-based occupancy detection using \acs*{ML}: \acs*{RF}, \acs*{KNN}, XGBoost, \acs*{LR}, \acs*{SVM}.  Higher accuracy than CSI-based approach. 
    & No    & No    & Both   & LOS   & E \\
    \cite{mohsen2023locfree} & 2023  & Indoor 
    & Indoor, device-free tracking system.  
    Limited to scenarios involving a single individual. & Yes    & No    & ML   & Both   & E \\
    \cite{nikseresht2023ftm} & 2023  & Indoor 
    & \acs*{FTM} for sensor-free, adaptable occupancy detection in various indoor scenarios.  \acs*{FTM} signal variance enhances occupancy detection accuracy. 
    & No    & No    & Both   & Both   & E \\
    \cite{pagliari2024wi} & 2024 & Outdoor & Comparison of RSSI- and FTM-based localization of \acp{UAV}. FTM-based tracking sufficient under low or no GNSS singal. & Yes & No & Yes & LOS & E  \\
    \cite{guan2024autonomous} & 2024 & Indoor &  
    Crowdsourced Wi-Fi fingerprinting and self-detected Wi-Fi FTM stations. PF  combines various location sources. 
    & Yes & No & Yes & Both & E\\
    \cite{mohsen2024timesense} & 2024 & Indoor & DL-based multi-person device-free indoor localization. 
    & Tracking & Yes & Yes & Both & E\\
    \bottomrule
    \end{tabularx}%
  \label{tab_enabl_tech}%
\end{table*}%

\section{FTM for Building User Applications}
\label{sec:enabling-technology}

\begin{figure}[t]
    \centering
    \includegraphics[width=\linewidth]{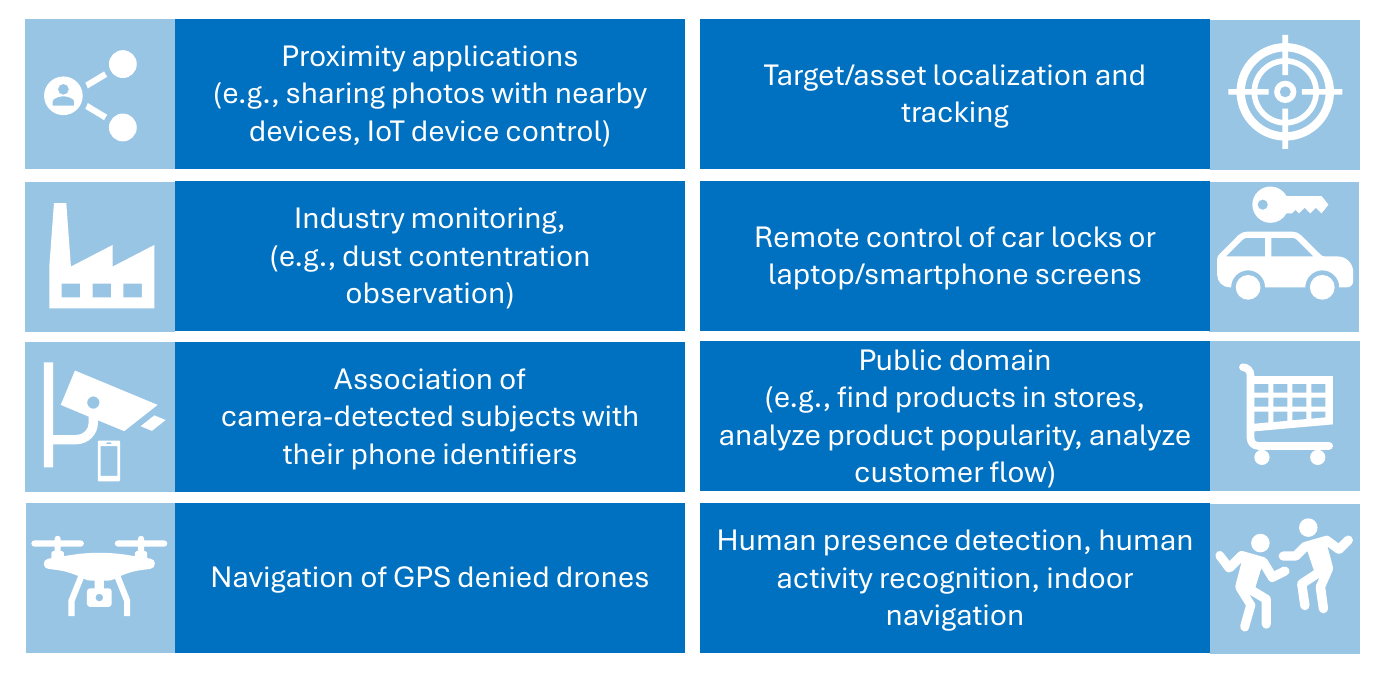}
    \caption{Exemplary FTM applications.}
    \label{fig:ftm_apps}
\end{figure}

\ac{FTM} ranging has also been described in the literature as an enabling technology for many \ac{ICT} and \ac{IoT} applications, as shown in Fig. \ref{fig:ftm_apps}. Most of the proposed solutions, summarized in Table~\ref{tab_enabl_tech}, involve indoor localization and positioning, and many of them involve \ac{ML} models trained with \ac{FTM}-derived data. They are typically validated in real indoor scenarios using only a single pair of devices. Therefore, one of the future challenges of FTM-based applications will be high-accuracy positioning in dense indoor environments (in which multiple stations want to localize themselves), while sustaining data communication services (regular data traffic may impact FTM frame exchanges). We next describe the details of the already identified FTM applications.

The positioning of user devices, based on the trilateration of the distance to several \acp{AP}, is the typical application of \ac{FTM}. 
However, the inverse, i.e., establishing the position of \acp{AP}, is also a valid application, e.g., when installing a new \ac{AP} \cite{chan2021transfer}.
Fig.~\ref{fig:gnssftm} provides an illustrative example.
\citet{manabe2023performance} determine the indoor positions of \acp{AP} and construct a local coordinate system based on \ac{FTM} measurements. Then, smartphones use this local system to position themselves within a building using the lateration of four or more \acp{AP}.

A natural extension of any localization system is tracking \cite{mohsen2024timesense}, i.e., following the movement of people or objects, e.g., to deliver warning messages to users.
\citet{cao2022vitag} combine visual data (camera) and motion and Wi-Fi data (from smartphones) to associate subjects detected by cameras with their corresponding phone identifiers. 
\citet{mohsen2023locfree} use a device-free localization system based on an \ac{FTM}-based fingerprinting approach, i.e., they locate device-less users in a room based on how their presence influences \ac{FTM} measurements. Before online operation, a deep \ac{MLP} network is trained on the \ac{RTT} data related to the ground truth position of a person. This network is then used to infer the position based only on the \ac{RTT} fingerprint of the in-room active devices. Additionally, \citet{shaikhanov2022falcon, shaikhanov2020autonomous} use an on-drone \ac{FTM} sensor network for accurate outdoor tracking of targets. 
Outdoor tracking is also analyzed by \citet{pagliari2024wi}. They compare RSSI- and FTM-based tracking of static and moving \acp{UAV}. In both cases, \ac{KF} is used to improve localization accuracy by removing outliers and measurement noise. FTM-based tracking in mobile conditions is shown to be sufficient for environments with GNSS-denied or weak GNSS conditions. Furthermore, in \cite{pagliari2024wi} RSSI-based and FTM-based UAV localization is compared. The FTM-based localization is shown to be more accurate, reducing the positioning error by 37\% compared to the RSSI-based positioning. 
Additionally, the authors of \cite{guan2024autonomous} propose an autonomous wireless positioning system that uses crowdsourced Wi-Fi fingerprinting and self-detected FTM stations in order to provide 3D in multi-storey environments.

We also identify three alternative uses of \ac{FTM} to support communications services. First, FTM-based ranging can deal with the impact of collisions on transmission rate selection, as opposed to traditional rate selection algorithms, e.g., Minstrel, which cannot distinguish between poor radio conditions and collisions. In particular, different statistical methods, \ac{KF}, \ac{PF}, and \ac{ES}, are used for \ac{RTT}-based channel quality estimation and MCS selection as shown in \cite{ciezobka2023ftmrate,ciezobka2024using}.
Secondly, \citet{mohan2023fine} propose hierarchical FTM-based time synchronization (with an accuracy below \SI{5}{\micro\second}) for multi-AP industrial networks. A designated master AP broadcasts as part of the beacon frames its ability to distribute time, which allows stations to use FTM requests to synchronize their clocks with the master AP.  
Thirdly, \citet{ashraf2022p2p} propose remote navigation of GPS denied drones (distress drones) with the help of FTM ranging. In this work, a helper drone (with GPS access) estimates the localization of distress drones using its own GPS location information. 

Finally, an emerging area of Wi-Fi application is sensing, i.e., detecting human presence, activity recognition, etc. \cite{tan2022commodity, nikseresht2023ftm, wang2023enabling}.
One of the examples is proposed by \citet{wang2023enabling}, who use \ac{FTM} measurements on different wireless channels to detect human presence: the variance is low (high) if the room is unoccupied (occupied). Based on this finding, they involve \ac{ML} classifiers (e.g., \ac{RF}, \ac{NN}) to detect the presence of humans. 
They observe better performance than using a \ac{CSI}-based approach and indicate \ac{SVM} as the best solution for this task. 
Furthermore, \citet{ando2021combining} propose an interesting sensing-based use case by combining an \ac{FTM} sensor with a \acl*{WPM} for dust concentration observation in a factory to detect unhealthy conditions.

\begin{table*}[t!]
\footnotesize
  \centering
  \caption{Summary of research in the area of FTM security. The columns names designate the following: L/P -- support for localization/positioning, ML/SM -- application of machine learning or statistical methods, EM -- method of evaluation. Evaluation methods: E -- experiments, S -- simulations.}
		\begin{tabularx}{\linewidth}{ccp{4em}p{28em}ccccc}
    \toprule
    \textbf{Key} & \textbf{Year} & \textbf{Scenario} & \textbf{Contribution} 
    & \textbf{L/P} & \textbf{Dense} & \textbf{ML} & \textbf{Conditions} & \textbf{EM} \\
    \midrule
    \cite{banin2019scalable} & 2019  & Indoor & Passive ranging for privacy-preserving localization. Distributed FTM supports more clients than legacy P2P FTM. &        Yes   & Yes   & ---    & NLOS  & E \\
    \cite{schepers2021here} & 2021  & Indoor 
    & Security recommendations against distance manipulation. FTM shown to be prone to multiple security attacks. 
    &        No    & No    & ---    & LOS   & E \\
    \cite{schepers2022privacypreserving} & 2022  & Indoor
    & Privacy-preserving indoor passive positioning. In legacy FTM implementations, passive adversaries can track any client. 
    &        Yes   & No    & ---    & LOS   & E \\
    \cite{mohsen2023privacypreserving} & 2023  & Indoor 
    & Privacy-preserving indoor passive positioning. DNN finds relationships between TDOA fingerprints and user locations. 
    & Yes   & Yes    & ML   & NLOS  & E \\
    \cite{sen2023goplaces} & 2023  & Indoor
    & Bidirectional LSTM-based model for trajectory detection combined with historical information-based prediction tree for indoor place prediction with privacy protection. 
    & Yes   & No    & ML   & Both  & E \\
    \cite{singh2023benchmarking} & 2023  & Indoor, outdoor 
    & Comparison of FTM operation in narrow (20 MHz) and wide (80 MHz) channels. Proposal of a threat-avoidance model. 
    &   Yes   & No    & ---    & LOS   & E \\
    \bottomrule
    \end{tabularx}%
  \label{tab_security}%
\end{table*}%

\section{Security Issues of 802.11mc FTM}
\label{sec:security}

Security and privacy of time-based distance measurements in indoor positioning is an important challenge. Even though \ac{ToF}-based distance estimation are inherently more secure than distance estimation based on RSS \cite{yang2024positioning, sartayeva2023survey}, current FTM implementations are prone to different attacks (Table~\ref{tab_security}).
Therefore, the security of FTM has been investigated by both researchers and the IEEE 802.11 Working Group.
In the following, we describe two security areas: attacks disrupting \ac{FTM} (e.g., affecting measurements) and privacy abuse attacks (e.g., disclosing location).
We first describe relevant academic research and then outline the security-related work presented during IEEE sessions.

In pre-\ac{FTM} research, 
\citet{glass2010insecurity} notice that adversaries can report incorrect ranging information for man-in-the-middle attacks. This problem is addressed in more detail in \citet{schepers2021here}. In particular, they observe that multiple distance reduction and enlargement attacks are possible, e.g., spoofing FTM responses, replaying whole FTM sessions, or single responses. 
As a countermeasure, they recommend using random time delays by \ac{FTM}-initiating stations as well as encryption and authentication of round-trip times. Additionally, they touch upon privacy issues and recommend using full MAC address randomization, i.e., after generating a new MAC address, its sequence number counters should be reset or randomized to prevent deanonymization. Furthermore, any kind of sequential counter should be avoided to minimize the risk of fingerprinting and identifying a given client.

\citet{schepers2022privacypreserving} continue the previously discussed work by identifying flaws in MAC address randomization: (i) FTM client's MAC address leaks, e.g., in case of wake-up frame transmission; (ii) sequence numbers may serve as a side-channel.
Additionally, they observe that different FTM initiating stations support different sets of FTM-related parameters (e.g., ASAP-capable, min delta FTM) which could be used by the attackers in client's device identification. 
As a remedy to observed problems, they propose a passive positioning system to avoid leaking client-specific information. In particular, hyperbolic localization is used for passive self-positioning supported by an infrastructure composed of FTM initiators and responders. 
A similar approach is used by \citet{banin2019scalable}.
Another privacy-preserving solution is proposed by \citet{mohsen2023privacypreserving}, in which user devices use \ac{DNN} agents, which find the relationships between passive \ac{TDoA} fingerprints and user locations. This ensures protection of user privacy by design.

\citet{sen2023goplaces} also address privacy protection by design. The proposed solution does not detect coordinate-level locations but rather identifies trajectories. It uses only a single FTM-enabled AP to estimate distance and uses smartphone sensors for place prediction. This allows preserving user privacy, since data is stored and computations are done locally at the smartphone. 

Additional security recommendations are given by \citet{singh2023benchmarking} to avoid both tracking user location and replay attacks. The authors recommend the following threat model: WPA3 or WPA2 with \ac{PMF} for improved privacy protection, packet number check in LMAC firmware to prevent replay attacks, re-keying for rogue device detection, and multi-factor authentication.

\section{Evolution of IEEE 802.11mc FTM}
\label{sec:ftm_successors}
IEEE 802.11mc (formally known as IEEE 802.11-2016) has two successors: IEEE 802.11az-2022 and IEEE 802.11bk-2025. The former introduces important changes in the operation of the FTM protocol (Table \ref{tab:az_wifi_features}). The latter extends 802.11az operation with 320~MHz channels in the 6~GHz band while supporting all the new modes of operation introduced in 802.11az \cite{henry2025wi} and described below.

\begin{table}[t]
\scriptsize
 \caption{Comparison of FTM protocol features in different IEEE~802.11 amendments \cite{80211-promo1}.}
    \centering
    \scriptsize
    \begin{tabular}{lccc}
    \toprule
    \textbf{Feature} & \textbf{802.11mc} & \textbf{802.11az} & \textbf{802.11bk} \\ 
    \midrule
    Channel width 
    [MHz] & 20-80  & 20-160  & 20-320  \\ 
    Operating bands [GHz] & 2.4, 5  & 2.4, 5, 6  & 2.4, 5, 6  \\ 
    Ranging station-to-station & $\checkmark$ & $\checkmark$ & $\checkmark$ \\ 
    Ranging station to multi-station
    & \xmark & $\checkmark$ & $\checkmark$ \\ 
    Multi-User MIMO 
    & \xmark & Up to 8x8 & Up to 8x8 \\ 
    Single TXOP range protocol 
    & \xmark & $\checkmark$ & $\checkmark$ \\ 
    LTF repetitions 
    & \xmark & $\checkmark$ & $\checkmark$ \\ 
    Protected Frames (MAC)
    & \xmark & $\checkmark$ & $\checkmark$ \\ 
    Protected LTF (PHY) 
    & \xmark & $\checkmark$ & $\checkmark$ \\ 
    Passive Ranging 
    & \xmark & $\checkmark$ & $\checkmark$ \\ 
    \midrule
    Resulting FTM accuracy & 1-2~m & $<1$~m & $<0.1$~m
    \\ \bottomrule
    \end{tabular}
   \label{tab:az_wifi_features}
\end{table}

\begin{figure*}
\captionsetup[subfigure]{justification=centering}
     \centering
     \begin{subfigure}[b]{0.62\textwidth}
    \includegraphics[width=\linewidth]{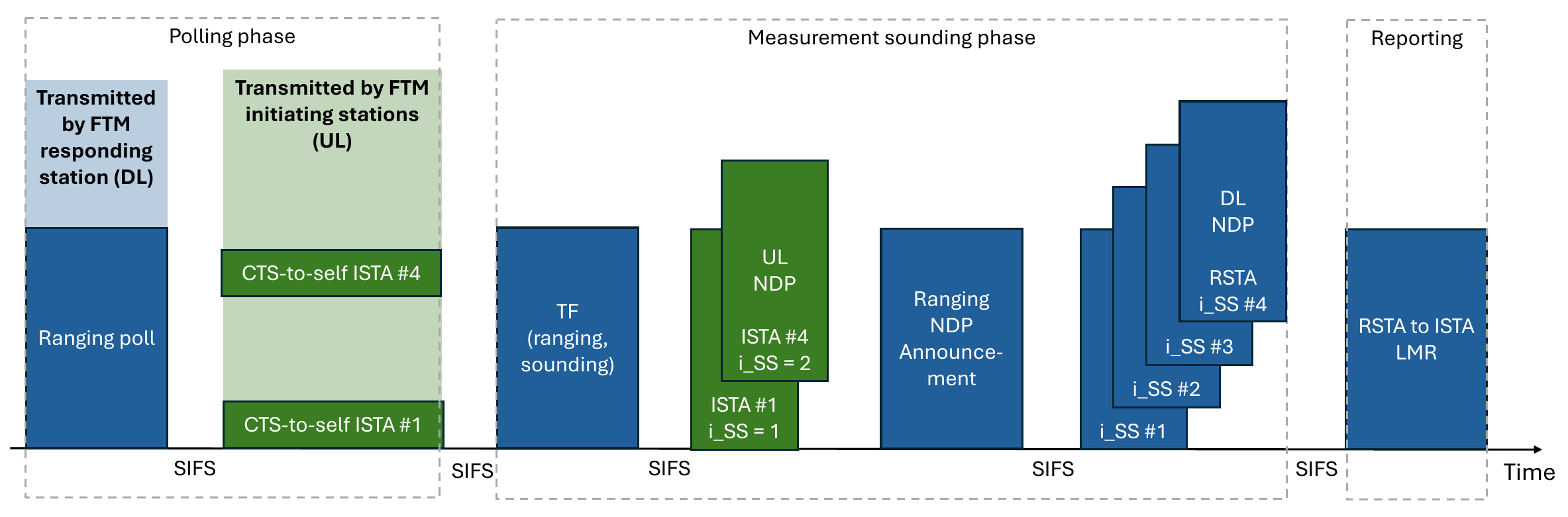}
    \caption{}
    \label{fig:tb-11az-ndp}
     \end{subfigure}
     \begin{subfigure}[b]{0.35\textwidth}
    \includegraphics[width=\linewidth]{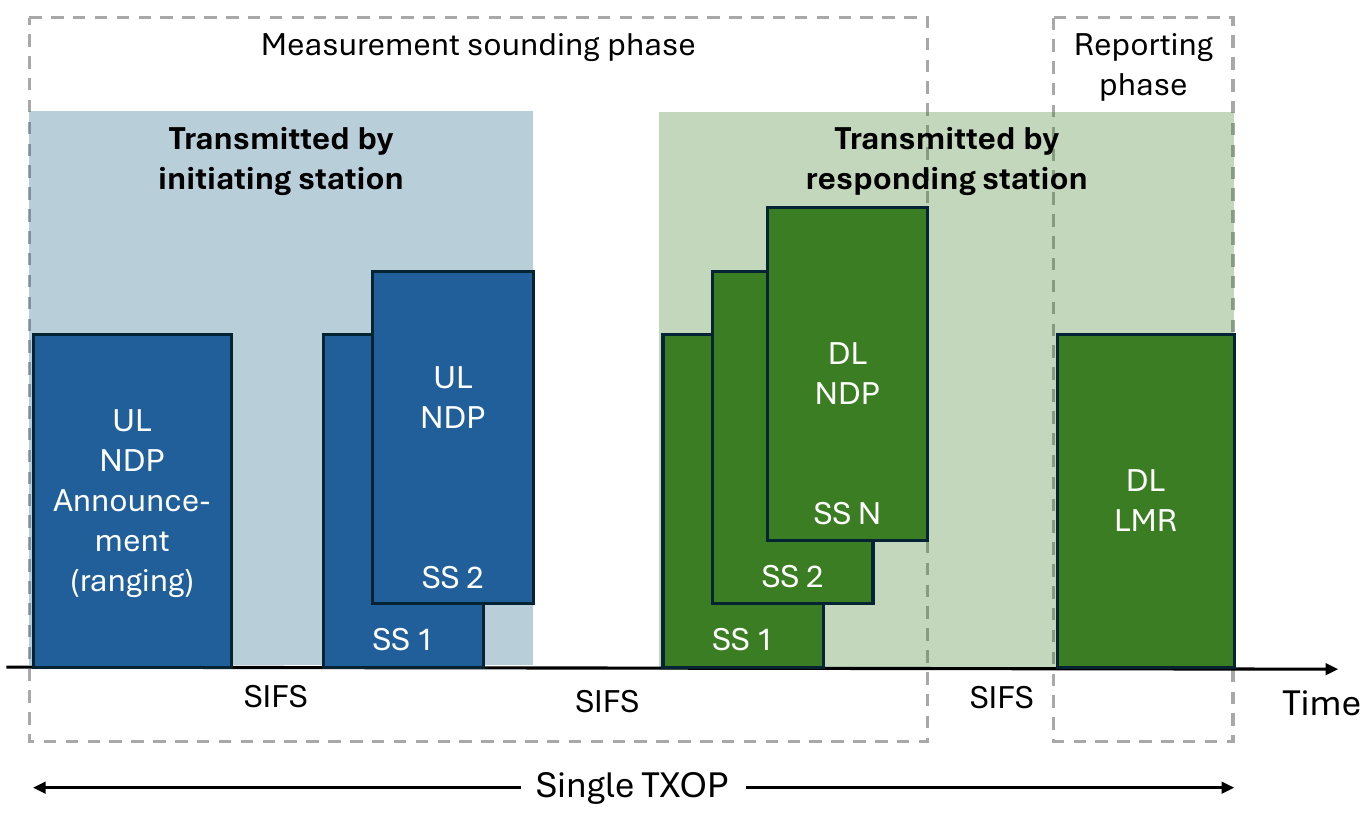}
    \caption{}
    \label{fig:11az-ndp}
     \end{subfigure}
-\caption{Frame exchange in 802.11az ranging \cite{ieee80211az}: (a) trigger-based, (b) non-trigger-based. ISTA -- initiating station, RSTA --responding station, LMR -- location measurement report.}
\label{fig:11az-tb-ntb}
\end{figure*}

\begin{figure}
    \centering
    \includegraphics[width=0.9\columnwidth]{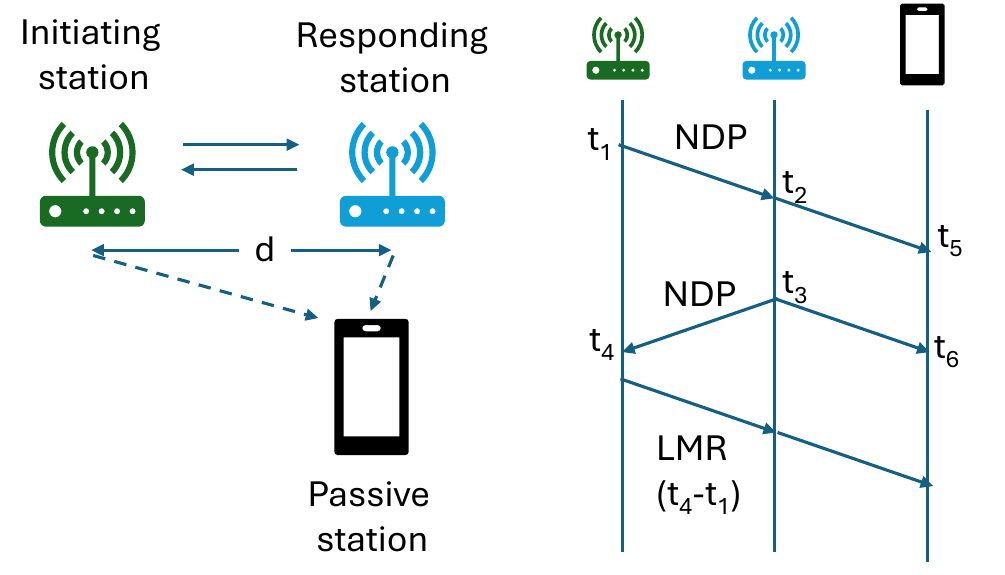}
    \caption{802.11az passive ranging \cite{80211-promo2}, where the passive station can calculate the
    TDoA \cite{martin-escalona2020passive} and use hyperbolic  positioning.
}
    \label{fig:ftm-passive}
\end{figure}

\begin{figure}
    \centering
    \includegraphics[width=0.9\columnwidth]{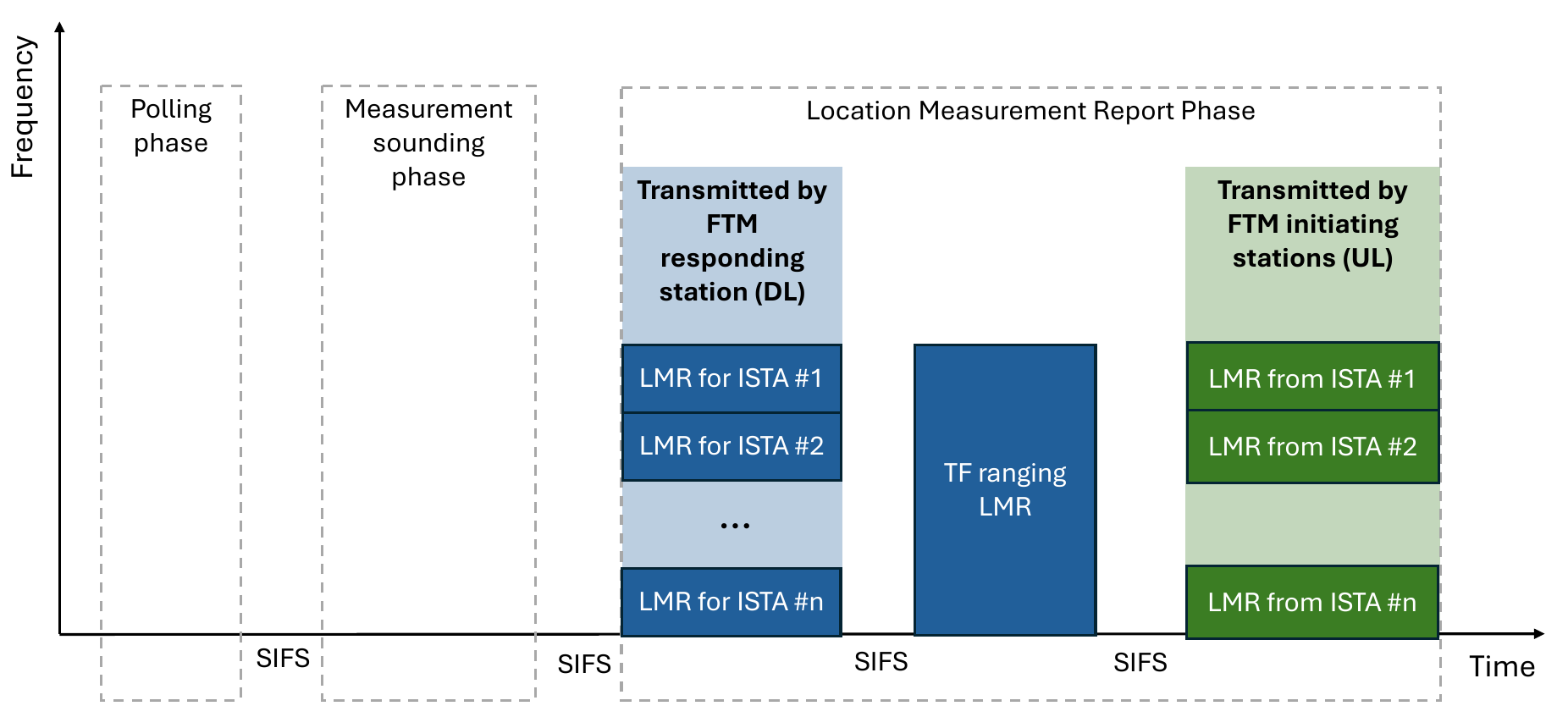}
    \caption{Trigger-based bidirectional location measurement for $n$ FTM initializing stations \cite{ieee80211az}.}
    \label{fig:tb-2wayLMR}
\end{figure}

The 802.11az amendment introduces the following new operation modes: trigger-based and non-trigger-based (for ranging from a station to multiple stations) as well as passive ranging. All three modes implement the exchange of encrypted \ac{ToF} estimations. The first two modes improve FTM scalability, by having the AP use 802.11ax MU-MIMO transmissions to provide simultaneous ranging for multiple stations (Fig.~\ref{fig:tb-11az-ndp}). 
In the passive mode, the AP broadcasts frames that can be detected by nearby stations (based on \ac{TDoA} and hyperbolic positioning, Fig.~\ref{fig:ftm-passive}).
This passive mode not only provides simultaneous ranging for an unlimited number of stations, but also preserves their privacy. 
Furthermore, 
full UL/DL MIMO is introduced, which improves FTM operation in NLOS conditions. 
Additionally, IEEE 802.11az introduces repetitions of the High Efficiency Long Training Field (HE-LTF) in the PHY header of FTM frames to improve accuracy (SNR can be increased with multiple LTFs since the noise between LTFs is not correlated) and scalability. 
Furthermore, IEEE 802.11az allows the exchange of FTM data within a single TXOP by using short PHY-based \ac{NDP} instead of longer MAC-based packets, previously used by the IEEE 802.11mc amendment, which improves scalability, saves energy, and occupies much less time on the channel (Fig.~\ref{fig:11az-ndp}). The new amendment supports wider channels for FTM ranging (up to 160 MHz). This change is important because the larger the channel, the better the accuracy of FTM ranging, according to the Cramer Rao lower bound formula  \cite{horn2022indoor}. 
Another interesting new feature is that the stations can share their timers or their computed ranges with the AP (Fig.~\ref{fig:tb-2wayLMR}). This enables new applications, such as private navigation within a store or asset tracking \cite{henry2025wi}.

A detailed analysis of the 802.11az features performed with the MATLAB next-generation positioning toolbox is available in \cite{Famili2025Unlocking}. The authors confirm that (i) larger bandwidths lead to higher ranging accuracy, (ii) 802.11az FTM combined with MIMO operation outperforms legacy 802.11mc SISO operation, and (iii) the introduced HE-LTF repetitions result in decreased ranging errors.

\begin{figure}
    \centering
    \includegraphics[width=0.9\columnwidth]{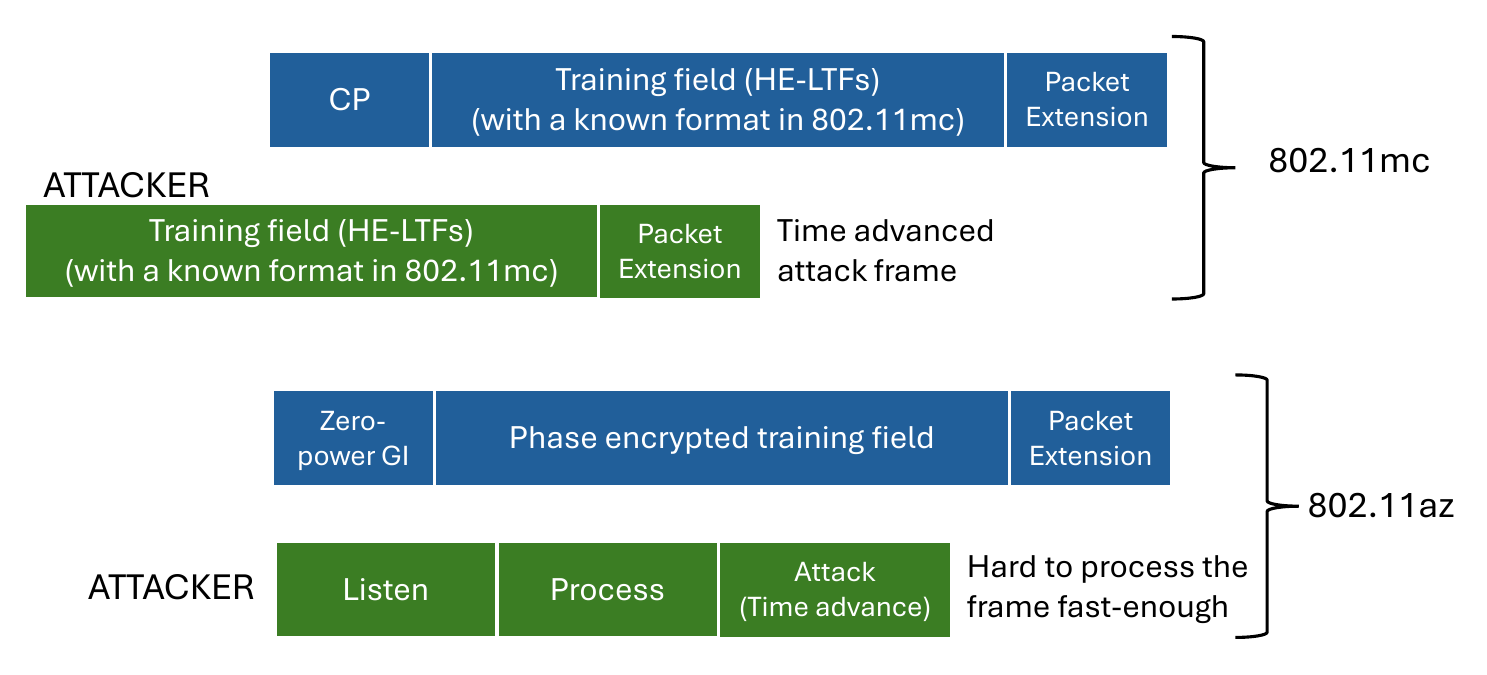}
    \caption{Time advance attack \cite{80211-promo2, ieee80211az}}
    \label{fig:time-advance-attack}
\end{figure}

IEEE 802.11az also addresses the security issues of FTM.  
For example, the original REVmc is vulnerable to eavesdropping and spoofing attacks \cite{abramovsky2017secured}.
In the former, an eavesdropper estimates the differential \ac{ToA} of multiple FTM measurement exchanges, and using its own known location and the known location of the \ac{AP}, solves the multilateration equation set used to place the station on a curve of equal time difference. The interception of multiple such curves provides the location of the attacked station. 
In the spoofing attack, an active attacker either imitates an AP and responds with false FTM time values or, during an ongoing FTM session, preemptively sends an ACK to disrupt $t_4$ calculation in (\ref{eq:rtt}).
In fact, instead of sending the whole ACK, transmitting the beginning of the frame, called the \ac{LTF}, can be sufficient.
Alternatively, attackers can collaborate to spoof the distance between the station and the AP \cite{wang2017relay}.
Therefore, as a remedy to different well-known attacks against time-based distance measurements, 802.11az defines multiple enhancements at the PHY and MAC layers: protected \ac{LTF} for defending against time-advance attacks, protected FTM frames for authenticity and privacy, and passive ranging for privacy protection. 
In case of protected LTF, single-use AES-256-based pseudo-random sequences are added in the range estimation. This protects the FTM range estimation from man-in-the-middle and other time-advance attacks (Fig.~\ref{fig:time-advance-attack}).
Further security analyses from the TGaz group can be found, e.g., in \cite{lindskog2017replay,tian2020secure,batra2020further,shellhammer2020secure}.

\section{FTM Datasets}
\label{sec:datasets}


The availability of publicly accessible datasets is crucial for advancing research in FTM-based positioning. Such datasets allow researchers to evaluate and compare different algorithms and approaches under controlled conditions. Currently, a limited number of datasets are available online.

One notable dataset\footnote{
\url{https://github.com/Fx386483710/WiFi-RTT-RSS-dataset}} encompasses FTM measurements collected in three distinct indoor environments: a building floor, an office space, and an apartment  \cite{feng2022wifi, feng2022analysis}. The dataset includes both \ac{RSS} and \ac{RTT} measurements for various reference points, along with an indicator specifying which \acp{AP} operate in \ac{LOS} conditions at each reference point. This additional information is valuable for researchers investigating the impact of channel propagation on FTM accuracy.
Another dataset\footnote{
\url{https://github.com/intel/Reference-PE-and-Measurements-DB-for-WiFi-Time-based-Scalable-Location}} offers time-delay measurements for an indoor environment. 
This dataset includes real-world measured data, along with simulated time-delay data and ground-truth client positions. Furthermore, it incorporates Matlab code for FTM positioning using a Kalman filter. However, it was gathered seven years ago, and it is no longer maintained. 
Furthermore, another dataset\footnote{
\url{https://github.com/WiFiLocalization/ESP32C3_WiFi_FTM_RSSI_Indoor_Localization}}
provides Wi-Fi FTM/RSSI measurements done at the AP, collected at three different locations, using eight different APs equipped with ESP32C3 \citet{yuen2022wi}.  
Further FTM results (for indoor, outdoor and mixed indoor/outdoor conditions)\footnote{
\url{https://github.com/IMDEANetworksWNG/MultiLoc}}, together with pcap files and Matlab scripts, are available as part of a work towards augmenting mmWave localization accuracy with the help of sub-6 GHz off-the-shelf devices \cite{blanco2022augmenting}.
Finally, more FTM results\footnote{
\url{https://github.com/IMDEANetworksWNG/UbiLocate}}, gathered with an off-the-shelf 802.11ac router, together with a guide on how to extract ToF using ASUS RT-AC86U routers with updated firmware, are available \cite{pizarro2021accurate}. This last repository is particularly useful because it allows researchers to easily perform their own tests and gather more FTM datasets in their own settings. Apart from measuring \ac{ToF} the provided implementation allows extracting \ac{CSI} information, which is an additional advantage of this resource.

Although the described datasets provide valuable resources for the research community, the limited number of currently available FTM-related datasets hinders a comprehensive exploration of FTM's potential. The development and dissemination of new, large-scale datasets that encompass diverse indoor environments and real-world deployment scenarios would significantly benefit future research efforts in FTM-based positioning. 
Additionally, completely novel types of datasets will be required to measure and validate the operation of the upcoming 802.11az/bk devices. They will surely increase the possible set of FTM-based applications, such as simultaneous, accurate, and secure localization of multiple users.



\section{Statistical Analysis}
\label{sec:statistical-analysis}


\begin{figure}
    \centering
\includegraphics[width=0.9\columnwidth]{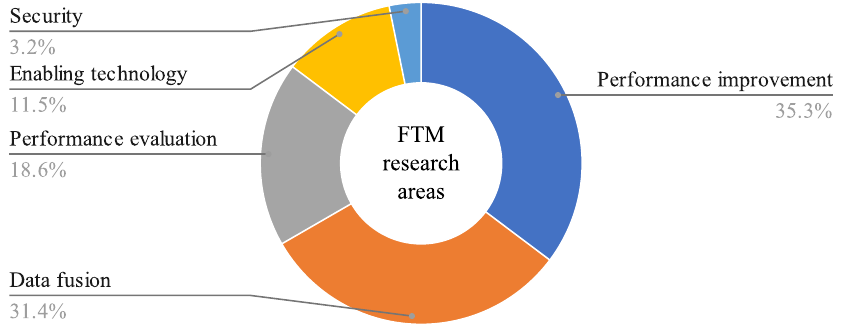}
    \caption{FTM research areas by number of related papers.}
    \label{fig:categories}
\end{figure}

Having concluded the literature review, we perform a statistical analysis of the surveyed papers.
Besides the substantial increase in \ac{FTM}-related research in recent years (Fig.~\ref{fig:cumulative_papers}), we note the following:
\begin{itemize}
\item The most active areas of research so far have been improving \ac{FTM} performance and combining \ac{FTM} with other positioning technologies (Fig.~\ref{fig:categories}). In the future, we expect more research in the area of \ac{FTM} security.
\item 
Almost all papers (about 93\%) focus on single device scenarios, i.e., the ranging or localization is performed for a single device (even if it connects with multiple \acp{AP}). 
From this, we conclude that dense FTM scenarios require further investigation.
\item 
Even though \ac{FTM} by itself provides only ranging information 
\cite{ieee802112020}, about 78\% of the research focuses on localization.
This confirms the applicability of using \ac{FTM} for indoor positioning.
\item 
About 89\% of papers include results from real trials while the remaining 11\% include simulation results. 
This is a positive trend as most Wi-Fi-related research is typically too simulation-oriented.
\item
A slight majority of works (55\%) analyze the operation of FTM in \ac{LOS} conditions, the rest considered either \ac{NLOS} conditions (18\%) or a mix of both (27\%). 
Support for \ac{NLOS} conditions is crucial for the success of \ac{FTM} and we acknowledge the importance of this research.
\end{itemize}

\begin{figure}
\captionsetup[subfigure]{justification=centering}
     \centering
     \begin{subfigure}[b]{0.3\textwidth}
          \centering
         \includegraphics[width=0.9\columnwidth]{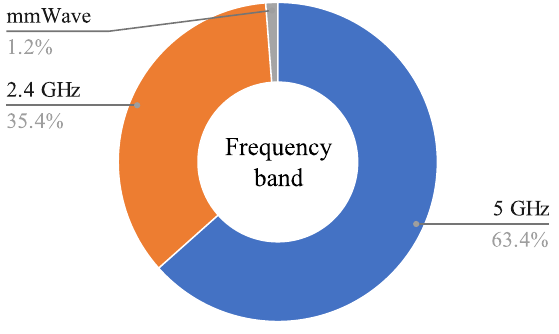}
         \caption{Frequency band}
         \label{fig:band}
     \end{subfigure}
     \begin{subfigure}[b]{0.3\textwidth}
          \centering
         \includegraphics[width=0.9\columnwidth]{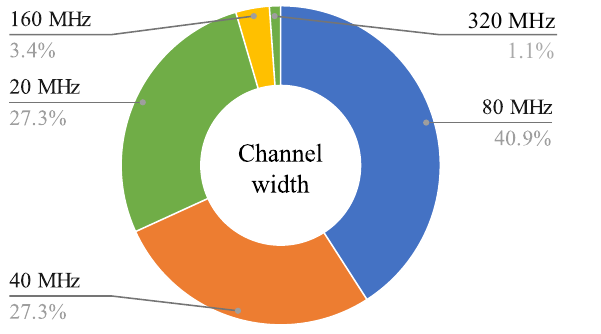}
         \caption{Channel width}
         \label{fig:channel}
     \end{subfigure} 
\caption{Configurations of FTM experiments.}
\label{fig:bands-and-channels}
\end{figure}

\begin{figure*}
\captionsetup[subfigure]{justification=centering}
     \centering
     \begin{subfigure}[b]{0.3\textwidth}
         \includegraphics[height=3cm]{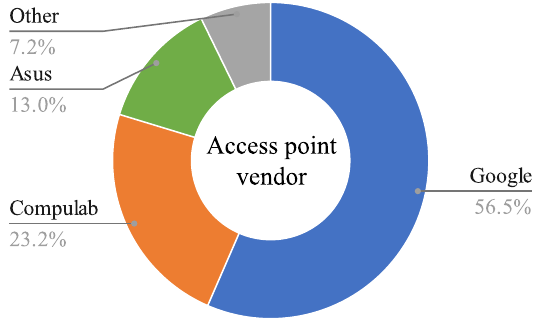}
         \caption{Access points}
         \label{fig:ap}
     \end{subfigure}
     \begin{subfigure}[b]{0.3\textwidth}
         \includegraphics[height=3cm]{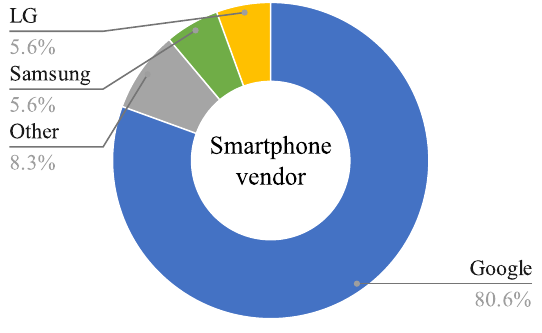}
         \caption{Smartphones}
         \label{fig:smart}
     \end{subfigure}
     \begin{subfigure}[b]{0.3\textwidth}
         \includegraphics[height=3cm]{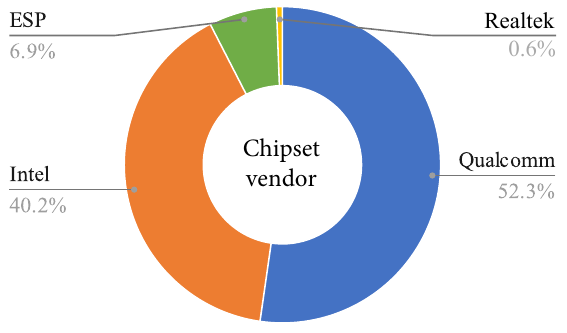}
         \caption{Wi-Fi chipsets}
         \label{fig:chip}
     \end{subfigure}
        
\caption{Most common FTM vendors used in experimental studies.}
\label{fig:devices}
\end{figure*}

We also conduct an extensive evaluation of the \ac{ML} techniques employed in the reviewed papers. We differentiate between \acf{SM}, defined as generally simpler and less resource-intensive methods, and \acf{ML}, which encompasses more advanced models whose performance typically improves with the availability of data. In this area, we arrive at the following conclusions:
\begin{itemize}
    \item The involvement of \ac{ML} is increasing in many network applications, and this trend is particularly pronounced in the context of \ac{FTM}: approximately 58.4\% of surveyed papers apply either \ac{ML} or \ac{SM}. We observe that \ac{ML} methods are typically used for data analysis, channel classification, and error correction, suggesting their growing importance in \ac{FTM} deployments.
    \item Among the surveyed papers which apply \ac{ML} or \ac{SM} methods, there is a fair division between the two groups, 
    with 28.9\% of the reviewed research papers utilizing \ac{ML} compared to 29.5\% employing \ac{SM}.
    \item The most popular \ac{SM} methods are \acl{KF}s, \acl{PF}s, and \acl{LR}, whereas the most popular \ac{ML} methods are \acl{NN}s, \acl{RF}s, and \acl{SVM}s. 
\end{itemize}
Regarding the experiments conducted within the surveyed research papers, we notice the following breakdown of configurations (Fig.~\ref{fig:bands-and-channels} and~\ref{fig:devices}):
\begin{itemize}
    \item 
In terms of frequency bands and channels, the \SI{5}{\giga\hertz} band is more frequently used than the \SI{2.4}{\giga\hertz} band. Meanwhile, the newly released \SI{6}{\giga\hertz} band has yet to receive attention.
\item
\ac{FTM} is evaluated for a number of channel widths, with only \SI{160}{\mega\hertz} and \SI{320}{\mega\hertz} channels being the least frequently selected.
In the future, we can expect more analyses of IEEE 802.11bk in \SI{320}{\mega\hertz} channels.
\item In terms of the most commonly used device and chipset vendors, we see that Google products are common (in particular different versions of Google Pixel smartphones and Google APs). This can be attributed to the early adoption of \ac{FTM} support in Android devices.
\end{itemize}


    
\section{Open Research Areas}
\label{sec:open-research}

In the previous sections, we systematically classified and discussed research papers related to Wi-Fi ranging and localization. 
All these works are related to the core idea of using the \ac{FTM} procedure of IEEE 802.11.
Although we have covered around 180 papers in this field, there are still several open research areas.
Therefore, based on our own experience and the conducted review, we outline below several future research directions related to using \ac{FTM} for indoor positioning.

\subsection{Operation in Dense Networks}
One of the main outstanding challenges in the field is the ``need for high-accuracy position estimation in dense indoor environments, while sustaining data communication services'' \cite{reddy2022multi}. 
Our survey confirms that current FTM implementations are better suited for supporting localization of single devices, which is mostly the result of the non-negligible signaling overhead. So far, researchers typically validate the existing localization techniques in non-dense scenarios and with no additional background traffic. 
Both IEEE 802.11az and 802.11bk bring new features to support positioning in such dense environments (such as support for multi-user \ac{MIMO} transmissions), but further analysis is required.

\subsection{Privacy and Security}
\citet{yang2024positioning} list the security and privacy of indoor positioning as one of the key challenges. 
Our survey shows that current FTM implementations are prone to different cyberattacks, such as the Cicada attack \cite{poturalski2010cicada}, in which the attacker predicts the preamble and payload data with 99\% accuracy, leveraging systems that rely on ``deterministic signaling with predefined data'' as well as ``early detect and late commit attacks'', in which the attacker learns the used symbol patterns. 
By-design privacy preservation would require changing FTM operation from \ac{P2P} to broadcasting or could be achieved with encryption and authentication of FTM frame exchanges. These considerations are part of the 802.11az amendment. 
However, even after devices supporting this amendment appear on the market, they will still need extensive tests against known and currently unidentified security vulnerabilities. 

Furthermore, the upcoming amendments 802.11bh and 802.11bi will also contribute to the improved security of Wi-Fi in general, including FTM \cite{ficara2023tutorial}. The 802.11bh amendment addresses the repercussions of implementing randomized and changing MAC addresses that were observed after 802.11aq formalized MAC privacy for 802.11 networks. The 802.11bi amendment provides enhanced data privacy, which is extremely important in the case of user and device identification, as often required by IoT applications \cite{wachter2018normative}.

\subsection{Impact of Puncturing}
IEEE 802.11 networks operate by default on \SI{20}{\mega\hertz} channels, which can be aggregated as a continuous frequency band up to \SI{320}{\mega\hertz} wide.
However, all constituent \SI{20}{\mega\hertz} channels remain independent, which means that, e.g., some of the 16 contiguous sub-channels forming a \SI{320}{\mega\hertz} channel allocation may be occupied at a given moment.
To maintain the throughput benefits of using wider channels without waiting for contiguous channels to be idle, IEEE 802.11be (\mbox{Wi-Fi 7}) allows non-contiguous channel allocations. This method is referred to as \textit{puncturing}.
IEEE 802.11bk, devoted to \SI{320}{\mega\hertz} positioning, defines only two static puncturing modes, removing \SI{80}{\mega\hertz} from either side of the \SI{320}{\mega\hertz} bandwidth.
Future Wi-Fi positioning systems will require greater flexibility, i.e., the support for dynamic puncturing patterns, which is particularly important in terms of coexistence, especially with narrowband technologies such as Bluetooth.
However, currently, the range performance of IEEE 802.11bk under such dynamic conditions remains unknown.

\subsection{Narrowband Operation}
\ac{FTM} measures \ac{ToF} using a wideband first-path detection approach.
An alternative is the phase-based approach used in narrowband technologies such as \ac{BLE} \cite{zand2019high}.
Meanwhile, an upcoming subset of new Wi-Fi devices based on 802.11bp (for ambient power \ac{IoT} communication) will also use narrowband operation, with channels as narrow as \SI{2}{\mega\hertz}, and will also benefit from ranging services.
Therefore, a suggested research area is the performance comparison of narrowband phase-based and wideband first-path detection \ac{ToF} measurements.

\subsection{Improving Communication Services}
\ac{FTM} provides more reliable measurements than RSS-based methods, typically used by Wi-Fi controllers, because the results are irrespective of device characteristics including antenna placement, gain, etc.
Therefore, \ac{FTM}-based measurements are more suitable to facilitate network-level decisions in such controllers.
For example, \ac{FTM} can be part of a method to determine whether devices are indoor or not.
This is crucial for operating in \SI{6}{\giga\hertz} bands, which require indoor operation for Wi-Fi devices \cite{dogan2023indoor}.
Therefore, an open research challenge is to determine how \ac{FTM} can improve this and other communication services.

\subsection{Residual Frequency Offset}
One major obstacle in \ac{FTM}'s \ac{ToF} estimation is the inaccuracy of the local electronic oscillator.
Since both communicating devices have independent oscillators, each with an accuracy of up to 20~ppm, this causes an error of up to \SI{30}{\centi\meter}.
\ac{FTM} deals with this issue by estimating the clock offset at the initiating station when considering reports from the responding station (Fig.~\ref{fig:ftm}).
This method is useful to a large extent; however, a residual frequency offset may still exist and may limit performance.
A method of canceling this offset is to perform three-way probing to fully eliminate the difference between the local oscillators.
Therefore, an important area of research is to evaluate the value of transitioning from single-sided (existing two-way sounding) to double-sided (three-way sounding) \ac{FTM} for improving the system's accuracy.

\section{Conclusions}
\label{sec:conclusions}

The IEEE 802.11 \ac{FTM} protocol, commercially known as Wi-Fi Location, has great potential to solve the challenge of indoor positioning.
One of its key strengths is that, unlike several competing technologies, it is not a dedicated technology which requires deployment of additional infrastructure and special radio interfaces.
Conversely, it is a service added to Wi-Fi, the most prominent indoor technology available, making it a cost-effective solution.
Furthermore, \ac{FTM} is being actively updated by the standardization community with 802.11az released in 2023 and 802.11bk already in development.
Both of these amendments are evolving together along with the updates brought by the mainstream developments (Wi-Fi 6 and 7, respectively).

\ac{FTM} has met with great interest by the research community as evidenced by the rapidly increasing number of published articles.
Having surveyed the related literature, we summarize the arising conclusions as follows.
First, we identify four main performance challenges for current devices: (i) ensuring proper calibration of \ac{FTM} to each new environment and for each device type, (ii) addressing the impact of NLOS paths, (iii) optimizing overhead in dense deployments, and (iv) providing localization privacy and FTM security.
Next, we identify two main gaps. There is a lack of 802.11az/bk performance analyses with only a few available papers. We expect this to change once the supporting devices become available.
We also note that there are only several available datasets, and therefore we encourage the research community to produce more to benchmark new solutions.
We also identify three arising opportunities which can positively impact Wi-Fi-based indoor positioning: 
(i) \ac{FTM} can potentially operate in unlicensed bands in the extremely high frequency range (mmWave), where beamforming allows higher accuracy \cite{picazo2023ieee},
(ii) \ac{ML} can improve \ac{FTM} performance but researchers should be wary of the cost of training and portability of solutions to new environments, and
(iii) integration with other positioning technologies, especially data fusion with \ac{INS}, is a promising approach to further improve accuracy, especially for handheld devices.
In summary, we believe that Wi-Fi offers a promising perspective for achieving seamless and ubiquitous indoor positioning.

\section{List of Acronyms}
\label{app_acronyms}
\printacronyms[name={}]

\section*{Acknowledgments}
This work was supported by the National Science Centre, Poland (2020/39/I/ST7/01457). 












\bibliographystyle{plainnat}
\bibliography{bibliography}

\begin{thebibliography}{232}
\providecommand{\natexlab}[1]{#1}
\providecommand{\url}[1]{\texttt{#1}}
\expandafter\ifx\csname urlstyle\endcsname\relax
  \providecommand{\doi}[1]{doi: #1}\else
  \providecommand{\doi}{doi: \begingroup \urlstyle{rm}\Url}\fi

\bibitem[Abdulrazzaq et~al.(2024)]{abdulrazzaq2024wireless}
Mohammed~Wahhab Abdulrazzaq et~al.
\newblock Wireless fine time measurement implementation and analysis in indoor factory environments.
\newblock In \emph{2024 ICETI}. IEEE, 2024.

\bibitem[Abramovsky et~al.(2017)Abramovsky, Bar-Shalom, and Ghosh]{abramovsky2017secured}
Benny Abramovsky, Ofer Bar-Shalom, and Chittabrata Ghosh.
\newblock Secured location threat model.
\newblock IEEE 802.11 TGaz IEEE 802.11-17/0120-02, Intel, 2017.
\newblock https://mentor.ieee.org/802.11/dcn/17/11-17-0120-02-00az-secured-location-threat-model.pptx.

\bibitem[Aggarwal et~al.(2022)]{aggarwal2022wifi}
Shivang Aggarwal et~al.
\newblock Is wifi 802.11mc fine time measurement ready for prime-time localization?
\newblock In \emph{Proc. of the 16th {{ACM Workshop}} on {{Wireless Network Testbeds}}, {{Experimental}} Evaluation \& {{CHaracterization}}}. {ACM}, 2022.
\newblock ISBN 978-1-4503-9527-4.

\bibitem[{\'A}lvarez-Merino et~al.(2022{\natexlab{a}}){\'A}lvarez-Merino, Khatib, et~al.]{alvarez-merino2022wifi}
Carlos~S. {\'A}lvarez-Merino, Emil~J. Khatib, et~al.
\newblock {{WiFi FTM}} and {{UWB Characterization}} for {{Localization}} in {{Construction Sites}}.
\newblock \emph{Sensors}, 22\penalty0 (14):\penalty0 5373, 7 2022{\natexlab{a}}.
\newblock ISSN 1424-8220.
\newblock \doi{10.3390/s22145373}.

\bibitem[{\'A}lvarez-Merino et~al.(2021)]{alvarez-merino2021wifi}
Carlos~S. {\'A}lvarez-Merino et~al.
\newblock {{WiFi FTM}}, {{UWB}} and {{Cellular-Based Radio Fusion}} for {{Indoor Positioning}}.
\newblock \emph{Sensors}, 21\penalty0 (21):\penalty0 7020, 10 2021.
\newblock ISSN 1424-8220.
\newblock \doi{10.3390/s21217020}.

\bibitem[{\'A}lvarez-Merino et~al.(2022{\natexlab{b}})]{alvarez2022victim}
Carlos~S {\'A}lvarez-Merino et~al.
\newblock Victim detection and localization in emergencies.
\newblock \emph{Sensors}, 22\penalty0 (21):\penalty0 8433, 2022{\natexlab{b}}.

\bibitem[Ando et~al.(2021)Ando, Sekoguchi, et~al.]{ando2021combining}
Hajime Ando, Shingo Sekoguchi, et~al.
\newblock Combining {{Indoor Positioning Using Wi-Fi Round Trip Time}} with {{Dust Measurement}} in the {{Field}} of {{Occupational Health}}.
\newblock \emph{Sensors}, 21\penalty0 (21):\penalty0 7261, 10 2021.
\newblock ISSN 1424-8220.
\newblock \doi{10.3390/s21217261}.

\bibitem[Asaad and Maghdid(2022)]{asaad2022comprehensive}
Safar~M Asaad and Halgurd~S Maghdid.
\newblock A comprehensive review of indoor/outdoor localization solutions in iot era: Research challenges and future perspectives.
\newblock \emph{Computer Networks}, 212:\penalty0 109041, 2022.

\bibitem[Ashraf and Ashok(2022)]{ashraf2022p2p}
Khadija Ashraf and Ashwin Ashok.
\newblock P2p-droneloc: Peer-to-peer localization for gps-denied drones using camera and wifi fine time measurement.
\newblock In \emph{2022 IEEE ANTS}, pages 1--6. IEEE, 2022.

\bibitem[Bahillo et~al.(2010)Bahillo, Mazuelas, et~al.]{bahillo2010accurate}
Alfonso Bahillo, Santiago Mazuelas, et~al.
\newblock Accurate and integrated localization system for indoor environments based on ieee 802.11 round-trip time measurements.
\newblock \emph{EURASIP Journal on Wireless Communications and Networking}, 2010, 2010.

\bibitem[Bai et~al.(2020)Bai, Kealy, et~al.]{bai2020comparative}
Yuntian~Brian Bai, Allison Kealy, et~al.
\newblock A comparative evaluation of wi-fi rtt and gps based positioning.
\newblock In \emph{Proceedings of the international global navigation satellite systems IGNSS 2020 conference, Sydney, Australia}, pages 5--7, 2020.

\bibitem[Banin et~al.(2013)Banin, Schtzberg, and Amizur]{banin2013next}
Leor Banin, U~Schtzberg, and Yuval Amizur.
\newblock Next generation indoor positioning system based on wifi time of flight.
\newblock In \emph{ION GNSS+ 2013}, pages 975--982, 2013.

\bibitem[Banin et~al.(2016)Banin, Schatzberg, and Amizur]{banin2016wifi}
Leor Banin, Uri Schatzberg, and Yuval Amizur.
\newblock {{WiFi FTM}} and {{Map Information Fusion}} for {{Accurate Positioning}}.
\newblock \emph{International Conference on Indoor Positioning and Indoor Navigation}, 2016.

\bibitem[Banin et~al.(2019)Banin, {Bar-Shalom}, Dvorecki, and Amizur]{banin2019scalable}
Leor Banin, Ofer {Bar-Shalom}, Nir Dvorecki, and Yuval Amizur.
\newblock Scalable {{Wi-Fi Client Self-Positioning Using Cooperative FTM-Sensors}}.
\newblock \emph{IEEE Transactions on Instrumentation and Measurement}, 68\penalty0 (10):\penalty0 3686--3698, 10 2019.
\newblock ISSN 0018-9456, 1557-9662.

\bibitem[Barral~Vales et~al.(2022)Barral~Vales, Fernandez, et~al.]{barralvales2022fine}
Valentin Barral~Vales, Omar~Campos Fernandez, et~al.
\newblock Fine {{Time Measurement}} for the {{Internet}} of {{Things}}: {{A Practical Approach Using ESP32}}.
\newblock \emph{IEEE Internet of Things Journal}, 9\penalty0 (19):\penalty0 18305--18318, 10 2022.
\newblock ISSN 2327-4662, 2372-2541.

\bibitem[Batra et~al.(2020)]{batra2020further}
Anuj Batra et~al.
\newblock Further updates on 11az secure ltf design.
\newblock IEEE 802.11 TGaz IEEE 802.11-20/1855-00, Apple, 2020.
\newblock https://mentor.ieee.org/802.11/dcn/20/11-20-1855-00-00az-further-updates-on-11az-secure-ltf-design.pptx.

\bibitem[Batstone et~al.(2016{\natexlab{a}})Batstone, Oskarsson, and {\AA}str{\"o}m]{batstone2016robust}
Kenneth Batstone, Magnus Oskarsson, and Kalle {\AA}str{\"o}m.
\newblock Robust time-of-arrival self calibration and indoor localization using wi-fi round-trip time measurements.
\newblock In \emph{2016 IEEE ICC}, pages 26--31. IEEE, 2016{\natexlab{a}}.

\bibitem[Batstone et~al.(2016{\natexlab{b}})Batstone, Oskarsson, and {\AA}str{\"o}m]{batstone2016robusta}
Kenneth Batstone, Magnus Oskarsson, and Kalle {\AA}str{\"o}m.
\newblock Robust time-of-arrival self calibration with missing data and outliers.
\newblock In \emph{2016 EUSIPCO}, pages 2370--2374. IEEE, 2016{\natexlab{b}}.

\bibitem[Biehl et~al.(2019)Biehl, Girgensohn, and Patel]{biehl2019achieving}
Jacob~T Biehl, Andreas Girgensohn, and Mitesh Patel.
\newblock Achieving accurate room-level indoor location estimation with emerging iot networks.
\newblock In \emph{Proceedings of the 9th International Conference on the Internet of Things}, pages 1--8, 2019.

\bibitem[Blanco et~al.(2022)]{blanco2022augmenting}
Alejandro Blanco et~al.
\newblock Augmenting mmwave localization accuracy through sub-6 ghz on off-the-shelf devices.
\newblock In \emph{Proceedings of the 20th Annual International Conference on Mobile Systems, Applications and Services}, 2022.

\bibitem[Bullmann et~al.(2020)Bullmann, Fetzer, et~al.]{bullmann2020comparison}
Markus Bullmann, Toni Fetzer, et~al.
\newblock Comparison of 2.4 {{GHz WiFi FTM-}} and {{RSSI-Based Indoor Positioning Methods}} in {{Realistic Scenarios}}.
\newblock \emph{Sensors}, 20\penalty0 (16):\penalty0 4515, 9 2020.
\newblock ISSN 1424-8220.
\newblock \doi{10.3390/s20164515}.

\bibitem[Bullmann et~al.(2022)Bullmann, Fetzer, et~al.]{bullmann2022data}
Markus Bullmann, Toni Fetzer, et~al.
\newblock Data {{Driven Sensor Model}} for {{Wi-Fi Fine Timing Measurement}}.
\newblock In \emph{2022 IEEE IPIN}, pages 1--8, {Beijing, China}, 8 2022. {IEEE}.
\newblock ISBN 978-1-72816-218-8.
\newblock \doi{10.1109/IPIN54987.2022.9918111}.

\bibitem[Busnel and Rivano(2023)]{busnel2023ftmbroadcast}
Yann Busnel and Herv{\'e} Rivano.
\newblock Ftm-broadcast: Efficient network-wide ranging.
\newblock In \emph{IEEE IPIN 2023}. IEEE, 2023.

\bibitem[Cao et~al.(2022)Cao, Alali, et~al.]{cao2022vitag}
Bryan~Bo Cao, Abrar Alali, et~al.
\newblock {{ViTag}}: {{Online WiFi Fine Time Measurements Aided Vision-Motion Identity Association}} in {{Multi-person Environments}}.
\newblock In \emph{2022 IEEE SECON}, pages 19--27, {Stockholm, Sweden}, 8 2022. {IEEE}.
\newblock ISBN 978-1-66548-643-9.

\bibitem[Cao et~al.(2020{\natexlab{a}})Cao, Wang, Bi, Xu, Qi, Si, and Yao]{cao2020wifi}
Hongji Cao, Yunjia Wang, Jingxue Bi, Shenglei Xu, Hongxia Qi, Minghao Si, and Guobiao Yao.
\newblock {{WiFi RTT Indoor Positioning Method Based}} on {{Gaussian Process Regression}} for {{Harsh Environments}}.
\newblock \emph{IEEE Access}, 8, 2020{\natexlab{a}}.
\newblock ISSN 2169-3536.

\bibitem[Cao et~al.(2021)Cao, Wang, and Bi]{cao2021smartphones}
Hongji Cao, Yunjia Wang, and Jingxue Bi.
\newblock Smartphones: {{3D Indoor Localization Using Wi-Fi RTT}}.
\newblock \emph{IEEE Communications Letters}, 25\penalty0 (4):\penalty0 1201--1205, 4 2021.
\newblock ISSN 1089-7798, 1558-2558, 2373-7891.
\newblock \doi{10.1109/LCOMM.2020.3044714}.

\bibitem[Cao et~al.(2024)Cao, Wang, et~al.]{cao2024los}
Hongji Cao, Yunjia Wang, et~al.
\newblock {LOS compensation and trusted NLOS recognition assisted WiFi RTT indoor positioning algorithm}.
\newblock \emph{Expert Systems with Applications}, 243:\penalty0 122867, 2024.

\bibitem[Cao et~al.(2020{\natexlab{b}})]{cao2020indoor}
Hongji Cao et~al.
\newblock Indoor {{Positioning Method Using WiFi RTT Based}} on {{LOS Identification}} and {{Range Calibration}}.
\newblock \emph{ISPRS International Journal of Geo-Information}, 9\penalty0 (11):\penalty0 627, 10 2020{\natexlab{b}}.
\newblock ISSN 2220-9964.
\newblock \doi{10.3390/ijgi9110627}.

\bibitem[Chan et~al.(2021)Chan, Lai, and Wu]{chan2021transfer}
Hao-Wei Chan, Alexander I-Chi Lai, and Ruey-Beei Wu.
\newblock Transfer {{Learning}} of {{Wi-Fi FTM Responder Positioning}} with {{NLOS Identification}}.
\newblock In \emph{2021 IEEE WiSNeT}, pages 23--26, {San Diego, CA, USA}, 1 2021. {IEEE}.
\newblock ISBN 978-1-66541-581-1.

\bibitem[Chan et~al.(2022)Chan, Wu, et~al.]{chan2022fusionbased}
Hao-Wei Chan, Pei-Yuan Wu, et~al.
\newblock Fusion-{{Based Smartphone Positioning Using Unsupervised Calibration}} of {{Crowdsourced Wi-Fi FTM}}.
\newblock \emph{IEEE Access}, 10:\penalty0 96260--96272, 2022.
\newblock ISSN 2169-3536.
\newblock \doi{10.1109/ACCESS.2022.3204799}.

\bibitem[Chataut et~al.(2024)Chataut, Nankya, and Akl]{chataut20246g}
Robin Chataut, Mary Nankya, and Robert Akl.
\newblock 6g networks and the ai revolution—exploring technologies, applications, and emerging challenges.
\newblock \emph{Sensors}, 24\penalty0 (6):\penalty0 1888, 2024.

\bibitem[Chen et~al.(2011)Chen, Pei, and Chen]{chen2011smart}
Ruizhi Chen, Ling Pei, and Yuwei Chen.
\newblock A smart phone based pdr solution for indoor navigation.
\newblock In \emph{ION GNSS 2011}, pages 1404--1408, 2011.

\bibitem[Chen et~al.(2015)Chen, Zou, Jiang, Zhu, Soh, and Xie]{chen2015fusion}
Zhenghua Chen, Han Zou, Hao Jiang, Qingchang Zhu, Yeng~Chai Soh, and Lihua Xie.
\newblock Fusion of wifi, smartphone sensors and landmarks using the kalman filter for indoor localization.
\newblock \emph{Sensors}, 15\penalty0 (1):\penalty0 715--732, 2015.

\bibitem[Chigullapally and Patnaik(2020)]{chigullapally2020wifi}
Mohith Chigullapally and Swabhiman Patnaik.
\newblock Wi-{{Fi}} 802.11mc {{Distance Classification}} and {{Error Reduction}} using {{Machine Learning}}.
\newblock In \emph{2020 ICCES}, pages 599--604, {Coimbatore, India}, 6 2020. {IEEE}.
\newblock ISBN 978-1-72815-371-1.
\newblock \doi{10.1109/ICCES48766.2020.9137922}.

\bibitem[Choi(2021)]{choi2021enhanced}
Jeongsik Choi.
\newblock Enhanced {{Wi-Fi RTT Ranging}}: {{A Sensor-Aided Learning Approach}}, 12 2021.

\bibitem[Choi and Choi(2021)]{choi2021calibrationfree}
Jeongsik Choi and Yang-Seok Choi.
\newblock Calibration-{{Free Positioning Technique Using Wi-Fi Ranging}} and {{Built-in Sensors}} of {{Mobile Devices}}.
\newblock \emph{IEEE Internet of Things Journal}, 8\penalty0 (1):\penalty0 541--554, 1 2021.
\newblock ISSN 2327-4662, 2372-2541.
\newblock \doi{10.1109/JIOT.2020.3004774}.

\bibitem[Choi et~al.(2019)Choi, Choi, and Talwar]{choi2019unsupervised}
Jeongsik Choi, Yang-Seok Choi, and Shilpa Talwar.
\newblock Unsupervised {{Learning Techniques}} for {{Trilateration}}: {{From Theory}} to {{Android APP Implementation}}.
\newblock \emph{IEEE Access}, 7:\penalty0 134525--134538, 2019.
\newblock ISSN 2169-3536.
\newblock \doi{10.1109/ACCESS.2019.2941657}.

\bibitem[Ciezobka et~al.(2023)Ciezobka, Wojnar, Kosek-Szott, Szott, and Rusek]{ciezobka2023ftmrate}
Wojciech Ciezobka, Maksymilian Wojnar, Katarzyna Kosek-Szott, Szymon Szott, and Krzysztof Rusek.
\newblock {{FTMRate}}: {{Collision-Immune Distance-based Data Rate Selection}} for {{IEEE}} 802.11 {{Networks}}.
\newblock In \emph{2023 {{IEEE}} WoWMoM}, pages 242--251, {Boston, MA, USA}, 6 2023. {IEEE}.

\bibitem[Ciezobka et~al.(2024)Ciezobka, Wojnar, Rusek, Kosek-Szott, Szott, Zubow, and Dressler]{ciezobka2024using}
Wojciech Ciezobka, Maksymilian Wojnar, Krzysztof Rusek, Katarzyna Kosek-Szott, Szymon Szott, Anatolij Zubow, and Falko Dressler.
\newblock {Using ranging for collision-immune IEEE 802.11 rate selection with statistical learning}.
\newblock \emph{Computer Communications}, 225:\penalty0 10--26, 2024.

\bibitem[Ciurana et~al.(2007)Ciurana, Barcelo-Arroyo, and Izquierdo]{ciurana2007ranginga}
M~Ciurana, F~Barcelo-Arroyo, and F~Izquierdo.
\newblock A ranging system with ieee 802.11 data frames.
\newblock In \emph{2007 IEEE radio and wireless symposium}, pages 133--136. IEEE, 2007.

\bibitem[Ciurana et~al.(2009)Ciurana, L{\'o}pez, and Barcel{\'o}-Arroyo]{ciurana2009softoa}
Marc Ciurana, David L{\'o}pez, and Francisco Barcel{\'o}-Arroyo.
\newblock Softoa: Software ranging for toa-based positioning of wlan terminals.
\newblock In \emph{Location and Context Awareness}, pages 207--221. Springer Berlin Heidelberg, 2009.
\newblock ISBN 978-3-642-01721-6.

\bibitem[Ciurana et~al.(2011)Ciurana, Barcel{\'o}-Arroyo, and Mart{\'\i}n-Escalona]{ciurana2011comparative}
Marc Ciurana, Francisco Barcel{\'o}-Arroyo, and Israel Mart{\'\i}n-Escalona.
\newblock {Comparative performance evaluation of IEEE 802.11v for positioning with time of arrival}.
\newblock \emph{Computer Standards \& Interfaces}, 33\penalty0 (3):\penalty0 344--349, 2011.

\bibitem[Dabove et~al.(2018)Dabove, Di~Pietra, et~al.]{dabove2018}
Paolo Dabove, Vincenzo Di~Pietra, et~al.
\newblock Indoor positioning using ultra-wide band (uwb) technologies: Positioning accuracies and sensors' performances.
\newblock In \emph{2018 IEEE/ION PLANS}, pages 175--184, 2018.
\newblock \doi{10.1109/PLANS.2018.8373379}.

\bibitem[Dai et~al.(2023)Dai, Wang, et~al.]{dai2023survey}
Jihan Dai, Maoyi Wang, et~al.
\newblock A survey of latest wi-fi assisted indoor positioning on different principles.
\newblock \emph{Sensors}, 23\penalty0 (18):\penalty0 7961, 2023.

\bibitem[Diggelen et~al.(2018)Diggelen, Want, and Wang]{diggelen2018how}
Frank~Van Diggelen, Roy Want, and Wei Wang.
\newblock How to achieve 1-meter accuracy in android, 2018.
\newblock URL \url{https://www.gpsworld.com/how-to-achieve-1-meter-accuracy-in-android/}.
\newblock Accessed on 2025-04-24.

\bibitem[Dogan-Tusha et~al.(2023)Dogan-Tusha, Tusha, et~al.]{dogan2023indoor}
Seda Dogan-Tusha, Armed Tusha, et~al.
\newblock Indoor and outdoor measurement campaign for unlicensed 6 ghz operation with wi-fi 6e.
\newblock In \emph{2023 WPMC}, pages 1--6. IEEE, 2023.

\bibitem[Domuta and Palade(2021)]{domuta2021two}
Ioan Domuta and Tudor~Petru Palade.
\newblock Two-way ranging algorithms for clock error compensation.
\newblock \emph{IEEE Transactions on Vehicular Technology}, 70\penalty0 (8):\penalty0 8237--8250, 2021.

\bibitem[Dong et~al.(2023)Dong, Rana, Cui, et~al.]{dong2023investigation}
Jiabin Dong, Lila Rana, Shuyu Cui, et~al.
\newblock Investigation on indoor positioning by improved rtt-rss fusion ranging method.
\newblock In \emph{2023 IEEE ICECE}, pages 49--53. IEEE, 2023.

\bibitem[Dong et~al.(2022{\natexlab{a}})Dong, Arslan, and Yang]{dong2022realtime}
Yinhuan Dong, Tughrul Arslan, and Yunjie Yang.
\newblock Real-{{Time NLOS}}/{{LOS Identification}} for {{Smartphone-Based Indoor Positioning Systems Using WiFi RTT}} and {{RSS}}.
\newblock \emph{IEEE Sensors Journal}, 22\penalty0 (6):\penalty0 5199--5209, 3 2022{\natexlab{a}}.
\newblock ISSN 1530-437X, 1558-1748, 2379-9153.

\bibitem[Dong et~al.(2022{\natexlab{b}})Dong, Shi, et~al.]{dong2022error}
Yinhuan Dong, Duanxu Shi, et~al.
\newblock Error {{Investigation}} on {{Wi-Fi RTT}} in {{Commercial Consumer Devices}}.
\newblock \emph{Algorithms}, 15\penalty0 (12):\penalty0 464, 12 2022{\natexlab{b}}.
\newblock ISSN 1999-4893.
\newblock \doi{10.3390/a15120464}.

\bibitem[D{\"u}mbgen et~al.(2019)D{\"u}mbgen, Oeschger, et~al.]{dumbgen2019multi}
Frederike D{\"u}mbgen, Cynthia Oeschger, et~al.
\newblock Multi-modal probabilistic indoor localization on a smartphone.
\newblock In \emph{2019 IPIN}, pages 1--8. IEEE, 2019.

\bibitem[Dvorecki et~al.(2019)Dvorecki, Bar-Shalom, et~al.]{dvorecki2019machine}
Nir Dvorecki, Ofer Bar-Shalom, et~al.
\newblock A machine learning approach for wi-fi rtt ranging.
\newblock In \emph{Proceedings of the 2019 international technical meeting of the institute of navigation}, pages 435--444, 2019.

\bibitem[Eleftherakis et~al.(2024)Eleftherakis, Santaromita, et~al.]{eleftherakis2024spring+}
Stavros Eleftherakis, Giuseppe Santaromita, et~al.
\newblock Spring+: Smartphone positioning from a single wifi access point.
\newblock \emph{IEEE Transactions on Mobile Computing}, 2024.

\bibitem[Famili et~al.(2025)Famili, Atalay, and Stavrou]{Famili2025Unlocking}
Alireza Famili, Tolga Atalay, and Angelos Stavrou.
\newblock Unlocking the potential of ieee 802.11az: A deep dive into ranging capabilities.
\newblock In \emph{2025 International Conference on Computing, Networking and Communications (ICNC)}, 2025.

\bibitem[Feng et~al.(2022{\natexlab{a}})Feng, Nguyen, and Luo]{feng2022wifi}
Xu~Feng, Khuong~An Nguyen, and Zhiyuan Luo.
\newblock {{WiFi Access Points Line-of-Sight Detection}} for {{Indoor Positioning Using}} the {{Signal Round Trip Time}}.
\newblock \emph{Remote Sensing}, 14\penalty0 (23):\penalty0 6052, 11 2022{\natexlab{a}}.
\newblock ISSN 2072-4292.
\newblock \doi{10.3390/rs14236052}.

\bibitem[Feng et~al.(2024)Feng, Nguyen, and Luo]{feng2024wifi}
Xu~Feng, Khuong~An Nguyen, and Zhiyuan Luo.
\newblock A wi-fi rss-rtt indoor positioning model based on dynamic model switching algorithm.
\newblock \emph{IEEE Journal of Indoor and Seamless Positioning and Navigation}, 2:\penalty0 151--165, 2024.

\bibitem[Feng et~al.(2022{\natexlab{b}})]{feng2022analysis}
Xu~Feng et~al.
\newblock An {{Analysis}} of the {{Properties}} and the {{Performance}} of {{WiFi RTT}} for {{Indoor Positioning}} in {{Non-line-of-sight Environments}}.
\newblock \emph{17th International Conference on Location Based Services}, 2022{\natexlab{b}}.

\bibitem[Fetzer et~al.(2023)Fetzer, Bullmann, et~al.]{fetzer2023interacting}
Toni Fetzer, Markus Bullmann, et~al.
\newblock Interacting {{Multiple Model Particle Filter}} for {{Indoor Positioning Applications}}.
\newblock In \emph{2023 {{International Technical Meeting}} of {{The Institute}} of {{Navigation}}}, pages 1089--1100, {Long Beach, California}, 2 2023.

\bibitem[Ficara et~al.(2024)Ficara, Garroppo, and Henry]{ficara2023tutorial}
Domenico Ficara, Rosario~G. Garroppo, and Jerome Henry.
\newblock A tutorial on privacy, rcm and its implications in wlan.
\newblock \emph{IEEE Communications Surveys \& Tutorials}, 26\penalty0 (2):\penalty0 1003--1040, 2024.
\newblock \doi{10.1109/COMST.2023.3345746}.

\bibitem[Fujii(2022)]{fujii2022study}
Masahiro Fujii.
\newblock A {{Study}} on high accuracy positioning by extended {{FTM}} network.
\newblock In \emph{2022 {{IEEE}} GCCE}, pages 34--35, {Osaka, Japan}, 10 2022. {IEEE}.
\newblock ISBN 978-1-66549-232-4.
\newblock \doi{10.1109/GCCE56475.2022.10014056}.

\bibitem[{Garcia-Fernandez} et~al.(2021){Garcia-Fernandez}, {Hoyas-Ester}, et~al.]{garcia-fernandez2021accuracy}
Miquel {Garcia-Fernandez}, Isaac {Hoyas-Ester}, et~al.
\newblock Accuracy in {{WiFi Access Point Position Estimation Using Round Trip Time}}.
\newblock \emph{Sensors}, 21\penalty0 (11):\penalty0 3828, 6 2021.
\newblock ISSN 1424-8220.
\newblock \doi{10.3390/s21113828}.

\bibitem[{Garcia-Fernandez} et~al.(2022){Garcia-Fernandez}, Siutkowska, and Tejada]{garcia-fernandez2022mitigation}
Miquel {Garcia-Fernandez}, Malgorzata Siutkowska, and Aleix Tejada.
\newblock Mitigation of {{GNSS Errors}} in {{Urban Canyon Using Wi-Fi RTT}}.
\newblock \emph{ICL-GNSS}, 2022.

\bibitem[Gentner et~al.(2020)Gentner, Ulmschneider, et~al.]{gentner2020wifirtt}
Christian Gentner, Markus Ulmschneider, et~al.
\newblock {{WiFi-RTT Indoor Positioning}}.
\newblock In \emph{2020 {{IEEE}}/{{ION}} PLANS}, pages 1029--1035, {Portland, OR, USA}, 4 2020. {IEEE}.
\newblock ISBN 978-1-72810-244-3.
\newblock \doi{10.1109/PLANS46316.2020.9110232}.

\bibitem[Girgensohn et~al.(2020)Girgensohn, Patel, and Biehl]{girgensohn2020radiofrequencybased}
Andreas Girgensohn, Mitesh Patel, and Jacob~T. Biehl.
\newblock Radio-frequency-based indoor-localization techniques for enhancing {{Internet-of-Things}} applications.
\newblock \emph{Personal and Ubiquitous Computing}, 10 2020.
\newblock ISSN 1617-4909, 1617-4917.

\bibitem[Github()]{ftmdevices}
Github.
\newblock Wi-fi ftm hardware.
\newblock URL \url{https://github.com/domienschepers/wifi-ftm/blob/master/HARDWARE.md}.
\newblock Accessed on 2025-04-24.

\bibitem[Glass et~al.(2010)]{glass2010insecurity}
Steve Glass et~al.
\newblock The {{Insecurity}} of {{Time-of-Arrival Distance-Ranging}} in {{IEEE}} 802.11 {{Wireless Networks}}.
\newblock In \emph{2010 {{IEEE}} 30th {{International Conference}} on {{Distributed Computing Systems Workshops}}}. {IEEE}, 6 2010.
\newblock ISBN 978-1-4244-7471-4.

\bibitem[Golden and Bateman(2007)]{golden2007sensor}
Stuart~A. Golden and Steve~S. Bateman.
\newblock Sensor {{Measurements}} for {{Wi-Fi Location}} with {{Emphasis}} on {{Time-of-Arrival Ranging}}.
\newblock \emph{IEEE Transactions on Mobile Computing}, \penalty0 (10):\penalty0 1185--1198, 10 2007.
\newblock ISSN 1536-1233.
\newblock \doi{10.1109/TMC.2007.1002}.

\bibitem[Gonzalez~D{\'\i}az et~al.(2023)Gonzalez~D{\'\i}az, Zola, and Martin-Escalona]{gonzalez2023assessing}
Nestor Gonzalez~D{\'\i}az, Enrica Zola, and Israel Martin-Escalona.
\newblock Assessing the impact of coupling rtt and rssi measurements in fingerprinting wi-fi indoor positioning.
\newblock In \emph{Proc. of ACM MSWiM}, 2023.

\bibitem[Guan et~al.(2024)]{guan2024autonomous}
Fangli Guan et~al.
\newblock Autonomous wireless positioning system using crowdsourced wi-fi fingerprinting and self-detected ftm stations.
\newblock \emph{Expert Systems with Applications}, 255:\penalty0 124566, 2024.

\bibitem[G{\"u}nther and Hoene(2005)]{gunther2005measuring}
Andr{\'e} G{\"u}nther and Christian Hoene.
\newblock Measuring round trip times to determine the distance between wlan nodes.
\newblock In \emph{International conference on research in networking}, pages 768--779. Springer, 2005.

\bibitem[Guo et~al.(2019)Guo, Chen, Ye, Peng, Liu, and Pan]{guo2019indoor}
Guangyi Guo, Ruizhi Chen, Feng Ye, Xuesheng Peng, Zuoya Liu, and Yuanjin Pan.
\newblock Indoor {{Smartphone Localization}}: {{A Hybrid WiFi RTT-RSS Ranging Approach}}.
\newblock \emph{IEEE Access}, 7:\penalty0 176767--176781, 2019.
\newblock ISSN 2169-3536.

\bibitem[Guo et~al.(2022{\natexlab{a}})Guo, Chen, et~al.]{guo2022robust}
Guangyi Guo, Ruizhi Chen, et~al.
\newblock A {{Robust Integration Platform}} of {{Wi-Fi RTT}}, {{RSS Signal}}, and {{MEMS-IMU}} for {{Locating Commercial Smartphone Indoors}}.
\newblock \emph{IEEE Internet of Things Journal}, 9\penalty0 (17):\penalty0 16322--16331, 8 2022{\natexlab{a}}.
\newblock ISSN 2327-4662, 2372-2541.

\bibitem[Guo et~al.(2023)]{guo2023factor}
Guangyi Guo et~al.
\newblock Factor {{Graph Framework}} for {{Smartphone Indoor Localization}}: {{Integrating Data-Driven PDR}} and {{Wi-Fi RTT}}/{{RSS Ranging}}.
\newblock \emph{IEEE Sensors Journal}, 23\penalty0 (11):\penalty0 12346--12354, 6 2023.
\newblock ISSN 1530-437X, 1558-1748, 2379-9153.

\bibitem[Guo and Wu(2023)]{guo2023framework}
Xiaochen Guo and Haitao Wu.
\newblock {Framework and Methods of State Monitoring-based Positioning System on WIFI-RTT Clock Drift Theory}.
\newblock \emph{IEEE Transactions on Aerospace and Electronic Systems}, 2023.

\bibitem[Guo et~al.(2022{\natexlab{b}})Guo, Wu, et~al.]{guo2022impact}
Xiaochen Guo, Haitao Wu, et~al.
\newblock Impact of clock drift on {{WiFi}} round-trip-time ranging and positioning.
\newblock \emph{GPS Solutions}, 26\penalty0 (4):\penalty0 123, 10 2022{\natexlab{b}}.
\newblock ISSN 1080-5370, 1521-1886.
\newblock \doi{10.1007/s10291-022-01313-4}.

\bibitem[Han et~al.(2019)Han, Yu, and Kim]{han2019smartphonebased}
Kyuwon Han, Seung~Min Yu, and Seong-Lyun Kim.
\newblock Smartphone-based {{Indoor Localization Using Wi-Fi Fine Timing Measurement}}.
\newblock In \emph{2019 IPIN}, pages 1--5, {Pisa, Italy}, 8 2019. {IEEE}.
\newblock ISBN 978-1-72811-788-1.
\newblock \doi{10.1109/IPIN.2019.8911751}.

\bibitem[Han et~al.(2021)Han, Yu, et~al.]{han2021exploiting}
Kyuwon Han, Seung~Min Yu, et~al.
\newblock Exploiting {{User Mobility}} for {{WiFi RTT Positioning}}: {{A Geometric Approach}}, 3 2021.

\bibitem[Hashem et~al.(2020{\natexlab{a}})Hashem, Harras, and Youssef]{hashem2020deepnar}
Omar Hashem, Khaled~A. Harras, and Moustafa Youssef.
\newblock {{DeepNar}}: {{Robust Time-based Sub-meter Indoor Localization}} using {{Deep Learning}}.
\newblock In \emph{2020 IEEE SECON}, pages 1--9, {Como, Italy}, 6 2020{\natexlab{a}}. {IEEE}.
\newblock ISBN 978-1-72816-630-8.
\newblock \doi{10.1109/SECON48991.2020.9158428}.

\bibitem[Hashem et~al.(2020{\natexlab{b}})Hashem, Youssef, and Harras]{hashem2020winar}
Omar Hashem, Moustafa Youssef, and Khaled~A. Harras.
\newblock {{WiNar}}: {{RTT-based Sub-meter Indoor Localization}} using {{Commercial Devices}}.
\newblock In \emph{2020 {{IEEE}} PerCom}, pages 1--10, {Austin, TX, USA}, 3 2020{\natexlab{b}}. {IEEE}.
\newblock ISBN 978-1-72814-657-7.
\newblock \doi{10.1109/PerCom45495.2020.9127363}.

\bibitem[Hashem et~al.(2021)Hashem, Harras, and Youssef]{hashem2021accurate}
Omar Hashem, Khaled~A. Harras, and Moustafa Youssef.
\newblock Accurate indoor positioning using {{IEEE}} 802.11mc round trip time.
\newblock \emph{Pervasive and Mobile Computing}, 75:\penalty0 101416, 9 2021.
\newblock ISSN 15741192.
\newblock \doi{10.1016/j.pmcj.2021.101416}.

\bibitem[Hayward et~al.(2022)Hayward, van Lopik, et~al.]{hayward2022survey}
SJ~Hayward, Katherine van Lopik, et~al.
\newblock A survey of indoor location technologies, techniques and applications in industry.
\newblock \emph{Internet of Things}, 20:\penalty0 100608, 2022.

\bibitem[Henry et~al.(2020)Henry, Montavont, et~al.]{henry2020sensor}
Jerome Henry, Nicolas Montavont, et~al.
\newblock Sensor {{Self-location}} with {{FTM Measurements}}.
\newblock In \emph{2020 WiMob}, pages 1--6, {Thessaloniki, Greece}, 10 2020. {IEEE}.
\newblock ISBN 978-1-72819-722-7.
\newblock \doi{10.1109/WiMob50308.2020.9253395}.

\bibitem[Henry et~al.(2021)Henry, Montavont, et~al.]{henry2021geometric}
Jerome Henry, Nicolas Montavont, et~al.
\newblock A {{Geometric Approach}} to {{Noisy EDM Resolution}} in {{FTM Measurements}}.
\newblock \emph{Computers}, 10\penalty0 (3):\penalty0 33, 3 2021.
\newblock ISSN 2073-431X.
\newblock \doi{10.3390/computers10030033}.

\bibitem[Hoene and Willmann(2008)]{hoene2008four}
Christian Hoene and Jorg Willmann.
\newblock Four-way toa and software-based trilateration of ieee 802.11 devices.
\newblock In \emph{2008 IEEE 19th International Symposium on Personal, Indoor and Mobile Radio Communications}. IEEE, 2008.

\bibitem[Horn(2020{\natexlab{a}})]{horn2020doubling}
Berthold Horn.
\newblock Doubling the {{Accuracy}} of {{Indoor Location}}: {{Frequency Diversity}}.
\newblock Preprint, {ENGINEERING}, 1 2020{\natexlab{a}}.

\bibitem[Horn(2020{\natexlab{b}})]{horn2020observation}
Berthold K.~P. Horn.
\newblock Observation {{Model}} for {{Indoor Positioning}}.
\newblock \emph{Sensors}, 20\penalty0 (14):\penalty0 4027, 7 2020{\natexlab{b}}.
\newblock ISSN 1424-8220.

\bibitem[Horn(2022)]{horn2022indoor}
Berthold K.~P. Horn.
\newblock Indoor {{Localization Using Uncooperative Wi-Fi Access Points}}.
\newblock \emph{Sensors}, 22\penalty0 (8):\penalty0 3091, 4 2022.
\newblock ISSN 1424-8220.

\bibitem[Horn(2024)]{horn2024round}
Berthold~KP Horn.
\newblock Round-trip time ranging to wi-fi access points beats gnss localization.
\newblock \emph{Applied Sciences}, 14\penalty0 (17):\penalty0 7805, 2024.

\bibitem[Huang et~al.(2020)Huang, Yu, et~al.]{huang2020hpips}
Lu~Huang, Baoguo Yu, et~al.
\newblock {{HPIPS}}: {{A High-Precision Indoor Pedestrian Positioning System Fusing WiFi-RTT}}, {{MEMS}}, and {{Map Information}}.
\newblock \emph{Sensors}, 20\penalty0 (23):\penalty0 6795, 11 2020.
\newblock ISSN 1424-8220.
\newblock \doi{10.3390/s20236795}.

\bibitem[Huilla et~al.(2020)]{huilla2020indoor}
Sami Huilla et~al.
\newblock Indoor {{Localization}} with {{Wi-Fi Fine Timing Measurements Through Range Filtering}} and {{Fingerprinting Methods}}.
\newblock In \emph{Proc. of IEEE PIMRC}. {IEEE}, 9 2020.
\newblock ISBN 978-1-72814-490-0.

\bibitem[Ibrahim et~al.(2018)Ibrahim, Liu, et~al.]{ibrahim2018verification}
Mohamed Ibrahim, Hansi Liu, et~al.
\newblock Verification: Accuracy evaluation of wifi fine time measurements on an open platform.
\newblock In \emph{Proceedings of the 24th Annual MobiCom}, pages 417--427, 2018.

\bibitem[Ibrahim et~al.(2020)]{ibrahim2020wigo}
Mohamed Ibrahim et~al.
\newblock Wi-{{Go}}: Accurate and scalable vehicle positioning using {{WiFi}} fine timing measurement.
\newblock In \emph{Proc. of the 18th {{International Conference}} on {{Mobile Systems}}, {{Applications}}, and {{Services}}}. {ACM}, 6 2020.
\newblock ISBN 978-1-4503-7954-0.

\bibitem[IEEE({\natexlab{a}})]{80211-promo}
IEEE.
\newblock Newly released ieee 802.11az standard improving wi-fi location accuracy is set to unleash a new wave of innovation, {\natexlab{a}}.
\newblock URL \url{https://standards.ieee.org/beyond-standards/newly-released-ieee-802-11az-standard-improving-wi-fi-location-accuracy-is-set-to-unleash-a-new-wave-of-innovation/}.
\newblock Accessed on 2023-11-17.

\bibitem[IEEE({\natexlab{b}})]{ieee80211-2016}
IEEE.
\newblock Ieee standard for information technology—telecommunications and information exchange between systems local and metropolitan area networks—specific requirements - part 11: Wireless lan medium access control (mac) and physical layer (phy) specifications.
\newblock \emph{IEEE Std 802.11-2016 (Revision of IEEE Std 802.11-2012)}, 2016{\natexlab{b}}.

\bibitem[{IEEE}(2020{\natexlab{a}})]{8021as}
{IEEE}.
\newblock {IEEE Standard for Local and Metropolitan Area Networks--Timing and Synchronization for Time-Sensitive Applications}.
\newblock Std 802.1as-2020, IEEE, 6 2020{\natexlab{a}}.

\bibitem[{IEEE}(2020{\natexlab{b}})]{ieee802112020}
{IEEE}.
\newblock {IEEE Standard for Information Technology--Telecommunications and Information Exchange between Systems - Local and Metropolitan Area Networks--Specific Requirements - Part 11: Wireless LAN Medium Access Control (MAC) and Physical Layer (PHY) Specifications}.
\newblock Std 802.11-2020, IEEE, 5 2020{\natexlab{b}}.

\bibitem[IEEE({\natexlab{c}})]{ieee80211az}
IEEE.
\newblock Ieee standard for information technology—telecommunications and information exchange between systems local and metropolitan area networks—specific requirements - part 11: Wireless lan medium access control (mac) and physical layer (phy) specifications - amendment 4: Enhancements for positioning.
\newblock \emph{IEEE Std 802.11az}, 2022{\natexlab{c}}.

\bibitem[Irshad et~al.(2023)Irshad, Rozner, Bhartia, and Chen]{irshad2023picture}
Shazal Irshad, Eric Rozner, Apurv Bhartia, and Bo~Chen.
\newblock {A Picture is Worth 1,000 Millimeters: Combining Vision and Wi-Fi to Improve Localization}.
\newblock In \emph{2023 IEEE WoWMoM}, pages 56--66, 2023.
\newblock \doi{10.1109/WoWMoM57956.2023.00020}.

\bibitem[Isaia and Michaelides(2023)]{isaia2023review}
Constantina Isaia and Michalis~P Michaelides.
\newblock A review of wireless positioning techniques and technologies: From smart sensors to 6g.
\newblock \emph{Signals}, 4\penalty0 (1):\penalty0 90--136, 2023.

\bibitem[Jathe et~al.(2019)Jathe, L{\"u}tjen, and Freitag]{jathe2019indoor}
Nicolas Jathe, Michael L{\"u}tjen, and Michael Freitag.
\newblock Indoor {{Positioning}} in {{Car Parks}} by using {{Wi-Fi Round-Trip-Time}} to support {{Finished Vehicle Logistics}} on {{Port Terminals}}.
\newblock \emph{IFAC-PapersOnLine}, 52\penalty0 (13):\penalty0 857--862, 2019.
\newblock ISSN 24058963.

\bibitem[Jerome et~al.(2025)Jerome, Hart, Gupta, and Smith]{henry2025wi}
Henry Jerome, Brian Hart, Binita Gupta, and Malcolm Smith.
\newblock \emph{Wi-Fi 7 in Depth: Your Guide to Mastering Wi-Fi 7, the 802.11 be Protocol, and Their Deployment}.
\newblock Addison-Wesley Professional, 2025.

\bibitem[Jin and Papadimitratos(2022)]{jin2022offtheshelf}
Hongyu Jin and Panos Papadimitratos.
\newblock Off-the-shelf {{Wi-Fi Indoor Smartphone Localization}}.
\newblock In \emph{2022 WONS}, pages 1--4, {Oppdal, Norway}, 3 2022. {IEEE}.
\newblock ISBN 978-3-903176-46-1.
\newblock \doi{10.23919/WONS54113.2022.9764448}.

\bibitem[Jiokeng et~al.(2020)Jiokeng, Jakllari, et~al.]{jiokeng2020when}
Kevin Jiokeng, Gentian Jakllari, et~al.
\newblock When {{FTM Discovered MUSIC}}: {{Accurate WiFi-based Ranging}} in the {{Presence}} of {{Multipath}}.
\newblock In \emph{{{IEEE INFOCOM}} 2020 - {{IEEE Conference}} on {{Computer Communications}}}, pages 1857--1866. {IEEE}, 7 2020.
\newblock ISBN 978-1-72816-412-0.

\bibitem[Jung et~al.(2021)Jung, Chung, et~al.]{jung2021learning}
Boo-Geum Jung, Byung~Chang Chung, et~al.
\newblock Learning based {{Wi-Fi RTT Range Estimation}}.
\newblock In \emph{2021 ICTC}, pages 1030--1032, {Jeju Island, Korea, Republic of}, 10 2021. {IEEE}.
\newblock ISBN 978-1-66542-383-0.
\newblock \doi{10.1109/ICTC52510.2021.9620218}.

\bibitem[Jurdi et~al.(2024)Jurdi, Chen, et~al.]{jurdi2024where}
Rebal Jurdi, Hao Chen, et~al.
\newblock Whereartthou: A wifi-rtt-based indoor positioning system.
\newblock \emph{IEEE Access}, 12:\penalty0 41084--41101, 2024.
\newblock \doi{10.1109/ACCESS.2024.3377237}.

\bibitem[Kang et~al.(2021)Kang, Lee, et~al.]{kang2021time}
Yoohwa Kang, Sunwoo Lee, et~al.
\newblock Time-sensitive networking technologies for industrial automation in wireless communication systems.
\newblock \emph{Energies}, 14\penalty0 (15):\penalty0 4497, 2021.

\bibitem[Khatib et~al.(2023)Khatib, Alvarez-Merino, et~al.]{khatib2023designing}
Emil~J Khatib, Carlos~Simon Alvarez-Merino, et~al.
\newblock Designing a 6g testbed for location: Use cases, challenges, enablers and requirements.
\newblock \emph{IEEE Access}, 11:\penalty0 10053--10091, 2023.

\bibitem[Kia et~al.(2021)Kia, Talvitie, and Ruotsalainen]{kia2021rssbased}
Ghazaleh Kia, Jukka Talvitie, and Laura Ruotsalainen.
\newblock {{RSS-Based Fusion}} of {{UWB}} and {{WiFi-Based Ranging}} for {{Indoor Positioning}}.
\newblock \emph{IPIN 2021 WiP Proceedings}, 2021.

\bibitem[Kia et~al.(2022)Kia, Ruotsalainen, and Talvitie]{kia2022accurate}
Ghazaleh Kia, Laura Ruotsalainen, and Jukka Talvitie.
\newblock Toward {{Accurate Indoor Positioning}}: {{An RSS-Based Fusion}} of {{UWB}} and {{Machine-Learning-Enhanced WiFi}}.
\newblock \emph{Sensors}, 22\penalty0 (9):\penalty0 3204, 4 2022.
\newblock ISSN 1424-8220.
\newblock \doi{10.3390/s22093204}.

\bibitem[Kim~Geok et~al.(2020)Kim~Geok, Zar~Aung, et~al.]{kimgeok2020review}
Tan Kim~Geok, Khaing Zar~Aung, et~al.
\newblock Review of indoor positioning: Radio wave technology.
\newblock \emph{Applied Sciences}, 11\penalty0 (1):\penalty0 279, 2020.

\bibitem[Koenig et~al.(2011)Koenig, Schmidt, and Hoene]{koenig2011multipath}
Stefan Koenig, Mark~T Schmidt, and Christian Hoene.
\newblock Multipath mitigation for indoor localization based on ieee 802.11 time-of-flight measurements.
\newblock In \emph{Proc. of IEEE WoWMoM}. IEEE, 2011.

\bibitem[Kov{\'a}csh{\'a}zy and Rusvai(2024)]{kovacshazy2024performance}
Tam{\'a}s Kov{\'a}csh{\'a}zy and Mikl{\'o}s Rusvai.
\newblock Performance evaluation of wi-fi ftm indoor positioning for embedded applications.
\newblock In \emph{2024 6th Global Power, Energy and Communication Conference (GPECOM)}, pages 771--776. IEEE, 2024.

\bibitem[Laoudias et~al.(2018)Laoudias, Moreira, et~al.]{laoudias2018survey}
Christos Laoudias, Adriano Moreira, et~al.
\newblock A survey of enabling technologies for network localization, tracking, and navigation.
\newblock \emph{IEEE Communications Surveys \& Tutorials}, 20\penalty0 (4):\penalty0 3607--3644, 2018.

\bibitem[Leitch et~al.(2023)Leitch, Ahmed, et~al.]{leitch2023indoor}
Samuel~G Leitch, Qasim~Zeeshan Ahmed, et~al.
\newblock On indoor localization using wifi, ble, uwb, and imu technologies.
\newblock \emph{Sensors}, 23\penalty0 (20):\penalty0 8598, 2023.

\bibitem[Li et~al.(2024)Li, Pang, et~al.]{li2024wireless}
Xingwang Li, Hua Pang, et~al.
\newblock Wireless positioning: Technologies, applications, challenges, and future development trends.
\newblock \emph{CMES-Computer Modeling in Engineering \& Sciences}, 2024.

\bibitem[Lin and Sundaresan(2023)]{lin2023new}
Yu-Tai Lin and Karthikeyan Sundaresan.
\newblock A {{New Paradigm}} of {{Communication-Aware Collaborative Positioning}} for {{FutureG Wireless Systems}}.
\newblock In \emph{Proc. of ACM MobiHoc}. {ACM}, 10 2023.
\newblock ISBN 978-1-4503-9926-5.

\bibitem[Lindskog et~al.(2017)Lindskog, Zhang, et~al.]{lindskog2017replay}
Erik Lindskog, Ning Zhang, et~al.
\newblock Cp replay attack protection.
\newblock IEEE 802.11 TGaz IEEE 802.11-17/1372-01, Qualcomm, 2017.
\newblock https://mentor.ieee.org/802.11/dcn/17/11-17-1372-01-00az-cp-replay-attack-protection.pptx.

\bibitem[Liu et~al.(2022{\natexlab{a}})Liu, Alali, Ibrahim, et~al.]{liu2022vi}
Hansi Liu, Abrar Alali, Mohamed Ibrahim, et~al.
\newblock Vi-fi: Associating moving subjects across vision and wireless sensors.
\newblock In \emph{2022 21st ACM/IEEE IPSN}, pages 208--219. IEEE, 2022{\natexlab{a}}.

\bibitem[Liu et~al.(2021)Liu, Zhou, et~al.]{liu2021kalman}
Xu~Liu, Baoding Zhou, et~al.
\newblock Kalman {{Filter-Based Data Fusion}} of {{Wi-Fi RTT}} and {{PDR}} for {{Indoor Localization}}.
\newblock \emph{IEEE Sensors Journal}, 21\penalty0 (6):\penalty0 8479--8490, 3 2021.
\newblock ISSN 1530-437X, 1558-1748, 2379-9153.
\newblock \doi{10.1109/JSEN.2021.3050456}.

\bibitem[Liu et~al.(2022{\natexlab{b}})Liu, Zhou, et~al.]{liu2022indoor}
Xu~Liu, Baoding Zhou, et~al.
\newblock An {{Indoor}} 3-{{D Quadrotor Localization Algorithm Based}} on {{WiFi RTT}} and {{MEMS Sensors}}.
\newblock \emph{IEEE Internet of Things Journal}, 9\penalty0 (21):\penalty0 20879--20888, 11 2022{\natexlab{b}}.
\newblock ISSN 2327-4662, 2372-2541.
\newblock \doi{10.1109/JIOT.2022.3175809}.

\bibitem[Liu et~al.(2023)Liu, Li, et~al.]{liu2023dqwiapdom}
Zihao Liu, Han Li, et~al.
\newblock {{DQ-WiAPDoM}}: {{A DQN-based AP Deployment Optimization Method}} for {{Wi-Fi FTM Positioning}}.
\newblock In \emph{2023 ICSTSN}, pages 1--5, {Villupuram, India}, 4 2023. {IEEE}.
\newblock \doi{10.1109/ICSTSN57873.2023.10151521}.

\bibitem[Llombart et~al.(2008)Llombart, Ciurana, and Barcelo-Arroyo]{llombart2008scalability}
Marc Llombart, Marc Ciurana, and Francisco Barcelo-Arroyo.
\newblock On the scalability of a novel wlan positioning system based on time of arrival measurements.
\newblock In \emph{5th Workshop on Positioning, Navigation and Communication}. IEEE, 2008.

\bibitem[{Lopez-Pastor} et~al.(2021)]{lopez-pastor2021wifi}
Jose~Antonio {Lopez-Pastor} et~al.
\newblock Wi-{{Fi RTT-Based Active Monopulse RADAR}} for {{Single Access Point Localization}}.
\newblock \emph{IEEE Access}, 9:\penalty0 34755--34766, 2021.
\newblock ISSN 2169-3536.
\newblock \doi{10.1109/ACCESS.2021.3062085}.

\bibitem[L{\'o}pez-Pastor et~al.(2023)]{lopez2023two}
Jos{\'e}~Antonio L{\'o}pez-Pastor et~al.
\newblock Two-dimensional localization system for mobile iot devices using a single wi-fi access point with a passive frequency-scanned antenna.
\newblock \emph{IEEE Internet of Things Journal}, 2023.

\bibitem[Lu et~al.(2024)Lu, Wang, Wen, and Zhang]{lu2024improving}
Bingxian Lu, Mengya Wang, Wu~Wen, and Yang Zhang.
\newblock Improving ftm ranging accuracy based on dnn for uav localization.
\newblock \emph{IEEE Internet of Things Journal}, 11\penalty0 (12), 2024.

\bibitem[Ma et~al.(2022)Ma, Wu, et~al.]{ma2022wifi}
Chengqi Ma, Bang Wu, et~al.
\newblock Wi-{{Fi RTT Ranging Performance Characterization}} and {{Positioning System Design}}.
\newblock \emph{IEEE Transactions on Mobile Computing}, 21\penalty0 (2):\penalty0 740--756, 2 2022.
\newblock ISSN 1536-1233, 1558-0660, 2161-9875.
\newblock \doi{10.1109/TMC.2020.3012563}.

\bibitem[Maduranga et~al.(2024)]{maduranga2024improved}
MWP Maduranga et~al.
\newblock Improved-rssi-based indoor localization by using pseudo-linear solution with machine learning algorithms.
\newblock \emph{Journal of Electrical Systems and Information Technology}, 11\penalty0 (1):\penalty0 10, 2024.

\bibitem[Maghdid et~al.(2016)Maghdid, Lami, et~al.]{maghdid2016seamless}
Halgurd~S Maghdid, Ihsan~Alshahib Lami, et~al.
\newblock Seamless outdoors-indoors localization solutions on smartphones: Implementation and challenges.
\newblock \emph{ACM Computing Surveys}, 48\penalty0 (4):\penalty0 1--34, 2016.

\bibitem[Makki et~al.(2015)Makki, Siddig, Saad, and Bleakley]{makki2015survey}
Ahmed Makki, Abubakr Siddig, Mohamed Saad, and Chris Bleakley.
\newblock Survey of wifi positioning using time-based techniques.
\newblock \emph{Computer Networks}, 88:\penalty0 218--233, 2015.

\bibitem[Manabe and Saba(2023)]{manabe2023performance}
Tetsuya Manabe and Kazuya Saba.
\newblock Performance {{Evaluation}} of {{Wi-Fi RTT Lateration}} without {{Pre-Constructing}} a {{Database}}.
\newblock \emph{IEICE Trans. on Fundamentals of Electronics, Communications and Computer Sciences}, E106.A, 2023.
\newblock ISSN 0916-8508, 1745-1337.

\bibitem[Marcaletti et~al.(2014)]{marcaletti2014filtering}
Andreas Marcaletti et~al.
\newblock Filtering {{Noisy}} 802.11 {{Time-of-Flight Ranging Measurements}}.
\newblock In \emph{Proc. of the 10th {{ACM International}} on {{Conference}} on Emerging {{Networking Experiments}} and {{Technologies}}}, pages 13--20. {ACM}, 12 2014.
\newblock ISBN 978-1-4503-3279-8.

\bibitem[{Martin-Escalona} and Zola(2020{\natexlab{a}})]{martin-escalona2020passive}
Israel {Martin-Escalona} and Enrica Zola.
\newblock Passive {{Round-Trip-Time Positioning}} in {{Dense IEEE}} 802.11 {{Networks}}.
\newblock \emph{Electronics}, 9\penalty0 (8):\penalty0 1193, 7 2020{\natexlab{a}}.
\newblock ISSN 2079-9292.
\newblock \doi{10.3390/electronics9081193}.

\bibitem[{Martin-Escalona} and Zola(2020{\natexlab{b}})]{martin-escalona2020ranging}
Israel {Martin-Escalona} and Enrica Zola.
\newblock Ranging {{Estimation Error}} in {{WiFi Devices Running IEEE}} 802.11mc.
\newblock In \emph{{{GLOBECOM}} 2020 - 2020 {{IEEE Global Communications Conference}}}, pages 1--7, {Taipei, Taiwan}, 12 2020{\natexlab{b}}. {IEEE}.
\newblock ISBN 978-1-72818-298-8.

\bibitem[Martin-Escalona and Zola(2022)]{martin2022improving}
Israel Martin-Escalona and Enrica Zola.
\newblock Improving fingerprint-based positioning by using ieee 802.11mc ftm/rtt observables.
\newblock \emph{Sensors}, 23\penalty0 (1):\penalty0 267, 2022.

\bibitem[Martinez et~al.(2009)]{martinez2009indoor}
Alfonso~Bahillo Martinez et~al.
\newblock Indoor location based on ieee 802.11 round-trip time measurements with two-step nlos mitigation.
\newblock \emph{Progress In Electromagnetics Research B}, 15:\penalty0 285--306, 2009.

\bibitem[Mathias et~al.(2008)Mathias, Leonardi, and Galati]{mathias2008efficient}
Adolf Mathias, Mauro Leonardi, and Gaspare Galati.
\newblock An efficient multilateration algorithm.
\newblock In \emph{2008 Tyrrhenian International Workshop on Digital Communications-Enhanced Surveillance of Aircraft and Vehicles}, pages 1--6. IEEE, 2008.

\bibitem[Mathworks()]{matlab}
Mathworks.
\newblock Ai, positioning, and sensing.
\newblock URL \url{https://www.mathworks.com/help/wlan/ai-positioning-and-sensing.html?s_tid=CRUX_lftnav}.
\newblock Accessed on 2025-04-24.

\bibitem[Mendoza-Silva et~al.(2019)]{mendoza2019meta}
Germ{\'a}n~Mart{\'\i}n Mendoza-Silva et~al.
\newblock A meta-review of indoor positioning systems.
\newblock \emph{Sensors}, 19\penalty0 (20):\penalty0 4507, 2019.

\bibitem[Mohan and Sofia(2023)]{mohan2023fine}
Sugandh~Huthanahally Mohan and Rute~C. Sofia.
\newblock Fine {{Time Measurement}} based {{Time Synchronization}} for {{Multi-AP Wireless Industrial Environments}}.
\newblock In \emph{2023 WiMob}, pages 399--404, {Montreal, QC, Canada}, 6 2023. {IEEE}.

\bibitem[Mohsen et~al.(2023{\natexlab{a}})Mohsen, Rizk, and Youssef]{mohsen2023privacypreserving}
Mohamed Mohsen, Hamada Rizk, and Moustafa Youssef.
\newblock Privacy-{{Preserving}} by {{Design}}: {{Indoor Positioning System Using Wi-Fi Passive TDOA}}, 6 2023{\natexlab{a}}.

\bibitem[Mohsen et~al.(2024)Mohsen, Rizk, Yamaguchi, and Youssef]{mohsen2024timesense}
Mohamed Mohsen, Hamada Rizk, Hirozumi Yamaguchi, and Moustafa Youssef.
\newblock Timesense: Multi-person device-free indoor localization via rtt.
\newblock \emph{IEEE Internet of Things Journal}, 2024.

\bibitem[Mohsen et~al.(2023{\natexlab{b}})]{mohsen2023locfree}
Mohamed Mohsen et~al.
\newblock Locfree: Wifi rtt-based device-free indoor localization system.
\newblock In \emph{Proc. of the 2nd ACM SIGSPATIAL International Workshop on Spatial Big Data and AI for Industrial Applications}, pages 32--40, 2023{\natexlab{b}}.

\bibitem[Najarro et~al.(2022)]{najarro2022fundamental}
Lismer Andres~Caceres Najarro et~al.
\newblock Fundamental limitations and state-of-the-art solutions for target node localization in wsns: a review.
\newblock \emph{IEEE Sensors Journal}, 22\penalty0 (24):\penalty0 23661--23682, 2022.

\bibitem[Naser et~al.(2023)]{naser2023smartphone}
Rana~Sabah Naser et~al.
\newblock Smartphone-based indoor localization systems: A systematic literature review.
\newblock \emph{Electronics}, 12\penalty0 (8):\penalty0 1814, 2023.

\bibitem[Neri et~al.(2019)]{neri2019indoor}
Alessandro Neri et~al.
\newblock Indoor vehicle localization based on wi-fi navigation beacons for multi-modal transportation applications.
\newblock In \emph{Proceedings of the ION 2019 Pacific PNT Meeting}, 2019.

\bibitem[Nessa et~al.(2020)Nessa, Adhikari, et~al.]{nessa2020survey}
Ahasanun Nessa, Bhagawat Adhikari, et~al.
\newblock A survey of machine learning for indoor positioning.
\newblock \emph{IEEE Access}, 8:\penalty0 214945--214965, 2020.

\bibitem[Nikseresht and Campbell(2023)]{nikseresht2023ftm}
Fateme Nikseresht and Bradford Campbell.
\newblock Ftm-sense: Robust sensor-free occupancy sensing leveraging wifi fine time measurement.
\newblock In \emph{Proc. of ACM BuildSys 2023}, 2023.

\bibitem[Nkrow et~al.(2023)]{nkrow2023wifi}
Raphael~Elikplim Nkrow et~al.
\newblock Wi-{{Fi Fine Time Measurement}}: {{Is}} it a {{Viable Alternative}} to {{Ultrawideband}} for {{Ranging}} in {{Industrial Environments}}?
\newblock \emph{IEEE Industrial Electronics Magazine}, 17\penalty0 (3):\penalty0 33--41, 8 2023.
\newblock ISSN 1932-4529, 1941-0115.

\bibitem[Numan et~al.(2022)Numan, Park, et~al.]{numan2022dnnbased}
Paulson~Eberechukwu Numan, Hyunwoo Park, et~al.
\newblock {{DNN-based Indoor Fingerprinting Localization}} with {{WiFi FTM}}.
\newblock In \emph{2022 MDM}, pages 367--371, {Paphos, Cyprus}, 6 2022. {IEEE}.
\newblock ISBN 978-1-66545-176-5.
\newblock \doi{10.1109/MDM55031.2022.00082}.

\bibitem[Numan et~al.(2023{\natexlab{a}})Numan, Park, et~al.]{numan2023dropout}
Paulson~Eberechukwu Numan, Hyunwoo Park, et~al.
\newblock Dropout {{Autoencoder Fingerprint Augmentation}} for {{Enhanced Wi-Fi FTM-RSS Indoor Localization}}.
\newblock \emph{IEEE Communications Letters}, 27\penalty0 (7):\penalty0 1759--1763, 7 2023{\natexlab{a}}.
\newblock ISSN 1089-7798, 1558-2558, 2373-7891.

\bibitem[Numan et~al.(2023{\natexlab{b}})Numan, Park, et~al.]{numan2023smartphonebased}
Paulson~Eberechukwu Numan, Hyunwoo Park, et~al.
\newblock Smartphone-{{Based Indoor Localization}} via {{Network Learning With Fusion}} of {{FTM}}/{{RSSI Measurements}}.
\newblock \emph{IEEE Networking Letters}, 5\penalty0 (1):\penalty0 21--25, 3 2023{\natexlab{b}}.
\newblock ISSN 2576-3156.

\bibitem[Ogawa and Choi(2020)]{ogawa2020measurement}
Masakatsu Ogawa and Hoyeon Choi.
\newblock Measurement accuracy of {{Wi-Fi FTM}} on actual devices.
\newblock \emph{IEICE Communications Express}, 9\penalty0 (12):\penalty0 567--572, 12 2020.
\newblock ISSN 2187-0136.
\newblock \doi{10.1587/comex.2020COL0001}.

\bibitem[Oguntala et~al.(2018)Oguntala, Abd-Alhameed, et~al.]{oguntala2018indoor}
George Oguntala, Raed Abd-Alhameed, et~al.
\newblock Indoor location identification technologies for real-time iot-based applications: An inclusive survey.
\newblock \emph{Computer Science Review}, 30:\penalty0 55--79, 2018.

\bibitem[Orfanos et~al.(2023)Orfanos, Perakis, et~al.]{orfanos2023testing}
Manos Orfanos, Harris Perakis, et~al.
\newblock Testing and {{Evaluation}} of {{Wi-Fi RTT Ranging Technology}} for {{Personal Mobility Applications}}.
\newblock \emph{Sensors}, 23\penalty0 (5):\penalty0 2829, 3 2023.
\newblock ISSN 1424-8220.
\newblock \doi{10.3390/s23052829}.

\bibitem[Pagliari et~al.(2024)Pagliari, Davoli, and Ferrari]{pagliari2024wi}
Emanuele Pagliari, Luca Davoli, and Gianluigi Ferrari.
\newblock Wi-fi-based real-time uav localization: a comparative analysis between rssi-based and ftm-based approaches.
\newblock \emph{IEEE Transactions on Aerospace and Electronic Systems}, 2024.

\bibitem[Panja et~al.(2022)Panja, Chowdhury, and Neogy]{panja2022survey}
Ayan~Kumar Panja, Chandreyee Chowdhury, and Sarmistha Neogy.
\newblock Survey on inertial sensor-based ils for smartphone users.
\newblock \emph{CCF Transactions on Pervasive Computing and Interaction}, 4\penalty0 (3):\penalty0 319--337, 2022.

\bibitem[Park et~al.(2023{\natexlab{a}})Park, Lee, et~al.]{park2023gpsaided}
Kyoung-Min Park, Byeong-ho Lee, et~al.
\newblock {{GPS-Aided Automatic Site Survey Method}} for {{WiFi RTT-Based Positioning}}.
\newblock \emph{IEEE Transactions on Vehicular Technology}, pages 1--10, 2023{\natexlab{a}}.
\newblock ISSN 0018-9545, 1939-9359.
\newblock \doi{10.1109/TVT.2023.3275203}.

\bibitem[Park et~al.(2023{\natexlab{b}})Park, Lee, et~al.]{park2023automated}
Kyoung-Min Park, Eunji Lee, et~al.
\newblock Automated {{Site Survey}} for {{Fingerprints}} in {{Fully Blind Indoor Environments Based}} on {{Sensor Integration}}.
\newblock \emph{IEEE Sensors Letters}, 7\penalty0 (1):\penalty0 1--4, 1 2023{\natexlab{b}}.
\newblock ISSN 2475-1472.
\newblock \doi{10.1109/LSENS.2022.3233577}.

\bibitem[Perdana et~al.(2022)Perdana, Pandya, and Akhyar]{perdana2022evaluation}
Doan Perdana, Favian~Hugo Pandya, and Fityanul Akhyar.
\newblock Evaluation of wifi rtt for indoor positioning system using classification algorithm.
\newblock \emph{Webology (ISSN: 1735-188X)}, 19\penalty0 (2), 2022.

\bibitem[Perdana et~al.(2023)Perdana, Indra~Tanaya, et~al.]{perdana2023evaluation}
Doan Perdana, I~Made~Arya Indra~Tanaya, et~al.
\newblock Evaluation of a {{High-Accuracy Indoor-Positioning System}} with {{Wi-Fi Time}} of {{Flight}} ({{ToF}}) and {{Deep Learning}}.
\newblock \emph{Journal of Computer Networks and Communications}, 2023, 4 2023.
\newblock ISSN 2090-715X, 2090-7141.

\bibitem[Picazo-Mart{\'\i}nez et~al.(2023)]{picazo2023ieee}
Pablo Picazo-Mart{\'\i}nez et~al.
\newblock Ieee 802.11az indoor positioning with mmwaves.
\newblock \emph{arXiv:2303.05996}, 2023.

\bibitem[Pizarro et~al.(2021)]{pizarro2021accurate}
Alejandro~Blanco Pizarro et~al.
\newblock Accurate ubiquitous localization with off-the-shelf {{IEEE}} 802.11ac devices.
\newblock In \emph{Proceedings of the 19th {{Annual International Conference}} on {{Mobile Systems}}, {{Applications}}, and {{Services}}}, pages 241--254, 6 2021.
\newblock ISBN 978-1-4503-8443-8.

\bibitem[Poturalski et~al.(2010)Poturalski, Flury, et~al.]{poturalski2010cicada}
Marcin Poturalski, Manuel Flury, et~al.
\newblock The cicada attack: degradation and denial of service in ir ranging.
\newblock In \emph{2010 IEEE International Conference on Ultra-Wideband}, volume~2, pages 1--4. IEEE, 2010.

\bibitem[{Poveda-Garcia} et~al.(2023){Poveda-Garcia}, {Lopez-Pastor}, et~al.]{poveda-garcia2023wifi}
Miguel {Poveda-Garcia}, Jose~Antonio {Lopez-Pastor}, et~al.
\newblock Wi-{{Fi Direction}} and {{Range Estimation}} with a {{Single Frequency-Scanned Antenna}} using {{RSSI}} and {{RTT}}.
\newblock In \emph{2023 EuCAP}, pages 1--5, {Florence, Italy}, 3 2023. {IEEE}.

\bibitem[Qiao et~al.(2023)Qiao, Cao, et~al.]{qiao2023trip}
Shuang Qiao, Chenhong Cao, et~al.
\newblock The trip to wifi indoor localization across a decade—a systematic review.
\newblock In \emph{2023 CSCWD}, pages 642--647. IEEE, 2023.

\bibitem[Raja and Groves(2023)]{raja2023wifi}
K~Jibran Raja and Paul~D Groves.
\newblock Wifi-rtt indoor positioning using particle, genetic and grid filters with rssi-based outlier detection.
\newblock In \emph{ION GNSS+ 2023}, pages 1644--1655, 2023.

\bibitem[Rana and Park(2024)]{rana2024enhanced}
Lila Rana and Joon~Goo Park.
\newblock An enhanced indoor positioning method based on rtt and rss measurements under los/nlos environment.
\newblock \emph{IEEE Sensors Journal}, 2024.

\bibitem[Ratnam et~al.(2024)Ratnam, Sadiq, et~al.]{ratnam2024widra}
Vishnu~V Ratnam, Bilal Sadiq, et~al.
\newblock Widra--enabling millimeter-level differential ranging accuracy in wi-fi using carrier phase.
\newblock \emph{arXiv preprint arXiv:2405.12168}, 2024.

\bibitem[Rea et~al.(2019)]{rea2019smartphone}
Maurizio Rea et~al.
\newblock Smartphone positioning with radio measurements from a single wifi access point.
\newblock In \emph{Proc. of the 15th International Conference on Emerging Networking Experiments and Technologies}, pages 200--206, 2019.

\bibitem[Reddy and St{\"u}ber(2022)]{reddy2022multi}
Varun~Amar Reddy and Gordon~L St{\"u}ber.
\newblock Multi-user position estimation and performance trade-offs in ieee 802.11az wlans.
\newblock In \emph{2022 IEEE VTC2022-Spring}, pages 1--5. IEEE, 2022.

\bibitem[Retscher(2022)]{retscher2022indoor}
G{\"u}nther Retscher.
\newblock Indoor navigation—user requirements, state-of-the-art and developments for smartphone localization.
\newblock \emph{Geomatics}, 3\penalty0 (1):\penalty0 1--46, 2022.

\bibitem[Retscher et~al.(2024)Retscher, Gikas, and Gabela]{retscher2024experiences}
G{\"u}nther Retscher, Vassilis Gikas, and Jelena Gabela.
\newblock Experiences with techniques and sensors for smartphone positioning.
\newblock \emph{Journal of Applied Geodesy}, \penalty0 (0), 2024.

\bibitem[Rizk et~al.(2022)Rizk, Elmogy, and Yamaguchi]{rizk2022robust}
Hamada Rizk, Ahmed Elmogy, and Hirozumi Yamaguchi.
\newblock A {{Robust}} and {{Accurate Indoor Localization Using Learning-Based Fusion}} of {{Wi-Fi RTT}} and {{RSSI}}.
\newblock \emph{Sensors}, 22\penalty0 (7):\penalty0 2700, 3 2022.
\newblock ISSN 1424-8220.
\newblock \doi{10.3390/s22072700}.

\bibitem[Sartayeva and Chan(2023)]{sartayeva2023survey}
Yerkezhan Sartayeva and Henry~CB Chan.
\newblock A survey on indoor positioning security and privacy.
\newblock \emph{Computers \& Security}, 131:\penalty0 103293, 2023.

\bibitem[Schepers and Ranganathan(2022)]{schepers2022privacypreserving}
Domien Schepers and Aanjhan Ranganathan.
\newblock Privacy-{{Preserving Positioning}} in {{Wi-Fi Fine Timing Measurement}}.
\newblock \emph{Proceedings on Privacy Enhancing Technologies}, 2022\penalty0 (2):\penalty0 325--343, 4 2022.
\newblock ISSN 2299-0984.
\newblock \doi{10.2478/popets-2022-0048}.

\bibitem[Schepers et~al.(2021)Schepers, Singh, and Ranganathan]{schepers2021here}
Domien Schepers, Mridula Singh, and Aanjhan Ranganathan.
\newblock Here, there, and everywhere: Security analysis of wi-fi fine timing measurement.
\newblock In \emph{Proceedings of ACM WiSec}, pages 78--89. {ACM}, 6 2021.
\newblock ISBN 978-1-4503-8349-3.

\bibitem[Segev and Berger()]{80211-promo1}
Jonathan Segev and Christian Berger.
\newblock Next generation wi-fi positioning. an overview of ieee 802.11az.
\newblock URL \url{https://www.computer.org/csdl/video-library/video/1UXPYs1JCrS}.
\newblock Accessed on 2025-04-24.

\bibitem[Sen et~al.(2023)Sen, Jiang, Wu, Talasila, Hsu, and Borcea]{sen2023goplaces}
Pritam Sen, Xiaopeng Jiang, Qiong Wu, Manoop Talasila, Wen-Ling Hsu, and Cristian Borcea.
\newblock Goplaces: An app for personalized indoor place prediction.
\newblock In \emph{2023 IEEE MASS}, pages 566--574. IEEE, 2023.

\bibitem[Seong et~al.(2021)Seong, Lee, et~al.]{seong2021highprecision}
Ju-Hyeon Seong, Soo-Hwan Lee, et~al.
\newblock High-{{Precision RTT-Based Indoor Positioning System Using RCDN}} and {{RPN}}.
\newblock \emph{Sensors}, 21\penalty0 (11):\penalty0 3701, 5 2021.
\newblock ISSN 1424-8220.
\newblock \doi{10.3390/s21113701}.

\bibitem[Sesyuk et~al.(2022)Sesyuk, Ioannou, and Raspopoulos]{sesyuk2022survey}
Andrey Sesyuk, Stelios Ioannou, and Marios Raspopoulos.
\newblock A survey of 3d indoor localization systems and technologies.
\newblock \emph{Sensors}, 22\penalty0 (23):\penalty0 9380, 2022.

\bibitem[Shaikhanov et~al.(2020)Shaikhanov, Boubrima, and Knightly]{shaikhanov2020autonomous}
Zhambyl Shaikhanov, Ahmed Boubrima, and Edward~W Knightly.
\newblock Autonomous drone networks for sensing, localizing and approaching rf targets.
\newblock In \emph{2020 IEEE VNC}, pages 1--8. IEEE, 2020.

\bibitem[Shaikhanov et~al.(2022)Shaikhanov, Boubrima, and Knightly]{shaikhanov2022falcon}
Zhambyl Shaikhanov, Ahmed Boubrima, and Edward~W. Knightly.
\newblock {{FALCON}}: {{A Networked Drone System}} for {{Sensing}}, {{Localizing}}, and {{Approaching RF Targets}}.
\newblock \emph{IEEE Internet of Things Journal}, 9\penalty0 (12):\penalty0 9843--9857, 6 2022.
\newblock ISSN 2327-4662, 2372-2541.

\bibitem[Shao et~al.(2020)Shao, Luo, et~al.]{shao2020accurate}
Wenhua Shao, Haiyong Luo, et~al.
\newblock Accurate {{Indoor Positioning Using Temporal}}\textendash{{Spatial Constraints Based}} on {{Wi-Fi Fine Time Measurements}}.
\newblock \emph{IEEE Internet of Things Journal}, 7\penalty0 (11):\penalty0 11006--11019, 11 2020.
\newblock ISSN 2327-4662, 2372-2541.
\newblock \doi{10.1109/JIOT.2020.2992069}.

\bibitem[Shao et~al.(2024)]{shao2024moc}
Wenhua Shao et~al.
\newblock Moc: Wi-fi ftm with motion observation chain for pervasive indoor positioning.
\newblock \emph{IEEE Transactions on Industrial Informatics}, 2024.

\bibitem[Shellhammer et~al.(2020)]{shellhammer2020secure}
Steve Shellhammer et~al.
\newblock Secure ltfs additional design details.
\newblock IEEE 802.11 TGaz IEEE 802.11-20/1863-01, Qualcomm, 2020.
\newblock https://mentor.ieee.org/802.11/dcn/20/11-20-1863-01-00az-secure-ltfs-additional-design-details.pptx.

\bibitem[Shi et~al.(2023)Shi, Li, et~al.]{shi2023decimeterlevel}
Fangzhan Shi, Wenda Li, et~al.
\newblock Decimeter-{{Level Indoor Localization Using WiFi Round-Trip Phase}} and {{Factor Graph Optimization}}.
\newblock \emph{IEEE Journal on Selected Areas in Communications}, pages 1--1, 2023.
\newblock ISSN 0733-8716, 1558-0008.
\newblock \doi{10.1109/JSAC.2023.3322812}.

\bibitem[Shida et~al.(2024)Shida, Aikawa, and Yamamoto]{shida2024correction}
Yusuke Shida, Satoru Aikawa, and Shinichiro Yamamoto.
\newblock Correction of round-trip time and selection of access points for estimating wireless lan locations by multilateration.
\newblock \emph{IEICE Communications Express}, 2024.

\bibitem[Si et~al.(2020)Si, Wang, et~al.]{si2020wifi}
Minghao Si, Yunjia Wang, et~al.
\newblock A wi-fi ftm-based indoor positioning method with los/nlos identification.
\newblock \emph{Applied Sciences}, 10\penalty0 (3):\penalty0 956, 2020.

\bibitem[Si et~al.(2022)Si, Wang, et~al.]{si2022adaptive}
Minghao Si, Yunjia Wang, et~al.
\newblock An {{Adaptive Weighted Wi-Fi FTM-Based Positioning Method}} in an {{NLOS Environment}}.
\newblock \emph{IEEE Sensors Journal}, 22\penalty0 (1):\penalty0 472--480, 1 2022.
\newblock ISSN 1530-437X, 1558-1748, 2379-9153.
\newblock \doi{10.1109/JSEN.2021.3124275}.

\bibitem[Singh et~al.(2023)Singh, Pandey, et~al.]{singh2023benchmarking}
Govind Singh, Anshul Pandey, et~al.
\newblock Benchmarking and security considerations of wi-fi ftm for ranging in iot devices.
\newblock In \emph{Proceedings of Cyber-Physical Systems and Internet of Things Week 2023}, pages 67--71. 2023.

\bibitem[Subedi and Pyun(2020)]{subedi2020survey}
Santosh Subedi and Jae-Young Pyun.
\newblock A survey of smartphone-based indoor positioning system using rf-based wireless technologies.
\newblock \emph{Sensors}, 20\penalty0 (24):\penalty0 7230, 2020.

\bibitem[Sugiyama et~al.(2023)Sugiyama, Kobayashi, and Chujo]{sugiyama2023study}
Yuichiro Sugiyama, Kentaro Kobayashi, and Wataru Chujo.
\newblock A study on indoor drone positioning using {{Wi-Fi RTT}} ranging.
\newblock In \emph{2023 APWCS}, pages 1--5, {Tainan city, Taiwan}, 9 2023. {IEEE}.
\newblock \doi{10.1109/APWCS60142.2023.10234042}.

\bibitem[Sun et~al.(2020)Sun, Wang, et~al.]{sun2020indoor}
Meng Sun, Yunjia Wang, et~al.
\newblock Indoor {{Positioning Tightly Coupled Wi-Fi FTM Ranging}} and {{PDR Based}} on the {{Extended Kalman Filter}} for {{Smartphones}}.
\newblock \emph{IEEE Access}, 8:\penalty0 49671--49684, 2020.
\newblock ISSN 2169-3536.
\newblock \doi{10.1109/ACCESS.2020.2979186}.

\bibitem[Sun et~al.(2022{\natexlab{a}})Sun, Wang, et~al.]{sun2022geomagnetic}
Meng Sun, Yunjia Wang, et~al.
\newblock Geomagnetic {{Positioning-Aided Wi-Fi FTM Localization Algorithm}} for {{NLOS Environments}}.
\newblock \emph{IEEE Communications Letters}, 26\penalty0 (5):\penalty0 1022--1026, 5 2022{\natexlab{a}}.
\newblock ISSN 1089-7798, 1558-2558, 2373-7891.
\newblock \doi{10.1109/LCOMM.2022.3155929}.

\bibitem[Sun et~al.(2022{\natexlab{b}})Sun, Wang, et~al.]{sun2022simultaneous}
Meng Sun, Yunjia Wang, et~al.
\newblock Simultaneous {{WiFi Ranging Compensation}} and {{Localization}} for {{Indoor NLoS Environments}}.
\newblock \emph{IEEE Communications Letters}, 26\penalty0 (9):\penalty0 2052--2056, 8 2022{\natexlab{b}}.
\newblock ISSN 1089-7798, 1558-2558, 2373-7891.
\newblock \doi{10.1109/LCOMM.2022.3187208}.

\bibitem[Sun et~al.(2022{\natexlab{c}})Sun, Wang, et~al.]{sun2022smartphonebased}
Meng Sun, Yunjia Wang, et~al.
\newblock Smartphone-based {{WiFi FTM Fingerprinting Approach}} with {{Map-aided Particle Filter}}.
\newblock In \emph{2022 {{IEEE}} IPIN}, pages 1--8, {Beijing, China}, 8 2022{\natexlab{c}}. {IEEE}.
\newblock ISBN 978-1-72816-218-8.
\newblock \doi{10.1109/IPIN54987.2022.9918110}.

\bibitem[Sun et~al.(2024)]{sun2024smartphone}
Meng Sun et~al.
\newblock Smartphone-based wifi rtt/rss/pdr/map indoor positioning system using particle filter.
\newblock \emph{IEEE Transactions on Instrumentation and Measurement}, 2024.

\bibitem[Tan et~al.(2022)Tan, Ren, Yang, and Chen]{tan2022commodity}
Sheng Tan, Yili Ren, Jie Yang, and Yingying Chen.
\newblock Commodity wifi sensing in ten years: Status, challenges, and opportunities.
\newblock \emph{IEEE Internet of Things Journal}, 9\penalty0 (18):\penalty0 17832--17843, 2022.
\newblock \doi{10.1109/JIOT.2022.3164569}.

\bibitem[Tekler et~al.(2020)Tekler, Low, Gunay, Andersen, and Blessing]{tekler2020}
Zeynep~Duygu Tekler, Raymond Low, Burak Gunay, Rune~Korsholm Andersen, and Lucienne Blessing.
\newblock A scalable bluetooth low energy approach to identify occupancy patterns and profiles in office spaces.
\newblock \emph{Building and Environment}, 171:\penalty0 106681, 2020.
\newblock ISSN 0360--1323.
\newblock \doi{https://doi.org/10.1016/j.buildenv.2020.106681}.

\bibitem[Tian et~al.(2020)]{tian2020secure}
Bin Tian et~al.
\newblock 11az secure ltf design.
\newblock IEEE 802.11 TGaz IEEE 802.11-20/0836-00, Qualcomm, 2020.
\newblock https://mentor.ieee.org/802.11/dcn/20/11-20-0836-00-00az-11az-secure-ltf-design.pptx.

\bibitem[Van~Marter et~al.(2024)Van~Marter, Ben-Shachar, Alpert, et~al.]{van2024multi}
Jayson~P Van~Marter, Matan Ben-Shachar, Yaron Alpert, et~al.
\newblock A multi-channel approach and testbed for centimeter-level wifi ranging.
\newblock \emph{IEEE Journal of Indoor and Seamless Positioning and Navigation}, 2024.

\bibitem[Vur and Jathe(2024)]{vur2024portable}
Burak Vur and Nicolas~andothers Jathe.
\newblock {A Portable Localization System for Dynamic AGV Positioning in Indoor Warehouses}.
\newblock In \emph{International Conference on Dynamics in Logistics}, pages 293--305. Springer, 2024.

\bibitem[Wachter(2018)]{wachter2018normative}
Sandra Wachter.
\newblock Normative challenges of identification in the internet of things: Privacy, profiling, discrimination, and the gdpr.
\newblock \emph{Computer law \& security review}, 34\penalty0 (3):\penalty0 436--449, 2018.

\bibitem[Wan et~al.(2022)Wan, Duan, et~al.]{wan2022self}
Qiao Wan, Xiaoqi Duan, et~al.
\newblock Self-calibrated multi-floor localization based on wi-fi ranging/crowdsourced fingerprinting and low-cost sensors.
\newblock \emph{Remote Sensing}, 14\penalty0 (21):\penalty0 5376, 2022.

\bibitem[Wang and Luan(2022)]{wang2022toward}
Hao Wang and Hao Luan.
\newblock Toward reliable localization with a single unaided receiver by aoa.
\newblock In \emph{2022 IEEE WoWMoM}, pages 319--324. IEEE, 2022.

\bibitem[Wang et~al.(2021)Wang, Cai, et~al.]{wang2021experimentations}
Lu~Wang, Xiaodong Cai, et~al.
\newblock Experimentations and analysis on indoor positioning through fusion with inertial sensors and dynamically calibrated wi-fi ftm ranging.
\newblock In \emph{2021 IEEE Sensors}, pages 1--4. IEEE, 2021.

\bibitem[Wang and Want(2017)]{wang2017relay}
Wei Wang and Roy Want.
\newblock Relay threat model for tgaz.
\newblock IEEE 802.11 TGaz IEEE 802.11-17/1118-03, Google, 2017.
\newblock https://mentor.ieee.org/802.11/dcn/17/11-17-1118-03-00az-relay-threat-model-for-tgaz.pptx.

\bibitem[Wang et~al.(2023)Wang, Nikseresht, et~al.]{wang2023enabling}
Wenpeng Wang, Fateme Nikseresht, et~al.
\newblock Enabling {{Ubiquitous Occupancy Detection}} in {{Smart Buildings}}: {{A WiFi FTM-Based Approach}}.
\newblock In \emph{2023 DCOSS-IoT}, pages 256--260, {Pafos, Cyprus}, 6 2023. {IEEE}.
\newblock \doi{10.1109/DCOSS-IoT58021.2023.00051}.

\bibitem[Wang et~al.(2024)Wang, Cao, et~al.]{wang2024survey}
Yizheng Wang, Zhimin Cao, et~al.
\newblock Survey of deep learning applications in wifi localization.
\newblock In \emph{2024 IEEE IAEAC}, volume~7, pages 517--522. IEEE, 2024.

\bibitem[Wang et~al.(2022)Wang, Yang, and Wang]{wang2022adaptive}
Zhongyuan Wang, Zhenyu Yang, and Zijian Wang.
\newblock An {{Adaptive Indoor Positioning Method Using Multisource Information Fusion Combing Wi-Fi}}/{{MM}}/{{PDR}}.
\newblock \emph{IEEE Sensors Journal}, 22\penalty0 (6):\penalty0 6010--6018, 3 2022.
\newblock ISSN 1530-437X, 1558-1748, 2379-9153.

\bibitem[Want and Raissinia.()]{80211-promo2}
R.~Want and A.~Raissinia.
\newblock Redefining wi-fi based location with ieee 802.11az.
\newblock URL \url{https://engagestandards.ieee.org/80211az-webinar-register.html.html}.
\newblock Accessed on 2025-04-24.

\bibitem[Wu et~al.(2022)Wu, Chen, et~al.]{wu2022indoor}
Yuan Wu, Ruizhi Chen, et~al.
\newblock Indoor positioning based on tightly coupling of {{PDR}} and one single {{Wi-Fi FTM AP}}.
\newblock \emph{Geo-spatial Information Science}, pages 1--16, 6 2022.
\newblock ISSN 1009-5020, 1993-5153.
\newblock \doi{10.1080/10095020.2022.2072776}.

\bibitem[Wu et~al.(2025)Wu, He, Li, Jian, Yu, Chen, and Chen]{Wu2025WiFi}
Yuan Wu, Mengyi He, Wei Li, Izzy~Yi Jian, Yue Yu, Liang Chen, and Ruizhi Chen.
\newblock Wi-fi fine time measurement–principles, applications, and future trends: A survey.
\newblock \emph{Information Fusion}, 118:\penalty0 102992, 2025.
\newblock ISSN 1566-2535.

\bibitem[Xu et~al.(2019)Xu, Chen, et~al.]{xu2019locating}
Shihao Xu, Ruizhi Chen, et~al.
\newblock Locating {{Smartphones Indoors Using Built-In Sensors}} and {{Wi-Fi Ranging With}} an {{Enhanced Particle Filter}}.
\newblock \emph{IEEE Access}, 7:\penalty0 95140--95153, 2019.
\newblock ISSN 2169-3536.
\newblock \doi{10.1109/ACCESS.2019.2927387}.

\bibitem[Yang et~al.(2021)Yang, Cabani, and Chafouk]{yang2021survey}
Tian Yang, Adnane Cabani, and Houcine Chafouk.
\newblock A survey of recent indoor localization scenarios and methodologies.
\newblock \emph{Sensors}, 21\penalty0 (23):\penalty0 8086, 2021.

\bibitem[Yang et~al.(2024)Yang, Chen, et~al.]{yang2024positioning}
Yang Yang, Mingzhe Chen, et~al.
\newblock Positioning using wireless networks: Applications, recent progress and future challenges.
\newblock \emph{arXiv preprint arXiv:2403.11417}, 2024.

\bibitem[Yang et~al.(2013)Yang, Zhou, and Liu]{yang2013from}
Zheng Yang, Zimu Zhou, and Yunhao Liu.
\newblock From rssi to csi: Indoor localization via channel response.
\newblock \emph{ACM Computing Surveys}, 46\penalty0 (2):\penalty0 1--32, 2013.

\bibitem[Yu et~al.(2019)Yu, Chen, et~al.]{yu2019robust}
Yue Yu, Ruizhi Chen, et~al.
\newblock A {{Robust Dead Reckoning Algorithm Based}} on {{Wi-Fi FTM}} and {{Multiple Sensors}}.
\newblock \emph{Remote Sensing}, 11\penalty0 (5):\penalty0 504, 3 2019.
\newblock ISSN 2072-4292.
\newblock \doi{10.3390/rs11050504}.

\bibitem[Yu et~al.(2020{\natexlab{a}})Yu, Chen, et~al.]{yu2020precise}
Yue Yu, Ruizhi Chen, et~al.
\newblock Precise 3-{{D Indoor Localization Based}} on {{Wi-Fi FTM}} and {{Built-In Sensors}}.
\newblock \emph{IEEE Internet of Things Journal}, 7\penalty0 (12):\penalty0 11753--11765, 12 2020{\natexlab{a}}.
\newblock ISSN 2327-4662, 2372-2541.
\newblock \doi{10.1109/JIOT.2020.2999626}.

\bibitem[Yu et~al.(2020{\natexlab{b}})Yu, Chen, et~al.]{yu2020wi}
Yue Yu, Ruizhi Chen, et~al.
\newblock Wi-fi fine time measurement: Data analysis and processing for indoor localisation.
\newblock \emph{The Journal of Navigation}, 73\penalty0 (5):\penalty0 1106--1128, 2020{\natexlab{b}}.

\bibitem[Yu et~al.(2022{\natexlab{a}})Yu, Chen, et~al.]{yu2022hwps}
Yue Yu, Ruizhi Chen, et~al.
\newblock H-{{WPS}}: {{Hybrid Wireless Positioning System Using}} an {{Enhanced Wi-Fi FTM}}/{{RSSI}}/{{MEMS Sensors Integration Approach}}.
\newblock \emph{IEEE Internet of Things Journal}, 9\penalty0 (14):\penalty0 11827--11842, 7 2022{\natexlab{a}}.
\newblock ISSN 2327-4662, 2372-2541.

\bibitem[Yu et~al.(2022{\natexlab{b}})Yu, Chen, et~al.]{yu2022precise}
Yue Yu, Ruizhi Chen, et~al.
\newblock Precise {{3D Indoor Localization}} and {{Trajectory Optimization Based}} on {{Sparse Wi-Fi FTM Anchors}} and {{Built-In Sensors}}.
\newblock \emph{IEEE Transactions on Vehicular Technology}, 71\penalty0 (4):\penalty0 4042--4056, 4 2022{\natexlab{b}}.
\newblock ISSN 0018-9545, 1939-9359.

\bibitem[Yu et~al.(2023)Yu, Zhang, et~al.]{yu2023intelligent}
Yue Yu, Yi~Zhang, et~al.
\newblock Intelligent {{Fusion Structure}} for {{Wi-Fi}}/{{BLE}}/{{QR}}/{{MEMS Sensor-Based Indoor Localization}}.
\newblock \emph{Remote Sensing}, 15\penalty0 (5):\penalty0 1202, 2 2023.
\newblock ISSN 2072-4292.
\newblock \doi{10.3390/rs15051202}.

\bibitem[Yu et~al.(2022{\natexlab{c}})]{yu2022ap}
Yue Yu et~al.
\newblock Ap detector: Crowdsourcing-based approach for self-localization of wi-fi ftm stations.
\newblock \emph{The International Archives of the Photogrammetry, Remote Sensing and Spatial Information Sciences}, 46:\penalty0 249--254, 2022{\natexlab{c}}.

\bibitem[Yuen et~al.(2022)Yuen, Bie, et~al.]{yuen2022wi}
Brosnan Yuen, Yifeng Bie, et~al.
\newblock Wi-fi and bluetooth contact tracing without user intervention.
\newblock \emph{IEEE Access}, 10:\penalty0 91027--91044, 2022.

\bibitem[Zand et~al.(2019)Zand, Romme, et~al.]{zand2019high}
Pouria Zand, Jac Romme, et~al.
\newblock A high-accuracy phase-based ranging solution with bluetooth low energy (ble).
\newblock In \emph{2019 IEEE WCNC}, pages 1--8. IEEE, 2019.

\bibitem[Zeng et~al.(2020)Zeng, Wang, and Liu]{zeng2020driver}
Xiaolu Zeng, Beibei Wang, and KJ~Ray Liu.
\newblock Driver arrival sensing for smart car using wifi fine time measurements.
\newblock In \emph{2020 APSIPA ASC}, pages 41--45. IEEE, 2020.

\bibitem[Zhou et~al.(2023)Zhou, Wu, et~al.]{zhou2023wifi}
Baoding Zhou, Zhiqian Wu, et~al.
\newblock Wi-{{Fi RTT}}/{{Encoder}}/{{INS-Based Robot Indoor Localization Using Smartphones}}.
\newblock \emph{IEEE Transactions on Vehicular Technology}, 72\penalty0 (5):\penalty0 6683--6694, 5 2023.
\newblock ISSN 0018-9545, 1939-9359.
\newblock \doi{10.1109/TVT.2023.3234283}.

\bibitem[Zola and {Martin-Escalona}(2021)]{zola2021ieee}
Enrica Zola and Israel {Martin-Escalona}.
\newblock {{IEEE}} 802.11mc {{Ranging Performance}} in a {{Real NLOS Environment}}.
\newblock In \emph{2021 WiMob}, pages 377--384, {Bologna, Italy}, 10 2021. {IEEE}.
\newblock ISBN 978-1-66542-854-5.
\newblock \doi{10.1109/WiMob52687.2021.9606323}.

\bibitem[Zola and Martin-Escalona(2025)]{zola2025assessing}
Enrica Zola and Israel Martin-Escalona.
\newblock Assessing the impact of the burst size in the ftm ranging procedure in cots wi-fi devices.
\newblock \emph{Computer Communications}, 229:\penalty0 107980, 2025.

\bibitem[Zubow et~al.(2022)Zubow, Laskos, and Dressler]{zubow2022ftmns3}
Anatolij Zubow, Christos Laskos, and Falko Dressler.
\newblock {{FTM-ns3}}: {{WiFi Fine Time Measurements}} for {{NS3}}.
\newblock In \emph{2022 WONS}, pages 1--7, {Oppdal, Norway}, 3 2022. {IEEE}.
\newblock ISBN 978-3-903176-46-1.
\newblock \doi{10.23919/WONS54113.2022.9764460}.

\bibitem[Zubow et~al.(2023)Zubow, Laskos, and Dressler]{zubow2023simulation}
Anatolij Zubow, Christos Laskos, and Falko Dressler.
\newblock Toward the simulation of {{WiFi Fine Time}} measurements in {{NS3}} network simulator.
\newblock \emph{Computer Communications}, 210:\penalty0 35--44, 10 2023.
\newblock ISSN 01403664.
\newblock \doi{10.1016/j.comcom.2023.07.028}.

\end{thebibliography}

\end{document}